\documentclass[a4paper,11pt]{article}
\pdfoutput=1
\usepackage[utf8]{inputenc}

\usepackage[UKenglish]{babel}
\usepackage{booktabs}
\usepackage{csquotes}
\usepackage{jheppub}
\usepackage[T1]{fontenc}
\usepackage{listings}
\usepackage{lmodern}
\usepackage[binary-units=true,detect-weight=true,group-digits=false,table-format=4.2]{siunitx}
\usepackage{xcolor}
\usepackage[percent]{overpic}
\usepackage{mathcomp}
\usepackage{tabularx}
\usepackage{textcomp}

\newcommand{\alphas}{\alpha_\mathrm{s}}
\DeclareSIUnit\permille{\text{\textperthousand}}

\preprint{Nikhef-2020-026 TIF-UNIMI-2020-22}

\title{PineAPPL: combining EW and QCD corrections\\
  for fast evaluation of LHC processes}

\author[a]{S. Carrazza,}
\author[b]{E.R. Nocera,}
\author[a]{C. Schwan}
\author[a]{and M. Zaro}

\affiliation[a]{Tif Lab, Dipartimento di Fisica, 
Universit\`a di Milano and INFN, Sezione di Milano, 20133 Milano, Italy}
\affiliation[b]{Nikhef Theory Group, Science Park 105, 1098 XG Amsterdam, 
The Netherlands}

\emailAdd{stefano.carrazza@mi.infn.it}
\emailAdd{e.nocera@nikhef.nl}
\emailAdd{christopher.schwan@mi.infn.it}
\emailAdd{marco.zaro@mi.infn.it}

\abstract{We introduce \textsc{PineAPPL}, a library that produces
  fast-interpolation grids of physical cross sections, computed with a 
  general-purpose Monte Carlo generator, accurate to fixed order in the
  strong, electroweak, and combined strong--electroweak couplings. We
  demonstrate this unique ability, that distinguishes \textsc{PineAPPL} from
  similar software available in the literature, by interfacing it to
  \textsc{MadGraph5\_aMC@NLO}. We compute fast-interpolation grids,
  accurate to next-to-leading order in the strong and electroweak couplings,
  for a representative set of LHC processes for which EW corrections may have
  a sizeable effect on the accuracy of the corresponding theoretical
  predictions. We formulate a recommendation on the format of the experimental
  deliverables in order to consistently compare them with computations
  that incorporate EW corrections, and specifically to determine
  parton distribution functions to the same accuracy.}

\arxivnumber{2008.12789}

\begin{document}

\maketitle
\flushbottom

\section{Introduction}
\label{sec:introduction}

With the recent completion of Run II, the Large Hadron Collider (LHC) has 
accumulated data from an integrated luminosity of approximately 
\SI{150}{\per\femto\barn}~\cite{Mangano:2020icy}. This represents only a small fraction of
the anticipated \SI{3000}{\per\femto\barn} that will eventually be recorded in the
forthcoming twenty years of LHC operation. Nevertheless the statistical 
uncertainty of the data has already shrunk to unprecedentedly small values,
typically \SI{1}{\percent} or less, a fact that will allow for precision tests of the
Standard Model (SM) and for indirect searches of New Physics only if 
theoretical predictions become comparatively precise. This entails the 
computation of additional higher-order contributions to the fixed-order 
perturbative expansion, on the one hand, and an increasingly
sophisticated determination of the Parton Distribution Functions (PDFs) of the 
proton~\cite{Gao:2017yyd,Ethier:2020way}, on the other hand. 

In the first respect, because Quantum Chromodynamics (QCD) dominates the 
interactions occurring within colliding protons at the LHC, much effort 
has been devoted to the computation of higher-order QCD corrections: 
fully-differential next-to-leading order (NLO) results, possibly matched to a
parton shower, are currently automated in various general-purpose
Monte Carlo generators~\cite{Gleisberg:2008ta,Alwall:2014hca,Bellm:2015jjp}
(see also ref.~\cite{Buckley:2011ms} for a review),
while an increasing number of next-to-next-to-leading order (NNLO) predictions 
are becoming available for processes with various degrees of inclusiveness
(see e.g.\ ref.~\cite{Amoroso:2020lgh} and references therein). The computation
of higher-order corrections in the electroweak (EW) and combined QCD+EW theory has
also witnessed a comprehensive progress. Frameworks 
were developed in which the QCD and EW couplings are simultaneously treated as 
small parameters in the perturbative expansion, and the computation of 
theoretical predictions, accurate to NLO in both (including multi-coupling
QCD--EW terms), is automated~\cite{Kallweit:2014xda,Biedermann:2017yoi,Frederix:2018nkq}. For an extensive and recent review, see ref.~\cite{Denner:2019vbn}.

In the second respect, contemporary PDF
sets~\cite{Harland-Lang:2014zoa,Ball:2017nwa,Hou:2019efy}
incorporate a significant amount of LHC data, which is analysed with NNLO QCD 
theory by default. No EW corrections are systematically included in the 
theoretical description of the experimental observables to which PDFs are 
optimised, except for QED effects if a photon PDF is 
determined~\cite{Martin:2004dh,Ball:2013hta,Schmidt:2015zda,Manohar:2016nzj,Manohar:2017eqh,Bertone:2017bme,Harland-Lang:2019pla}. The QED contribution is
included in the matrix element (at the lowest order for the LHC
processes considered) and in the DGLAP splitting
functions at leading order~\cite{Bertone:2013vaa}. Inclusion of NLO QCD--QED
combined corrections in the evolution is in principle also
possible~\cite{deFlorian:2015ujt}, and should be required to analyse the data
within consistent perturbative accuracy if higher-order EW and QCD corrections
are also included in partonic cross sections.
The resulting relative PDF uncertainty --- which accounts only for the
uncertainty of the data and of residual methodological inefficiencies inherent to 
each PDF determination --- can be as low as \SI{1}{\percent} at the EW
scale~\cite{Ball:2017nwa}. Theoretical uncertainties, possibly of comparable 
size (e.g.\ from missing higher-order terms in the QCD perturbative
expansion), have started to be represented 
into PDF uncertainties only very 
recently~\cite{AbdulKhalek:2019bux,AbdulKhalek:2019ihb}.

The two respects are intertwined, as they both concur to determine the accuracy
of the theoretical predictions that are matched to the precision of the data.
In particular, taking advantage of the automation pioneered
in refs.~\cite{Kallweit:2014xda,Biedermann:2017yoi,Frederix:2018nkq},
perturbative corrections that arise from the simultaneous expansion in both the
QCD and EW couplings should start to be
incorporated in calculations for LHC processes as standard. The reason is
twofold. First, one expects NNLO QCD and NLO EW corrections to be of
comparable size because, at the EW scale, the QCD and EW running couplings 
become similar, $\alphas^2\sim \alpha$. If NNLO QCD corrections are included
by default in the computations, NLO EW corrections should be
taken into account as well. Second, the virtual exchange of soft or 
collinear weak bosons leads to Sudakov 
logarithms~\cite{Denner:2000jv,Denner:2001gw}
(see also ref.~\cite{Denner:2019vbn} and references therein),
which can make the coefficients of the EW series grow faster than 
their QCD counterparts. This behaviour is relevant in
phase-space regions associated with large mass scales (roughly of the order
of a \si{\tera\electronvolt}), where several LHC data sets (e.g.\ Z-boson
transverse-momentum distributions) enter both the validation of the SM and the search
for new physics.

The consistent inclusion of QCD and EW corrections in precision computations for
LHC processes entails
the solution of two separate problems. First, a problem of efficiency:
fast-interpolation grids should be constructed, whereby partonic matrix 
elements, accurate to NLO QCD+EW, are precomputed in such a way that the 
numerical convolution with generic input PDFs can be efficiently approximated
by means of interpolation techniques. Such grids are essential
whenever the evaluation of the hadronic cross section needs to 
be performed a large number of times, as is the case in the evaluation of
scale variations or of PDF fits. While two formats already exist for
these grids, \textsc{APPLgrid}~\cite{Carli:2010rw} and
\textsc{fastNLO}~\cite{Kluge:2006xs,Wobisch:2011ij,Britzger:2012bs}, none of them supports the inclusion of EW
corrections nor the interface to a Monte Carlo generator accurate to 
NLO QCD+EW\@. Second, a problem of consistency: the way in which EW effects may
(or may not) be folded into the data varies across different experimental 
analyses, a fact that challenges their consistent theoretical interpretation. 
Examples are the subtraction of background processes which should not be 
considered as such (e.g.\ the $t$-channel photon-induced component in
neutral-current Drell--Yan, which is not a separate process beyond leading
order) or of just a part of the EW effects (e.g.\ multiple-photon
radiation from light particles in the final state of neutral- or charged-current
Drell--Yan, especially with electrons). Be that as it may, if EW effects are
systematically included in theoretical predictions, they should not be subtracted
from experimental results, otherwise they will be double counted.

In this paper we address the first of these two problems. Specifically, 
we develop \textsc{PineAPPL}, a library that allows any user to generate
fast-interpolation grids, accurate to any fixed order in the QCD and
EW couplings. The library supports variations of the factorisation
and renormalisation scales, and can be extended to include
resummation, and matching with a photon- and/or parton-shower.
The grids in the new \textsc{PineAPPL} format complement those
(accurate to fixed order in the strong coupling only) that can be generated
in the \textsc{APPLgrid} and \textsc{FastNLO} formats. The \textsc{PineAPPL}
library is interfaced to \textsc{MadGraph5\_aMC@NLO} (\textsc{mg5\_aMC}
henceforth), with which it has been developed and tested,
however it can be easily used with any Monte Carlo generator, 
e.g.\ \textsc{SHERPA}~\cite{Biedermann:2017yoi}. In this respect,
\textsc{PineAPPL} extends to EW corrections the scope of 
\textsc{aMCfast}~\cite{Bertone:2014zva} and
\textsc{MCgrid}~\cite{DelDebbio:2013kxa,Bothmann:2015dba}.

The paper is organised as follows. In section~\ref{sec:pineappl} we introduce
\textsc{PineAPPL}, we describe its features, we illustrate its operation, and we
assess its performance. In section~\ref{sec:results} we validate
\textsc{PineAPPL} and demonstrate its capabilities by computing fast-interpolation grids, accurate to NLO QCD and NLO QCD+EW, for a representative
set of LHC processes for which EW corrections may
have a sizeable effect on the accuracy of the theoretical predictions.
In section~\ref{sec:doublecounting} we try to detail in a more
comprehensive manner the double-counting problem sketched above, the solution of which,
however, remains beyond the scope of the current work. We provide our
conclusions and an outlook in section~\ref{sec:conclusion}. Examples of
usage and the installation of \textsc{PineAPPL} are provided in
appendix~\ref{app:pineappl}; appendix~\ref{app:lumis} collects the parton
luminosities for each process considered in section~\ref{sec:results}; and
appendix~\ref{app:add_plots} complements some of the results presented
in section~\ref{sec:results}.

\section{PDF-independent storage of phase-space weights with PineAPPL}
\label{sec:pineappl}

In this section we introduce \textsc{PineAPPL}.
We first describe the general scope and features of the library in comparison to \textsc{APPLgrid} and \textsc{fastNLO} in section~\ref{sec:library}.
Section~\ref{sec:multi-coupling-expansion} gives the details and, in particular, describes the problem of computing Monte Carlo weights for cross sections in a multi-coupling expansion; this section may be read first by readers unfamiliar with the programs mentioned previously.
In section~\ref{sec:perturbative-orders} we discuss a few properties of multi-coupling-expanded predictions.

\subsection{The PineAPPL library}
\label{sec:library}

\textsc{PineAPPL} is a new library which stores phase-space weights of a Monte Carlo (MC) integration of a fixed-order calculation independently from the chosen PDFs.
The task of computing predictions for physical observables is therefore split
into two steps: 1) the generation of the \emph{grids}, i.e.\ the files in
which the MC weights are stored, and 2) the convolution of these grids with a set of PDFs.
The advantage of this method is that the time-consuming step 1) has to be performed only once, and that step 2) is reduced to a fast convolution of a given grid with one (or more) PDF set(s).

The convolution is typically done in few seconds or less, a fact that offers at least two applications:
\begin{enumerate}
\item the study of the PDF-dependence of the observables; e.g.\ PDF set comparisons, PDF uncertainty computations, $\alphas$ variations, etc., and
\item the determination of PDF sets themselves; together with the corresponding experimental data, the grids constitute two important ingredients for a PDF fit.
\end{enumerate}
These features are common to both \textsc{PineAPPL} and \textsc{APPLgrid}~\cite{Carli:2010rw} or \textsc{FastNLO}~\cite{Kluge:2006xs,Wobisch:2011ij,Britzger:2012bs}. However, in comparison to the last two pieces of code,
\textsc{PineAPPL} allows one to include also higher-order corrections due to EW, and in general combined QCD--EW, effects for the first time.
Documenting this extension, and how to interface \textsc{PineAPPL} with a general-purpose matrix-element generator, is the main goal of this paper.

In particular, \textsc{PineAPPL} supports the following features.
\begin{itemize}
\item The inclusion of perturbative corrections (fixed-order, i.e.\ without parton-shower matching) with any given set of powers of $\alpha$, $\alphas$, in particular including combined QCD--EW corrections.
\item The support for non-coloured initial-state partons, such as photon-initiated contributions, and, more generally, arbitrary initial-state combinations, e.g.\ leptonic initial states~\cite{Bertone:2015lqa,Buonocore:2020nai}.
\item The estimate of theory uncertainties via variations of the renormalisation and factorisation scale (the electroweak coupling is assumed to be scale-independent, consistently with the most common renormalisation schemes).
\end{itemize}
On a more technical level, we point out the following additional features of \textsc{PineAPPL}.

\begin{itemize}

\item \textsc{PineAPPL} comes with the shell command \texttt{pineappl}, which performs convolutions on the command line, without requiring the user to write a new program.
In addition to convolutions, the shell command can also print how the luminosity function is constructed, which perturbative orders are stored in the grids, their size, the size of each partonic channel, etc., separately for each bin. See appendix~\ref{app:pineappl-demo} for details.
  
\item \textsc{PineAPPL} offers an easy-to-use interface written in the C programming language, that allows MC integrators to read and write \textsc{PineAPPL} grids.
C was chosen because it can be easily interfaced with both Fortran and C++, the two main programming languages in which most MC integrators are written.
The interface consists of roughly thirty functions, among which only a handful are needed in practice.
We also provide and support a Python package based on this C interface.
See appendix~\ref{app:example-program} for an example.

\item \textsc{PineAPPL} has been explicitly interfaced to \textsc{mg5\_aMC} (v3+)~\cite{Alwall:2014hca,Frederix:2018nkq},\footnote{A version of \textsc{mg5\_aMC} interfaced to \textsc{PineAPPL} can be downloaded from \url{https://launchpad.net/mg5amc-pineappl/trunk}, and it will be included as standard starting from the 3.0.4 release of \textsc{mg5\_aMC}.} see appendix~\ref{app:sample-runcard} for an example of a runcard.
The usage of \textsc{mg5\_aMC}(v3+) + \textsc{PineAPPL} is similar to that of \textsc{mg5\_aMC}(v2) + \textsc{aMCfast} + \textsc{APPLgrid}.
While the latter toolchain proved to be sufficiently efficient for NLO QCD calculations of typical kinematic distributions (with a few bins), this is no longer so when taking into account NLO EW calculations and, with more precise data, distributions that are typically more complex and have more bins.
Taking into account these two reasons, we find that the memory usage of \textsc{APPLgrid} is no longer satisfiable.
In particular, using \textsc{APPLgrid} for NLO QCD+EW\footnote{For this a modified version of \textsc{APPLgrid} was developed which added support for photon-initiated processes and EW corrections.} typically requires close to \SI{120}{\giga\byte} for high-mass Drell--Yan with 13 bins (see section~\ref{sec:dy-lepton-pair-production}).
The memory usage is substantially reduced to roughly \SI{1}{\giga\byte} after \enquote{optimisation} of the grids, i.e.\ an optimisation of the grid representation in memory.
However, this requires a two-step procedure: first, to produce an unoptimised grid in order to identify those parts where the cross section is either zero or extremely suppressed; second, after those parts are removed, to fill an optimised grid with a small number of bins.
\textsc{PineAPPL} avoids this by using a more space-efficient representation from the start. This leads to substantially faster run times, in particular for simple processes, because the grids do not need to be optimised 
and their combination is faster.

\end{itemize}

\subsection{Cross sections in a multi-coupling expansion}
\label{sec:multi-coupling-expansion}

Fixed-order partonic cross sections $a + b \to X$ supported by \textsc{PineAPPL} are written as an expansion in powers of the strong coupling $\alphas$, the electromagnetic coupling $\alpha$, and the logarithms of $\xi_\mathrm{R} = \mu_\mathrm{R} / Q$ and $\xi_\mathrm{F} = \mu_\mathrm{F} / Q$,
\begin{multline}
\frac{\mathrm{d} \sigma_{ab}}{\mathrm{d} \mathcal{O}} (x_1, x_2, \mathcal{O}, \xi_\mathrm{R}, \xi_\mathrm{F}) \\
= \sum_{k,l,m,n} \alphas^k \left( \xi_\mathrm{R}^2 Q^2 \right) \alpha^l \log^m ( \xi_\mathrm{R}^2 ) \log^n ( \xi_\mathrm{F}^2 ) W_{ab}^{(k,l,m,n)} \left( x_1, x_2, Q^2, \mathcal{O} \right) \text{.}
\label{eq:expansion}
\end{multline}
This cross section is differential with respect to the observable $\mathcal{O}$, which, in general, is a function of phase space and subject to the usual conditions (soft and collinear safety, etc.).

In experiments, but also for many calculations where the phase-space integration is performed using MC techniques, finite statistics does not allow for the exact reconstruction of the dependence of the cross section on the observable $\mathcal{O}$.
Instead, it is sufficient to approximate the derivative using a piecewise-constant function,
\begin{equation}
W_{ab}^{(k,l,m,n)} \left( x_1, x_2, Q^2, \mathcal{O} \right) \approx \sum_{o=1}^M \frac{\Theta (\mathcal{O}_o^\mathrm{min} \le \mathcal{O} < \mathcal{O}_o^\mathrm{max})}{\mathcal{O}_o^\mathrm{max} - \mathcal{O}_o^\mathrm{min}} w_{ab}^{(k,l,m,n,o)} \left( x_1, x_2, Q^2 \right) \text{,}
\label{eq:differential-cross-section}
\end{equation}
which uses $M$ bins with limits $\{ \mathcal{O}_o^\mathrm{min}, \mathcal{O}_o^\mathrm{max} \}_{o=1}^M$ to partition a finite range of the observable,
\begin{equation}
\mathcal{O}_0^\mathrm{min} < \mathcal{O}_0^\mathrm{max} = \mathcal{O}_1^\mathrm{min} < \ldots < \mathcal{O}_{M-1}^\mathrm{max} = \mathcal{O}_M^\mathrm{min} < \mathcal{O}_M^\mathrm{max} \text{.}
\label{eq:bins-of-diff-xsection}
\end{equation}
The Ellis--Sexton scale $Q^2$, if chosen dynamically, depends on the phase space, however we assume the fractions $\xi_\mathrm{R}$ and $\xi_\mathrm{F}$ to be  phase-space constants in any case.
This allows for variations around the central scale choice $\xi_\mathrm{R} = \xi_\mathrm{F} = 1$, but it does not otherwise allow for arbitrary changes of the scale.
The terms with powers $m > 0$ and $n > 0$ vanish for the central scale choice and are only required for variations of the factorisation and renormalisation scales.
To estimate the perturbative QCD uncertainty --- no EW uncertainty is covered by this method --- one typically uses scale variations. Common prescriptions are
7-point and 9-points scale variations, which evaluate the cross section using
respectively the following values
  \begin{align}
    (\xi_\mathrm{R}, \xi_\mathrm{F})_{\mathrm{7-pt}}
    & \in
    \left\{ \bigl( 1, 1 \bigr), \bigl( \tfrac{1}{2}, \tfrac{1}{2} \bigr), \bigl( 2, 2 \bigr), \bigl( \tfrac{1}{2}, 1 \bigr), \bigl( 1, \tfrac{1}{2} \bigr), \bigl( 2, 1 \bigr), \bigl( 1, 2 \bigr) \right\} \text{;}
    \label{eq:7pt}\\
    (\xi_\mathrm{R}, \xi_\mathrm{F})_{\mathrm{9-pt}}
    & \in
    \left\{ \bigl( 1, 1 \bigr), \bigl( \tfrac{1}{2}, \tfrac{1}{2} \bigr), \bigl( 2, 2 \bigr), \bigl( \tfrac{1}{2}, 1 \bigr), \bigl( 1, \tfrac{1}{2} \bigr), \bigl( 2, 1 \bigr), \bigl( 1, 2 \bigr), \bigl( \tfrac{1}{2}, 2 \bigr), \bigl(2, \tfrac{1}{2} \bigr) \right\}  \text{.}
    \label{eq:9pt}
   \end{align}
The (asymmetric) uncertainties are then given as the minimum and maximum value (the envelope), measured from the central value $(1, 1)$.
As is clear from eq.~\eqref{eq:expansion}, the EW coupling $\alpha$ is assumed not to be a dynamically varying coupling, but instead a constant over phase space.
This, however, includes the most common choices of the coupling, which are (not necessarily in this order), $\alpha (0)$, $\alpha (M_\mathrm{Z})$, and $\alpha_{G_\mu}$.

The problem that \textsc{PineAPPL} solves can now be described: approximately reconstruct the functions
\begin{equation}
w_{ab}^{(k,l,m,n,o)} \left( x_1, x_2, Q^2 \right)
\label{eq:weight-map}
\end{equation}
of eq.~\eqref{eq:differential-cross-section} from a set of $N$ function evaluations for specific momentum fractions, scales, and values of the observable
\begin{equation}
\left\{ x_1^{(i)}, x_2^{(i)}, Q^2_i, \mathcal{O}_i \right\}_{i=1}^N \text{,} \label{eq:phase-space-weights}
\end{equation}
given by the MC integrator together with the corresponding value of the weights, eq.~\eqref{eq:weight-map}.
This problem is solved by finding an appropriate representation of eq.~\eqref{eq:phase-space-weights} and is described in section~\ref{sec:grid-representation}.
Using eq.~\eqref{eq:expansion} and
\begin{multline}
\frac{\mathrm{d} \sigma}{\mathrm{d} \mathcal{O}} (\mathcal{O}, \xi_\mathrm{R}, \xi_\mathrm{F}) \\
= \sum_{a,b} \int_0^1 \mathrm{d} x_1 \int_0^1 \mathrm{d} x_2 \int_{Q^2_\mathrm{min}}^{Q^2_\mathrm{max}} \mathrm{d} Q^2 \, f_a (x_1, \xi_\mathrm{F}^2 Q^2) f_b (x_2, \xi_\mathrm{F}^2 Q^2) \sigma_{ab} (x_1, x_2, Q^2, \xi_\mathrm{R}, \xi_\mathrm{F}) \text{,}
\label{eq:pineappl-convolution}
\end{multline}
\textsc{PineAPPL} can then quickly calculate hadronic cross sections for an arbitrary number of PDF sets and perform scale variations.

Note that we have omitted the dependence of the weights $w$, the observable $\mathcal{O}$, and the scales $\mu_\mathrm{F}, \mu_\mathrm{R}, Q^2$ on the specific kinematics for which they are computed.
Indeed, beyond LO, different kinematic contributions have to be considered (in ref.~\cite{Bertone:2014zva}, for example, they are labelled with an index $\alpha$, see eq.~(12) therein).
In the FKS subtraction scheme~\cite{Frixione:1995ms,Frixione:1997np} employed in \textsc{mg5\_aMC} one type of kinematics for each counterterm (soft, collinear, and soft-collinear) is needed, however this is not the general case.
In Catani--Seymour subtraction~\cite{Catani:1996jh,Catani:2002hc}, for example, different dipoles have different phase spaces and therefore different scales.
\textsc{PineAPPL} remains completely blind to this fact, and a consistent treatment is ensured by filling each event into a grid using the \emph{numerical value} of $Q^2$.

\subsubsection{Grid representations}
\label{sec:grid-representation}

We now explain the details of how the phase-space weights $w_{ab}$ in eq.~\eqref{eq:bins-of-diff-xsection} are represented.

\paragraph{4-tuples.}
A straightforward representation is given by 4-tuples, i.e.\ a list of the momentum fractions $x_1$ and $x_2$, the scale $Q^2$, and the phase-space weight $w$ for each phase-space point; 4-tuples are sufficient to reconstruct the differential cross sections.
For each combination $(a,b,k,l,m,n,o)$ (see section~\ref{sec:multi-coupling-expansion} for their definition) we save the following 4-tuples,
\begin{equation}
\left\{ x_1^i, x_2^i, Q^2_i, w^{(k,l,m,n,o)}_{ab} (x_1^i, x_2^i, Q^2_i, \mathcal{O}_i) \right\}_{i=1}^N \text{.} \label{eq:four-tuples}
\end{equation}
The reconstruction of the differential cross sections are then done by simply multiplying the phase-space weights $w$ with PDFs evaluated with the correct arguments given in the 4-tuple and summing over all indices $a$, $b$, $k$, $l$, $m$, $n$, $o$, and $i$.

The 4-tuple representation is very easy to implement and test.
Furthermore, it reproduces the exact numerical value that is also calculated by the MC integrator.
However, the price one has to pay is the size of the 4-tuples.
For example, NLO QCD+EW Drell--Yan lepton-pair production (see section~\ref{sec:pineappl-example} and section~\ref{sec:dy-lepton-pair-production}) needs \SI{159}{\mega\byte} of storage for a target precision of \SI{1}{\percent} of the integrated cross section.
While this is an acceptable size, increasing the precision by an order of magnitude would require roughly 100 times the size, due to the Monte Carlo convergence that goes as $1/\sqrt{N}$ with $N$ being the number of 4-tuples.
With increasing size also the speed of the convolution degrades, because it basically becomes bound by the speed with which the 4-tuples can be read from disk.
However, due to the uncompressed nature of this representation it can serve as an intermediate format to develop and quickly cross check more space-efficient representations, one of which we will discuss next.

\paragraph{Lagrange-interpolation grid.}
A different strategy is to partition a subset $H$ of the $(x_1, x_2, Q^2)$ space,
\begin{equation}
H = [x_\mathrm{min},x_\mathrm{max}]^2 \times [Q^2_\mathrm{min}, Q^2_\mathrm{max}] \ni (x_1, x_2, Q^2)
\end{equation}
along each axis into a small number of bins and to insert the phase-space weights $w$ into the corresponding discrete bin.
Using the bin centres and their values one already has a straightforward representation of eq.~\eqref{eq:weight-map}, but given a finite number of bins this approach usually yields an insufficient approximation for the cross section.
Increasing the number of bins improves the precision, but it also increases the space requirements.
This problem in turn is solved using interpolation methods, which increase the precision using the same number of bins.

We use the \enquote{Lagrange-interpolation grid} method presented in ref.~\cite{Carli:2010rw} with the parameters published in ref.~\cite{Bertone:2014zva}, which give sufficient precision (see section~\ref{sec:results}).
For the sake of completeness, the following summarises the interpolation algorithm.

This method first maps $(x_1, x_2, Q^2) \mapsto (y_1, y_2, \tau)$, with
\begin{equation}
y(x) = 5 (1-x) - \log x \text{,} \qquad \tau (Q^2) = \log \log \frac{Q^2}{(\SI{0.25}{\giga\electronvolt})^2} \text{.}
\label{eq:maps}
\end{equation}
The function $y(x)$ maps events with large $x$ effectively linearly and small $x$ effectively logarithmically onto $y$.
This reflects our knowledge of PDFs, which behave differently in those two regions, and thereby increases the precision of the interpolation.
For the convolution of a grid with a PDF set also the inverse functions are needed, which are
\begin{equation}
x(y) = \frac{1}{5} \operatorname{W}_0 (5 \exp (5-y)) \text{,} \qquad Q^2 (\tau) = (\SI{0.25}{\giga\electronvolt})^2 \exp (\exp (\tau)) \text{,}
\end{equation}
where $\operatorname{W}_0 (x)$ is (the principle branch of) the Lambert W function or product logarithm, which satisfies the relation $\operatorname{W} (x) \exp (\operatorname{W} (x)) = x$.

Following ref.~\cite{Carli:2010rw} (see eq.~(17) therein), we furthermore divide the weights, before filling them into the grid, by the function
\begin{equation}
\omega (x_1, x_2) = \left( \frac{\sqrt{x_1}}{1 - 0.99 x_1} \right)^3 \left( \frac{\sqrt{x_2}}{1 - 0.99 x_2} \right)^3 \text{.}
\end{equation}
This flattens the interpolated function in the region $x \to 1$, where the PDFs are small and tend towards zero, and enhances the function in the small-$x$ region.
The effect of this step is an improvement of the precision that depends on the initial states and the process, but it can be as large as a factor of \numrange{10}{100} (one or two more correct digits compared to the MC result).
Before performing a convolution this step is inverted by simply multiplying the interpolated grid values with $\omega (x_1, x_2)$.

The final step is filling the weights into the grid, which maps the variables $(y_1, y_2, \tau)$ onto the 3-dimensional Lagrange-interpolation grid with $N_y = 50$ points in each $y$ direction and $N_\tau = 30$ points in the $\tau$ direction.
The interpolation orders $s_y$ and $s_\tau$ are 3 for each dimension, and only the subspace $[\num{2e-7},1] \times [\num{2e-7},1] \times [\num{e2},\num{e6}] \subset H$\footnote{In ref.~\cite{Bertone:2014zva} the upper limit for $Q$ is given as \SI{3162}{\giga\electronvolt} (which corresponds to $Q_\mathrm{max}^2 \approx \SI{e7}{\giga\electronvolt}$), but in the code we found the value $Q_\mathrm{max}^2 = \SI{e6}{\giga\electronvolt\squared}$.} is mapped.

\begin{figure}[!t]
\centering
\parbox{0.6\textwidth}{\includegraphics{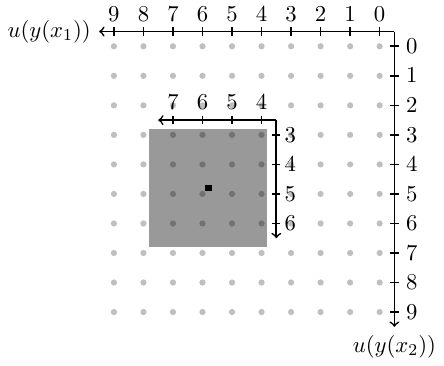}}
\begin{tabular}{ll}
\toprule
$u_1$/$u_2$ & $x_1$/$x_2$ \\
\midrule
0 & \num{1.00e0} \\
1 & \num{6.36e-1} \\
2 & \num{3.20e-1} \\
3 & \num{9.96e-2} \\
4 & \num{1.57e-2} \\
5 & \num{1.74e-3} \\
6 & \num{1.81e-4} \\
7 & \num{1.87e-5} \\
8 & \num{1.93e-6} \\
9 & \num{2.00e-7} \\
\bottomrule
\end{tabular}
\caption{Example of a 2-dimensional $10 \times 10$ grid, which is being filled at the location marked with the small black square at $(5.8,4.8)$.
Each side of the grey square starting at $k_i = 4$ and $k_j = 3$ has a length of $N_y + 1 = 4$.
This square marks the grid values (grey dots) that are being updated using eq.~\eqref{eq:interpolation}.
The table on the right-hand side gives the parton-momentum fractions for each grid point according to eq.~\eqref{eq:maps}.
Note that for $u \in [0, 3]$ the values are roughly linearly distributed, then logarithmically.}
\label{fig:grid}
\end{figure}

To illustrate the filling step we give an example in figure~\ref{fig:grid}, where, for simplicity, we have chosen a static scale, so that we do not need to interpolate in the $\tau$ direction, and where we also limited the number of grid points to $N_y = 10$.
Each grid point has a numerical value $a_{i,j}$ associated, and the set of all numerical values $\{ a_{i,j} \}$ for all grid indices $(i,j) \in [0,N_y) \times [0,N_y)$ constitute \enquote{the grid}.
Inserting a specific weight $W = w (x_1, x_2) / \omega(x_1, x_2)$ into the grid is shown in figure~\ref{fig:grid} as a small black square, inside the larger grey one.
We have defined the grid points at specific positions, but the points given by the MC will land somewhere between them.
The interpolation order $s_y$ then defines a square with length $s_y + 1$ around its centre $(u(y(x_1)),u(y(x_2)))$, given by the MC\@.
All grid points with indices $(i,j)$ covered by the grey square are then updated according to the following formula,
\begin{equation}
a_{i,j} \leftarrow a_{i,j} + I_i(u(y(x_1))) I_j(u(y(x_2))) W(x_1,x_2) / \omega(x_1,x_2) \text{,}
\label{eq:interpolation}
\end{equation}
with the Lagrange basis functions,
\begin{equation}
I_i (u) = \prod_{\substack{k=k_i \\ k \neq i}}^{k_i + s_y} \frac{u-k}{i-k} \text{,}
\end{equation}
where the product runs over all indices of the grid points covered by the grey square in figure~\ref{fig:grid}, starting from the smallest index in the square, $k_i$ and $k_j$.
Finally, we remap
\begin{equation}
u(y) = \frac{y-y_\text{min}}{\Delta y} \text{,} \quad
\end{equation}
using $y_\text{min} = y(x_\text{max})$ and $y_\text{max} = y(x_\text{min})$ and the grid spacing $\Delta y = (y_\text{max} - y_\text{min})/(N_y-1)$, so that the integer part of $u(y)$ gives the grid point index, e.g.\ $u(y_\text{min}) = 0$ and $u(y_\text{max}) = N_y - 1$, and the fractional part of $u(y)$ gives the relative location between the nearest grid points.

\subsection{Perturbative orders}
\label{sec:perturbative-orders}

In section~\ref{sec:multi-coupling-expansion} we labelled the different perturbative orders using the indices $k,l,m,n$; their values are process specific.
However, in general we define as leading order (LO) the set of contributions for all possible initial states $a b$ that lead to the same final state $X$, for which the sum of the coupling exponents in eq.~\eqref{eq:expansion} is smallest, i.e.\ $k + l = p$, where $p = \min (k+l)$.
This number is usually determined by the number of external particles.
For many processes there is only one contribution at LO (in terms of $k$ and $l$), but this is not true in general. Indeed, when a process has multiple quark lines, colourless (photons, \dots) and coloured particles (gluons, \ldots) can be exchanged between them, making it possible to have more than one combination of $\alphas$ and $\alpha$.
A typical example at the LHC is (on-shell) top-pair production, which has three different contributions at LO: $\mathcal{O} (\alpha^2)$, $\mathcal{O} (\alphas \alpha)$, and $\mathcal{O} (\alpha^2)$.
Each contribution receives a higher-order correction with an additional power of $\alphas$ or $\alpha$, which in general leads to at least two next-to-leading order (NLO) corrections.
The correction with the largest power in $\alphas$ can be unambiguously called \enquote{the} QCD correction, and the one with the largest power in $\alpha$ \enquote{the} EW correction.
All remaining corrections are of combined type, meaning that, in general, they cannot be attributed to either one of strong or electroweak origin. However, for the sake of simplicity and with a slight abuse of notation, it is customary to call
NLO QCD (EW) corrections those
corrections of order $\alphas$ ($\alpha$) times the couplings of the LO contribution 
with the largest power of $\alphas$. In the example above, NLO QCD and EW corrections to top pair production will denote
respectively those at $\mathcal{O} (\alphas^3 )$ and $\mathcal{O} (\alphas^2 \alpha)$. In this paper, in particular
when discussing results in sec.~\ref{sec:results}, we will explicitly list the orders that are considered in the cross
section at LO, NLO QCD and NLO QCD+EW accuracy.

Due to the typical sizes of the couplings $\alphas^2 \sim \alpha$, it is naively expected that within the same order, i.e.\ for fixed $k + l$, terms with larger powers $\alphas^k$ dominate over those with smaller powers (and larger powers of $\alpha^l$).
In practice, however, this naive expectation is not always true due to dynamic effects.
Some examples are vector-boson scattering processes~\cite{Biedermann:2017bss,Denner:2019tmn}, top-pair production with a W boson and four-top production~\cite{Frederix:2017wme}, and Higgs production with a bottom-pair~\cite{Pagani:2020rsg}.

\subsubsection{Example: Drell--Yan lepton-pair production at the LHC}
\label{sec:pineappl-example}

To give an example of eq.~\eqref{eq:expansion}, the following shows Drell--Yan lepton-pair production up to terms at NLO (with some arguments suppressed for the phase-space weights):
\begin{equation}
\begin{split}
\sigma_{ab}
    &= \alpha^2 W_{ab}^{(0,2,0,0)} \\
    &+ \alphas \left( \xi_\mathrm{R}^2 Q^2 \right) \alpha^2 W_{ab}^{(1,2,0,0)} (Q^2) + \alphas \left( \xi_\mathrm{R}^2 Q^2 \right) \log (\xi_\mathrm{F}^2) \alpha^2 W_{ab}^{(1,2,0,1)} \\
    &+ \alpha^3 W_{ab}^{(0,3,0,0)} (Q^2) + \log (\xi_\mathrm{F}^2) \alpha^3 W_{ab}^{(0,3,0,1)} \text{.}
\end{split}
\end{equation}
The term in the first line is the LO term, the second line shows the NLO QCD correction, and the third line the NLO EW correction.
Note that all terms depend on the renormalisation scale only indirectly through $\alphas$, because 1) higher-order terms in $\alpha$ never generate a renormalisation scale dependence (in the $\alpha$ schemes that are valid according to section~\ref{sec:multi-coupling-expansion}) and 2) higher-order QCD corrections only introduce an explicit renormalisation scale dependence in counterterms with vertices with more than two gluons.
At NLO these terms are not present for this process so that terms proportional to $\log (\xi_\mathrm{R}^2)$ vanish.
Both NLOs, however, have contributions from a collinear counterterm that depends on the factorisation scale.
Since this process has a single LO, combined QCD--EW corrections first appear at next-to-next-to-leading order (NNLO), which include the QCD correction at $\mathcal{O} (\alphas^2 \alpha^2)$, the EW correction $\mathcal{O} (\alpha^4)$, and a single combined correction at $\mathcal{O} (\alphas \alpha^3)$.

Note that all initial states have to be taken into account that lead to the same final state.
This includes the photon--photon initial state, which appears already at LO\@.
In the corresponding Feynman diagrams all particles are colourless, so that the photon--photon initiated contributions only receive EW corrections.
The EW corrections also introduce quark--photon contributions, in analogy of QCD corrections introducing quark--gluon contributions.

\section{Validation and interpretation of PineAPPL grids}
\label{sec:results}

In this section we demonstrate the capabilities of the \textsc{PineAPPL}
library by interfacing it to \textsc{mg5\_aMC}, and by
computing fast-interpolation grids, accurate to NLO QCD and NLO QCD+EW,
for a common set of LHC processes in which EW corrections may sizeably
affect the accuracy of the theoretical predictions.
In order to consider realistic kinematics for these
processes, we rely on a representative set of LHC measurements. Our aim is twofold. First, we want to validate the results
obtained with \textsc{PineAPPL}; second, we want to assess the
impact of the EW corrections for the specific experimental setups. We describe
the settings employed for the computations, the corresponding results for
each process, and possible implications for the determination of PDFs. It is worth to mention
that, in those phase-space regions where both NLO QCD and EW corrections are sizeable, the NNLO
correction at $\mathcal O (\alphas \alpha)$ is also expected to be important, and has either to
be included, when it is known exactly or at least in some approximate way, or to be accounted for 
as an extra uncertainty. We leave this aspect to future studies.

\subsection{Computational settings}
\label{subsec:computational_settings}

\textsc{mg5\_aMC} makes it possible
to compute predictions including NLO QCD and EW corrections for arbitrary processes
in an automated manner. It employs the FKS subtraction scheme~\cite{Frixione:1995ms,Frixione:1997np}
as automated in \textsc{MadFKS}~\cite{Frederix:2009yq,Frederix:2016rdc} to deal with IR singularities. One-loop
amplitudes are computed by \textsc{MadLoop}~\cite{Hirschi:2011pa}, which employs different
numerical techniques~\cite{Passarino:1978jh,Davydychev:1991va,Denner:2005nn,Ossola:2006us,Cascioli:2011va,Mastrolia:2012bu} implemented in the corresponding computer libraries~\cite{Ossola:2007ax,Peraro:2014cba,Hirschi:2016mdz,Denner:2016kdg}. Matching
with parton showers is available only for pure-QCD corrections via the MC@NLO method~\cite{Frixione:2002ik}, and 
will not be employed in the following.

We generate each process by means of the Universal FeynRules Output
(UFO)~\cite{Degrande:2011ua} model {\tt loop\_qcd\_qed\_sm\_Gmu},
included as standard in {\sc mg5\_aMC}. It contains the UV and $R_2$
counterterms relevant to NLO QCD and EW corrections, the latter in the
$\overline{G}_\mu$ scheme. The model features five massless quark flavours,
sets the CKM matrix equal to the identity, and is compatible with the usage of
the complex mass (CM) scheme~\cite{Denner:1999gp,Denner:2005fg} for all massive particles, see
ref.~\cite{Frederix:2018nkq} for details. We use this scheme
for all processes that involve only massless particles in the final state.
The photon is always considered as part of the proton in the initial state and
of any hadronic jet produced in the final state: 
 in particular, photon-induced (PI) contributions are consistently included at LO and NLO.\footnote{We employ the $\overline{G}_\mu$ scheme also for the QED coupling entering vertices involving initial-state photons, see section~4.3.3 of ref.~\cite{Denner:2019vbn}.}
We use a PDF set that contains a photon PDF, namely
{\tt NNPDF31\_nlo\_as\_0118\_luxqed}~\cite{Bertone:2017bme}. We evaluate the PDF
uncertainty associated to the theoretical predictions a posteriori, that is,
we convolve the fast-interpolation grid generated with {\sc PineAPPL} with
each member in the PDF set, and we compute the associated standard deviation. Monte Carlo
weights are stored as Lagrange-interpolation grids.

The central values of the renormalisation and factorisation scales, $\mu_\mathrm{R}$ and
$\mu_\mathrm{F}$, are chosen in a process-specific way, as discussed in sec.~\ref{subsec:processes_and_measurements}. In order to estimate the missing higher-order uncertainty,
we allow the events to be reweighted on-the-fly in the Monte Carlo generation upon scale
variations, with the technique presented in ref.~\cite{Frederix:2011ss}. To this purpose, we use the default \textsc{mg5\_aMC}
implementation, whereby the factorisation and renormalisation scales
are varied down to a factor $1/2$ and up to a factor $2$, and the envelope
from the nine-point scale variations is constructed,
see equation~\eqref{eq:9pt}. However, we note that
\textsc{PineAPPL} allows the user to determine the envelope with any point
prescription, see appendix~\ref{app:pineappl-demo} for an example.

The values of the relevant physical parameters are chosen as
\begin{equation}
\begin{aligned}
M_\mathrm{W} &= \SI{80.419}{\giga\electronvolt} \text{,} \quad &
\Gamma_\mathrm{W} &= \SI{2.09291}{\giga\electronvolt} \text{,} &
m_\mathrm{t} &= \SI{172.5}{\giga\electronvolt} \text{,} \quad \\
M_\mathrm{Z} &= \SI{91.176}{\giga\electronvolt} \text{,} \quad &
\Gamma_\mathrm{Z} &= \SI{2.49877}{\giga\electronvolt} \text{,} &
G_\mu &= \SI{1.16639e-5}{\per\giga\electronvolt\squared} \text{,} \\
M_\mathrm{H} &= \SI{125}{\giga\electronvolt} \text{,} &
\Gamma_\mathrm{H} &= \SI{4.074680e-03}{\giga\electronvolt} \text{,}
\end{aligned}
\label{eq:parameters}
\end{equation}
where $M_\mathrm{Z}$, $M_\mathrm{W}$, $M_\mathrm{H}$, $m_\mathrm{t}$ are the values of the Z-, W-, Higgs-boson
and top-quark masses, respectively, $\Gamma_\mathrm{Z}$, $\Gamma_\mathrm{W}$, $\Gamma_\mathrm{H}$ are the widths of
the Z, W and Higgs bosons, and $G_\mu$ is the value of the Fermi coupling. The value of the strong
coupling is chosen consistently with the PDF set, $\alphas(M_\mathrm{Z})=0.118$.

The definition of observables and cuts is process-specific, and it follows the corresponding experimental measurements,
see section~\ref{subsec:processes_and_measurements}.
When relevant, final-state photons and massless charged fermions (leptons and light quarks) are recombined together 
if they satisfy the condition $\Delta R_{f \gamma}<0.1$, where $\Delta R_{f \gamma}$
is the fermion-photon distance. In this case the
sum of their momenta is assigned to the charged fermion, and the photon is removed
from the event. Kinematic observables and cuts are defined starting from recombined momenta. If we were interested also in jet-related observables,
photons surviving the recombination would have to be clustered together with coloured
partons.\footnote{For issues related to the definition of jets in presence of EW corrections, in particular
    about the fragmentation of partons into photons and vice-versa,
see refs.~\cite{Glover:1993xc,Frederix:2016ost,Denner:2019zfp}.} Finally, although contributions corresponding to the radiation
of a heavy boson are formally of the same perturbative order of the EW corrections, they are not included in our
computations. In fact, while nothing prevents one to include these contributions \emph{a posteriori}, as they are finite,
their impact is either smaller than the one of \enquote{standard} EW corrections, or anyway negligible with respect to
the total cross section (see refs.~\cite{Frixione:2014qaa,Frixione:2015zaa,Pagani:2016caq} for some process-specific cases).

\subsection{Results for specific processes and measurements}
\label{subsec:processes_and_measurements}

We focus on the following three processes: Drell--Yan lepton-pair production,
top-quark pair production, and Z-boson (lepton-pair) production with non-zero
transverse momentum in proton-proton collisions. These are some of the most
commonly and most precisely measured processes at the LHC, which are widely
studied to test the SM and/or search for new physics. We therefore expect them
to allow us to clearly show the benefit of being able to make fast and
reliable theoretical predictions accurate to NLO QCD+EW with \textsc{PineAPPL}.
In the following, we shall
present the experimental measurements, the process-specific settings, and
the phenomenological results for each of these processes.

When presenting the results, for each of the processes and measurements considered, we compute differential
cross sections for the observables defined in the experimental analyses in two different
ways: directly, by means of \textsc{mg5\_aMC}, and a posteriori, by convolving
the fast-interpolation grid produced by \textsc{PineAPPL} with the PDF set
specified in section~\ref{subsec:computational_settings}. We
refer to the first result as the MC result, and to the second as
the \textsc{PineAPPL} result. We repeat the computation for theories accurate
to NLO QCD and to NLO QCD+EW, respectively. The corresponding orders of the
strong and EW couplings that we consider are specified for each process. In each case, we determine the
PDF uncertainty (coming from the PDF ensemble), the scale uncertainty (coming
from variations of the factorisation and renormalisation scales), and the Monte
Carlo uncertainty (coming from the finite number of events generated). In this
last respect, we consider by default high-statistics computations, whereby we
require a relative Monte Carlo precision of \SI{0.1}{\permille} on the integrated cross section. While
this choice does not affect the validation of the \textsc{PineAPPL} result
against the MC result, it ensures that the statistical uncertainty of
the computation remains negligible in comparison to the PDF and scale
uncertainties, as we will explicitly demonstrate. This is a desirable feature
to correctly interpret the size of the EW corrections. An example that
validates the \textsc{PineAPPL} result in the case of a low-statistic run is
nevertheless provided in appendix~\ref{app:add_plots}.

Our goal is indeed twofold. On the one hand, we aim to validate the
interpolation grids generated with \textsc{PineAPPL}: to this purpose we shall
verify that the MC and the \textsc{PineAPPL} results are identical up to
numerical inaccuracies due to the grid interpolation. This equivalence must
hold for any choice of renormalisation and factorisation scale and should not
depend on the MC uncertainty of the binned cross section. On the other
hand, we aim to study the size of the EW corrections, in particular with
respect to the kinematics of each process, and to three kinds of uncertainties:
the PDF uncertainty, the scale uncertainty, and the uncertainty of the
experimental data.

We present these comparisons in
figures~\ref{fig:atlaszhighmass49fb}, \ref{fig:cmsdy2d11_bins3456}, \ref{fig:atlastop}, and \ref{fig:cmsZ13TeV}.
The format of the plots is the same across all figures. The first panel
displays the relative difference (in per mille) between the {\sc PineAPPL} and
the MC results for the central scale choice and upper/lower edges of the
scale-uncertainty envelope, for
theories accurate to both NLO QCD and NLO QCD+EW\@. The following three panels
present the theoretical predictions, accurate to either NLO QCD or NLO QCD+EW,
always normalised to the former; on top of the theoretical predictions, the
PDF uncertainty, the scale uncertainty and the Monte Carlo uncertainty are
displayed in turn. The relative uncertainty of the experimental data is
also shown for comparison. We shall now discuss the results for each
process and data set.

\subsubsection{Drell--Yan lepton pair production.}
\label{sec:dy-lepton-pair-production}

\paragraph{Experimental measurements and process features.}
We select the single-differential invariant mass distribution of the
lepton pair, $M_{\ell \bar\ell}$, measured by the ATLAS experiment at a centre-of-mass
energy of \SI{7}{\tera\electronvolt} in the high-mass region
($M_{\ell\bar\ell}>\SI{116}{\giga\electronvolt}$)~\cite{Aad:2013iua}.
We also select the single-differential rapidity distribution, $y_{\ell\bar\ell}$, in slices of
the invariant mass of the lepton pair, $M_{\ell\bar\ell}$,
measured by the CMS experiment at a centre-of-mass energy of
\SI{7}{\tera\electronvolt}~\cite{Chatrchyan:2013tia}.
These measurements are currently included as standard in the
NNPDF3.1~\cite{Ball:2017nwa} and MMHT2014~\cite{Harland-Lang:2014zoa} PDF sets,
although with appropriate kinematic cuts that remove the bins at the largest
values of invariant mass, where EW corrections become sizeable. As explained in
section~\ref{sec:pineappl-example}, 
the process has a single LO, $\mathcal{O}(\alpha^2)$; at NLO, the
QCD contribution is $\mathcal{O}(\alphas\alpha^2)$, while the EW contribution
is $\mathcal{O}(\alpha^3)$. Our NLO QCD computation includes the
$\mathcal{O}(\alpha^2)$ and $\mathcal{O}(\alphas\alpha^2)$ contributions, while
our NLO QCD+EW computation includes the $\mathcal{O}(\alpha^2)$, 
$\mathcal{O}(\alphas\alpha^2)$ and $\mathcal{O}(\alpha^3)$ contributions.
Combined QCD--EW corrections occur only at NNLO~\cite{Buccioni:2020cfi}, and are therefore not
considered here. EW corrections for this process were computed in
refs.~\cite{Baur:2001ze,Arbuzov:2007db,Dittmaier:2009cr,Frederix:2018nkq} (see also~\cite{Alioli:2016fum} and references therein). The process receives contributions
from 13 (35) parton luminosities at NLO QCD (NLO QCD+EW),
see appendix~\ref{app:lumis} for details.

\paragraph{Process-specific settings.}
We use a fixed value for the renormalisation and factorisation scales $\mu_\mathrm{R}=\mu_\mathrm{F}=M_\mathrm{Z}$, where $M_\mathrm{Z}$ is
the mass of the Z boson, for the ATLAS measurement, and the scale
$\mu_\mathrm{R}=\mu_\mathrm{F}=M_{\ell\bar\ell}$, where $M_{\ell\bar\ell}$ is the central value of each
invariant mass slice, for the CMS one. In the case of ATLAS, we require $p_\mathrm{T}^\ell>\SI{25}{\giga\electronvolt}$,
$|\eta_\ell|<2.5$ and $\SI{116}{\giga\electronvolt}<M_{\ell\bar\ell}<\SI{1500}{\giga\electronvolt}$ for the transverse
momentum and the pseudorapidity of each lepton and for the invariant mass of the
lepton pair, respectively. Conversely, in the case of CMS, we require $p_\mathrm{T}^{\ell_1}>\SI{14}{\giga\electronvolt}$, $p_\mathrm{T}^{\ell_2}>\SI{9}{\giga\electronvolt}$,
$|\eta_\ell|<2.4$, $|y_{\ell\bar\ell}|<2.4$ and $\SI{20}{\giga\electronvolt}<M_{\ell\bar\ell}<\SI{1500}{\giga\electronvolt}$
for the transverse momentum and the pseudorapidity of each lepton, and for the
rapidity and the invariant mass of the lepton pair.

\begin{figure}[!t]
    \centering
    \includegraphics[width=0.5\textwidth]{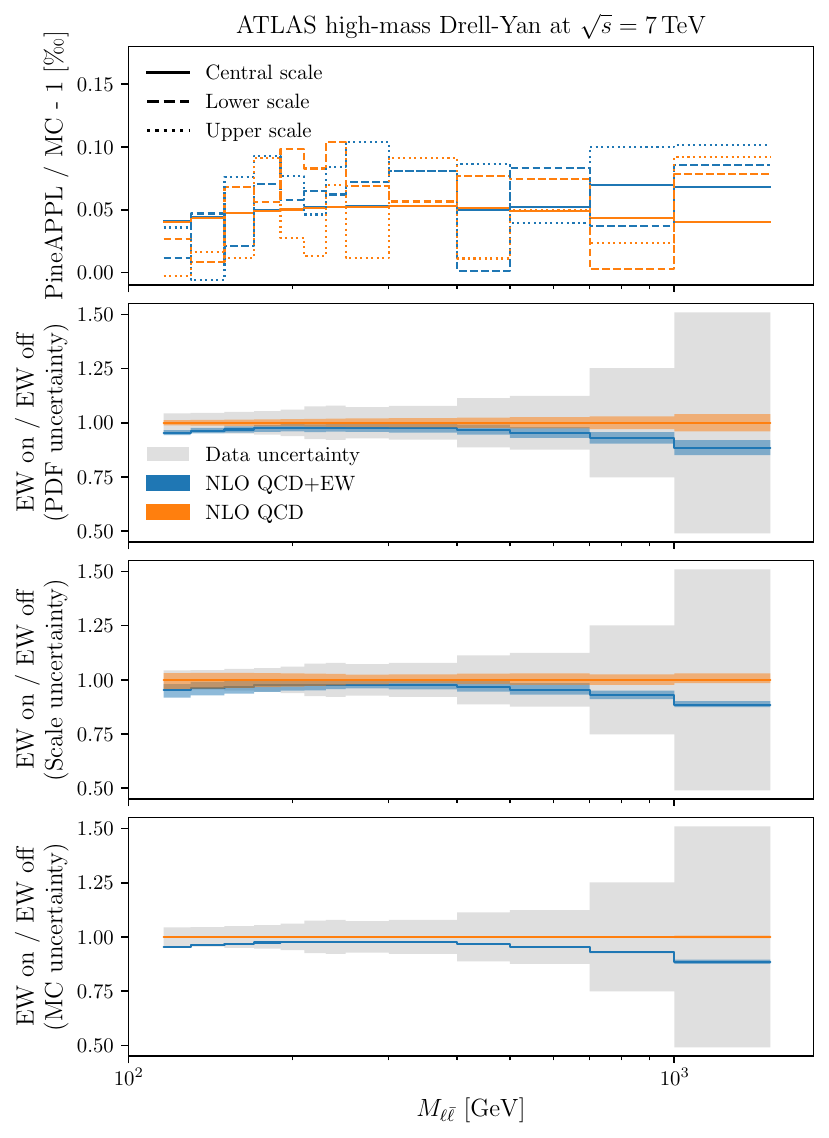}\\
    \caption{Validation and test of the \textsc{PineAPPL} grid for the ATLAS
      Drell--Yan lepton pair measurement in the high-mass region at
      a centre-of-mass energy of \SI{7}{\tera\electronvolt}~\cite{Aad:2013iua}. The first panel
      displays the relative difference (in per mille) between the {\sc PineAPPL}
      and the MC results for the central, upper and lower scale choices,
      for theories accurate to both NLO QCD and NLO QCD+EW\@. The second, third
      and fourth panels present the theoretical predictions, accurate to either
      NLO QCD or NLO QCD+EW, always normalised to the former; on top of the
      theoretical predictions, the PDF uncertainty, the scale uncertainty and
      the Monte Carlo uncertainty are displayed in turn. The relative
      uncertainty of the experimental data is also shown for comparison.}
    \label{fig:atlaszhighmass49fb}
\end{figure}

\paragraph{Numerical results.}
We first consider the single-differential measurement of a lepton-pair
for large invariant masses performed by the ATLAS experiment at \SI{7}{\tera\electronvolt}.
From figure~\ref{fig:atlaszhighmass49fb} we immediately
observe that the validation of the \textsc{PineAPPL} result against the MC
result is successful. The relative difference between the two is of the order of
\SI{0.1}{\permille} at most, with negligible fluctuations across different
invariant mass bins. The agreement is similarly good irrespective of the
perturbative accuracy of the theory (NLO QCD or NLO QCD+EW) or of the scale
choice. As explicitly demonstrated in appendix~\ref{app:add_plots}, the good
agreement is also independent from the numerical precision of the Monte Carlo
run.

The measurement is mainly driven by $\mathrm{q}\bar{\mathrm{q}}$ scattering: specifically, the
leading (next-to-leading) contribution comes from a $\mathrm{u}\bar{\mathrm{u}}$/$\mathrm{c}\bar{\mathrm{c}}$ ($\mathrm{d}\bar{\mathrm{d}}$/$\mathrm{s}\bar{\mathrm{s}}$)
parton luminosity, which accounts for about \SI{55}{\percent} (\SI{49}{\percent}) of the cross section
for the lowest invariant mass bins, and \SI{68}{\percent} (\SI{22}{\percent}) for the largest invariant
mass bins.\footnote{The size of these contributions may depend on the input PDF
  set. Here and in the following, we always quote results obtained from the
  \texttt{NNPDF31\_nlo\_as\_0118\_luxqed} PDF set.}
The PI contribution raises from about \SI{1.3}{\percent} in the lowest bin to
about \SI{3.6}{\percent} in the highest bin.\footnote{For the \texttt{MRST2004qed} PDF set~\cite{Martin:2004dh}, which is used to subtract PI contributions (see section~\ref{sec:doublecounting}), the value in the highest bin is roughly twice as large, \SI{6.9}{\percent} (also in absolute numbers).}
Overall, the EW corrections range between \SI{-5}{\percent}
around $M_{\ell\bar\ell}\lesssim \SI{150}{\giga\electronvolt}$, \SIrange{-2}{-3}{\percent} for intermediate invariant mass
values, $\SI{150}{\giga\electronvolt}\lesssim M_{\ell\bar\ell}\lesssim \SI{700}{\giga\electronvolt}$, and
\SIrange{-6}{-10}{\percent} for the largest invariant mass bin, $M_{\ell\bar\ell}>\SI{1000}{\giga\electronvolt}$,
see figure~\ref{fig:atlaszhighmass49fb}. For this reason, the data points with
$M_{\ell\bar\ell}>\SI{210}{\giga\electronvolt}$ were not included in the NNPDF3.1
analysis~\cite{Ball:2017nwa}.

The NLO QCD+EW corrections always lead to a reduction of the cross section in
comparison to the NLO QCD prediction. The size of this shift is comparable to
the data uncertainty at small values of $M_{\ell\bar\ell}$, and rapidly becomes
negligible with respect to it as the value of the invariant mass increases
and the data uncertainty blows up. This fact suggests a couple of observations
in light of the inclusion of EW corrections in a fit of PDFs. First, the
description of the more precise bins in the low invariant mass range is likely
to change, and will possibly become more accurate should the inclusion of EW
corrections improve the data/theory agreement. Second, the kinematic cut that
excludes any data point at large $M_{\ell\bar\ell}$ can be safely removed: any
shift in the predictions induced by the more accurate NLO QCD+EW theory is
likely to be easily accommodated by the large data uncertainty. However, 
due to the increased Run-II LHC luminosity, data will become more precise.

In comparison to the PDF uncertainty, the size of the EW corrections is
always larger, especially at the boundaries of the distribution. This fact
suggests that, once included in a global fit, EW corrections will make PDFs
more accurate. In comparison to the scale
uncertainty, the size of the EW correction is similar, except for the four bins
at the largest invariant mass, where the latter is significantly larger than
the former. This fact suggests that the impact of NNLO QCD corrections~\cite{Anastasiou:2003yy,Catani:2009sm,Gavin:2010az,Li:2012wna,Boughezal:2016wmq} is comparable
to the one of NLO QCD+EW, except at very large values of the invariant mass,
where the EW correction still dominates. This result stresses the need to
include the EW corrections in order to obtain an accurate description of the
large invariant mass bins. Finally, the Monte Carlo 
uncertainty remains negligible in comparison to the data, PDF and scale
uncertainties, and to the size of the EW correction. Our conclusions should
therefore not be affected by a generation of too few Monte Carlo events.

We then turn our attention to the double-differential measurement performed by
the CMS experiment at \SI{7}{\tera\electronvolt}. For illustrative purposes, we report only four out
of the six invariant mass bins, respectively below the Z-boson mass peak,
$\SI{45}{\giga\electronvolt}<M_{\ell\bar\ell}<\SI{60}{\giga\electronvolt}$, on the Z-boson mass peak,
$\SI{60}{\giga\electronvolt}<M_{\ell\bar\ell}<\SI{120}{\giga\electronvolt}$, above the mass peak,
$\SI{120}{\giga\electronvolt}<M_{\ell\bar\ell}<\SI{200}{\giga\electronvolt}$, and at very high invariant masses,
$\SI{200}{\giga\electronvolt}<M_{\ell\bar\ell}<\SI{1500}{\giga\electronvolt}$, see figure~\ref{fig:cmsdy2d11_bins3456}.
Analogous plots for the remaining low invariant mass bins are collected in
appendix~\ref{app:add_plots}. From figure~\ref{fig:cmsdy2d11_bins3456},
first of all we validate the \textsc{PineAPPL} result: its relative difference
with respect to the MC result is always below a fraction of per mille,
again irrespective of the accuracy of the theory, of the choice of scale, and
of the kinematic bin considered. 

\begin{figure}[!p]
    \centering
    \includegraphics[width=0.46\textwidth]{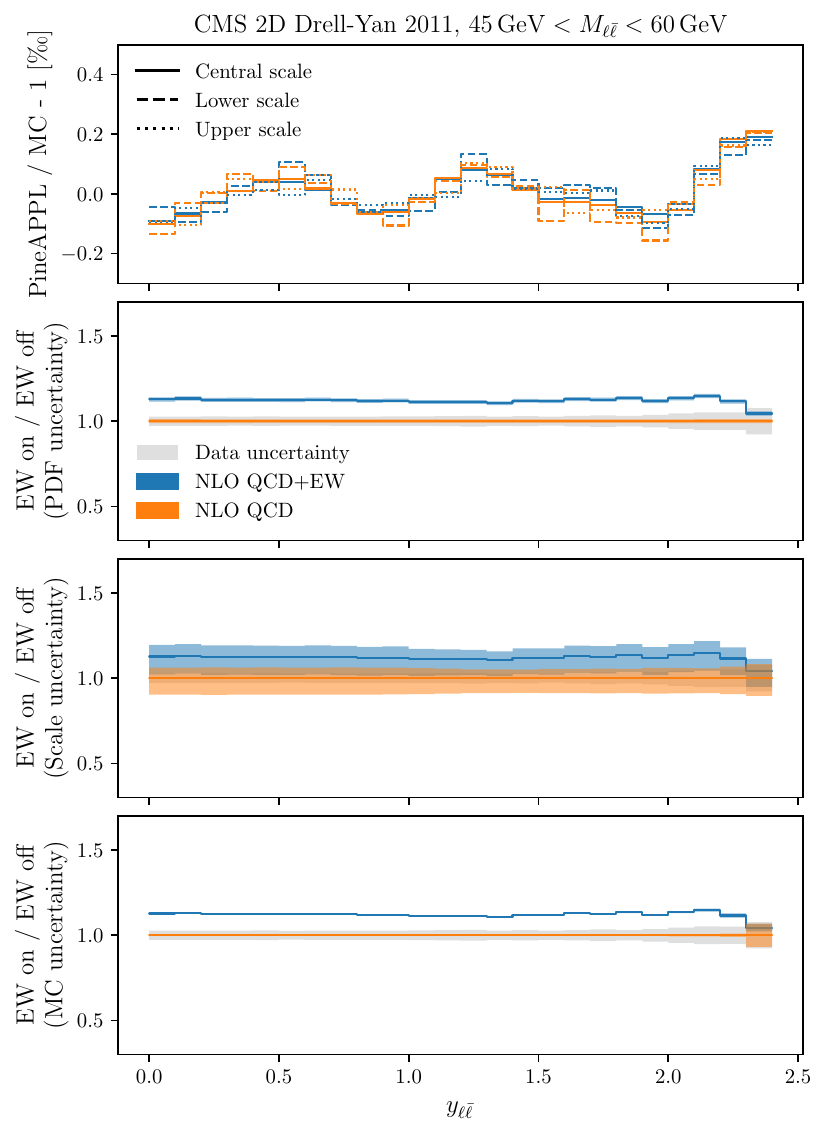}%
    \includegraphics[width=0.46\textwidth]{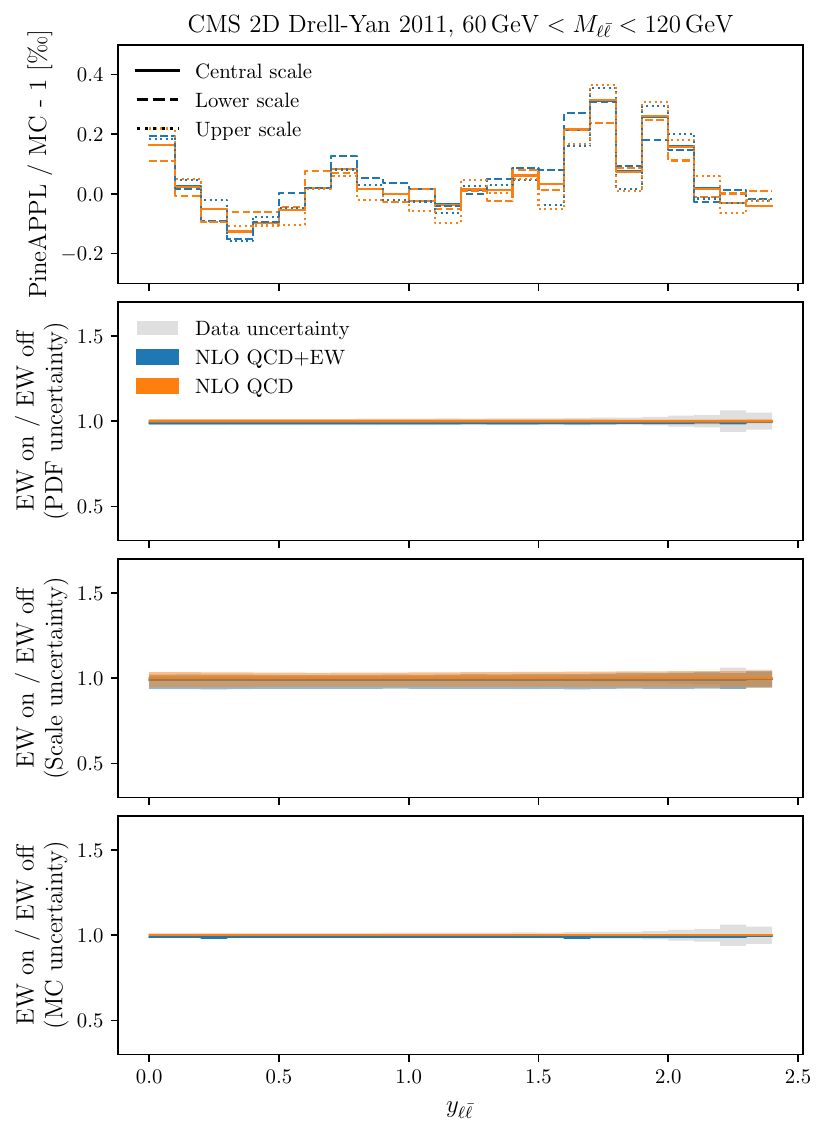}\\
    \includegraphics[width=0.46\textwidth]{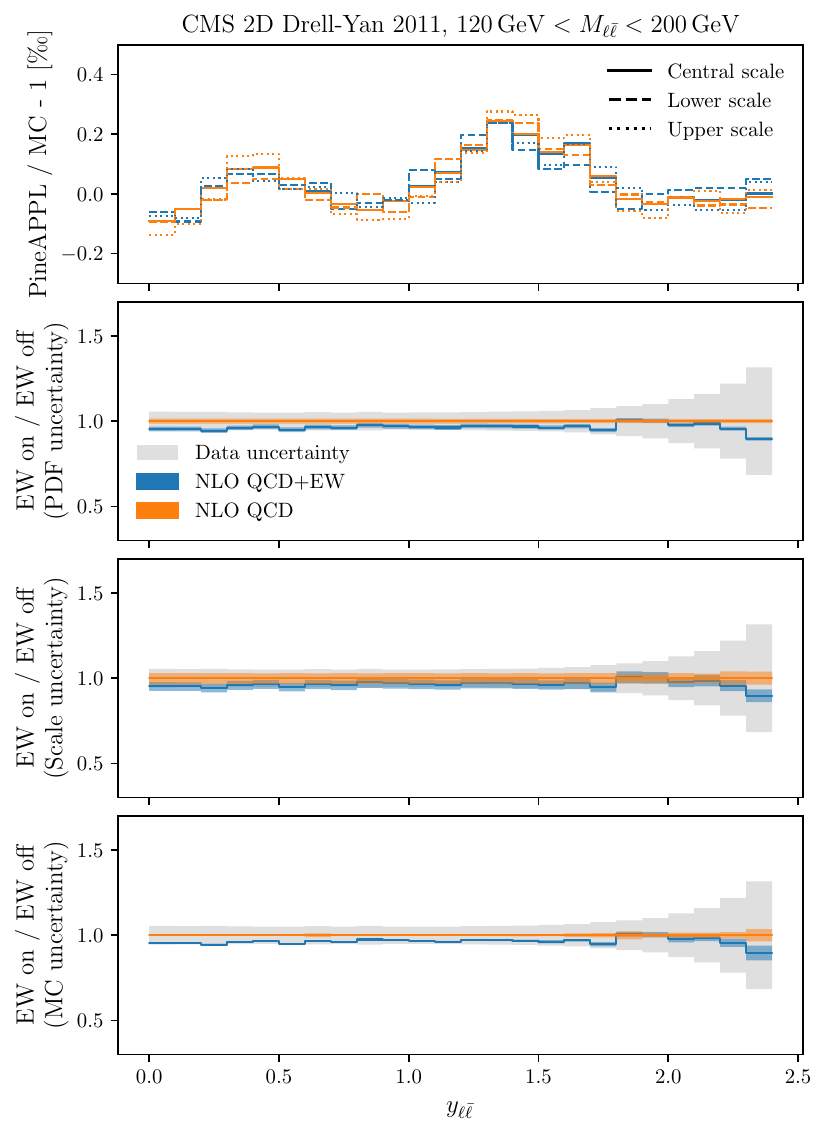}%
    \includegraphics[width=0.46\textwidth]{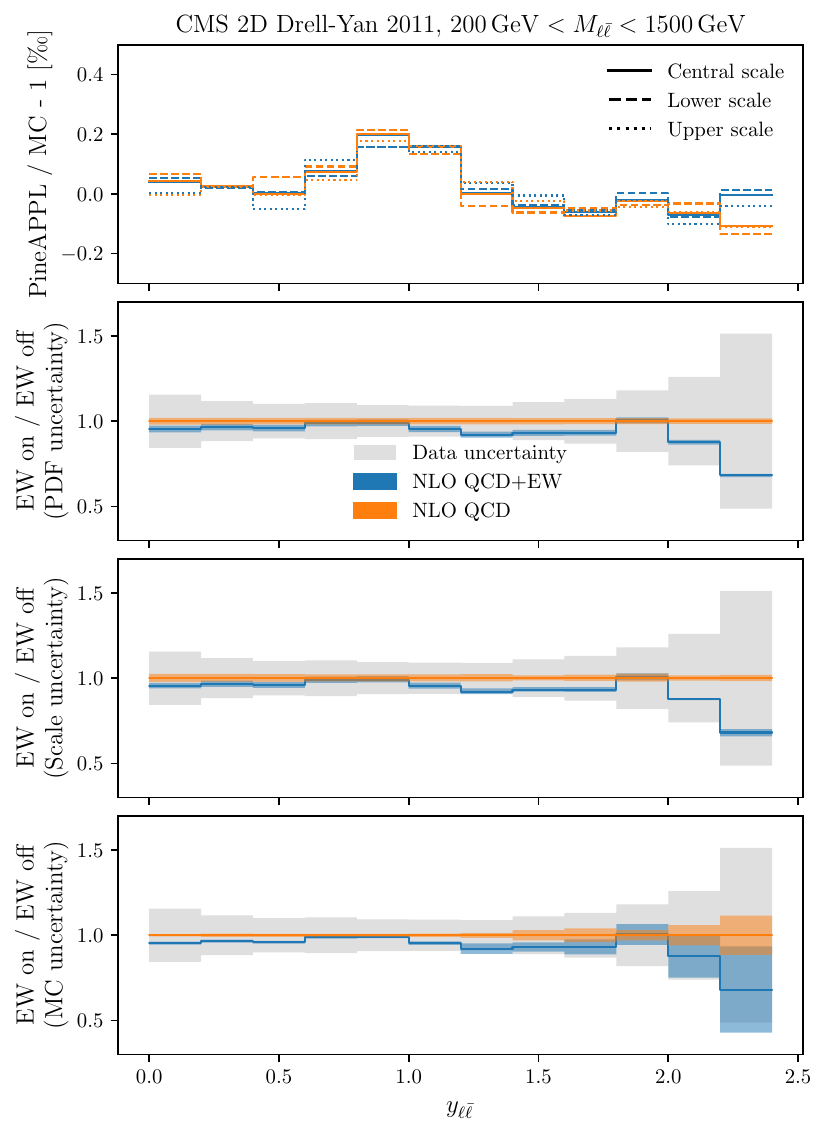}\\
    \caption{Same as figure~\ref{fig:atlaszhighmass49fb}, but for the CMS
      double-differential Drell--Yan lepton pair measurement at a
      centre-of-mass energy of \SI{7}{\tera\electronvolt}~\cite{Chatrchyan:2013tia}. Displayed are
      only four of the six invariant mass bins available, respectively below the
      Z-boson mass peak, $\SI{45}{\giga\electronvolt}<M_{\ell\bar\ell}<\SI{60}{\giga\electronvolt}$, on the Z-boson mass
      peak, $\SI{60}{\giga\electronvolt}<M_{\ell\bar\ell}<\SI{120}{\giga\electronvolt}$, above the mass peak,
      $\SI{120}{\giga\electronvolt}<M_{\ell\bar\ell}<\SI{200}{\giga\electronvolt}$, and at very high invariant masses,
      $\SI{200}{\giga\electronvolt}<M_{\ell\bar\ell}<\SI{1500}{\giga\electronvolt}$.
      Results for the slices $\SI{45}{\giga\electronvolt}<M_{\ell\bar\ell}<\SI{60}{\giga\electronvolt}$ and $\SI{45}{\giga\electronvolt}<M_{\ell\bar\ell}<\SI{60}{\giga\electronvolt}$ can be found in figure~\ref{fig:cmsdy2d11_bins12}.}
    \label{fig:cmsdy2d11_bins3456}
\end{figure}

As in the case of the ATLAS high-mass Drell--Yan measurement, the CMS
measurement is also dominated by $\mathrm{q}\bar{\mathrm{q}}$ scattering. The leading
(next-to-leading) contribution to the $\SI{45}{\giga\electronvolt}<M_{\ell\bar\ell}<\SI{60}{\giga\electronvolt}$ invariant
mass bin comes from
the $\mathrm{u}\bar{\mathrm{u}}$/$\mathrm{c}\bar{\mathrm{c}}$ ($\mathrm{d}\bar{\mathrm{d}}$/$\mathrm{s}\bar{\mathrm{s}}$) parton luminosity, which accounts for about \SI{70}{\percent}
(\SI{22}{\percent}) of the double differential cross section, with small fluctuations across
the rapidity range. The PI contribution decreases from about \SI{4}{\percent} at zero
rapidity to \SI{1.5}{\percent} in the largest rapidity bin. The situation is slightly
different in the $\SI{60}{\giga\electronvolt}<M_{\ell\bar\ell}<\SI{120}{\giga\electronvolt}$ invariant mass bin, where the
leading (next-to-leading) contribution comes instead from the $\mathrm{d}\bar{\mathrm{d}}$/$\mathrm{s}\bar{\mathrm{s}}$
($\mathrm{u}\bar{\mathrm{u}}$/$\mathrm{c}\bar{\mathrm{c}}$) parton luminosity, which accounts for about \SI{60}{\percent} (\SI{44}{\percent}) of the
double differential cross section at small rapidities, and for about \SI{56}{\percent} (\SI{50}{\percent})
at large rapidities. In the remaining two invariant mass bins, the leading
(next-to-leading) contribution comes again from the $\mathrm{u}\bar{\mathrm{u}}$/$\mathrm{c}\bar{\mathrm{c}}$ ($\mathrm{d}\bar{\mathrm{d}}$/$\mathrm{s}\bar{\mathrm{s}}$)
parton luminosity, which accounts for about \SIrange{69}{95}{\percent} (\SIrange{38}{30}{\percent}) and
\SIrange{57}{70}{\percent} (\SIrange{48}{34}{\percent}) of the cross section, respectively for
$\SI{120}{\giga\electronvolt}<M_{\ell\bar\ell}<\SI{200}{\giga\electronvolt}$ and $\SI{200}{\giga\electronvolt}<M_{\ell\bar\ell}<\SI{1500}{\giga\electronvolt}$ in the
corresponding rapidity intervals; PI contributions range between \SIrange{3.7}{0.6}{\percent}
and \SIrange{7.3}{1.6}{\percent} in the two invariant mass bins, respectively, for increasing
rapidity.

The way in which NLO QCD+EW corrections affect the theoretical prediction for
the double differential cross section (with respect to its counterpart accurate
to NLO QCD) depends on the invariant mass bin. In the
$\SI{45}{\giga\electronvolt}<M_{\ell\bar\ell}<\SI{60}{\giga\electronvolt}$ region, they enhance the value of the cross
section by about \SI{11}{\percent} across all the rapidity range. This is mostly due to photon-radiation effects
on events with $M_{\ell\bar\ell}\simeq M_\mathrm{Z}$ at the Born, for which the invariant mass is shifted to lower 
values. In the 
$\SI{60}{\giga\electronvolt}<M_{\ell\bar\ell}<\SI{120}{\giga\electronvolt}$ region, EW corrections suppress the value of the cross
section by about \SI{2}{\percent}, again across all the rapidity range; in the
$\SI{120}{\giga\electronvolt}<M_{\ell\bar\ell}<\SI{200}{\giga\electronvolt}$ bin, the suppression is around \SIrange{4}{5}{\percent}; and in
the $\SI{200}{\giga\electronvolt}<M_{\ell\bar\ell}<\SI{1500}{\giga\electronvolt}$ bin, the suppression increases further to
about \SIrange{6}{7}{\percent} for rapidities $y_{\ell\bar\ell}<2.0$. For this reason, for instance, the data points with
$M_{\ell\bar\ell}>\SI{200}{\giga\electronvolt}$ and $y_{\ell\bar\ell}>2.2$ were not included in the
NNPDF3.1 analysis~\cite{Ball:2017nwa}.

In general, the size of the EW corrections is comparable to or slightly larger
than the data uncertainty, except for the invariant mass bin 
$\SI{45}{\giga\electronvolt}<M_{\ell\bar\ell}<\SI{60}{\giga\electronvolt}$,
where the shift due to the EW correction overshoots the data uncertainty by
about a factor of ten, and at large rapidities, where the shift due to the EW
correction, although it can become large, is always a fraction of the data
uncertainty. Because EW effects are subtracted from the data used
in PDF fits (see section~\ref{sec:doublecounting}), a good agreement between
data and theory is usually achieved without the inclusion of EW corrections.
However, as already observed in the case of the ATLAS measurement, should EW
corrections be included in a fit of PDFs, the latter are likely to become more
accurate: even though the apparent description of the data will not
improve, by including the more precisely predicted bins in the low invariant mass range, PDFs
will resemble more closely the underlying truth. Furthermore, the kinematic cut
that excludes any data point at large invariant mass and/or rapidity can be
safely removed.

In comparison to the PDF uncertainty, the size of the EW corrections is always
larger. We therefore anticipate that, even if the agreement between the more
accurate theory (including EW corrections) and the data will remain the same,
the PDFs will however become overall more accurate. In comparison to the scale
uncertainty, the size of the EW correction is similar, except on the Z-boson
mass peak, $\SI{60}{\giga\electronvolt}<M_{\ell\bar\ell}<\SI{120}{\giga\electronvolt}$, where the scale uncertainty exceeds
the size of the EW correction by about a factor of five. This is due to the choice
of invariant mass window around the Z peak, in which positive and negative EW corrections almost cancel.
Finally, the Monte Carlo uncertainty remains negligible
in comparison to the data, PDF and scale uncertainties, and to the size of the
EW correction, except for a couple of bins at forward/backward rapidity in the highest
invariant mass bins. Improving the Monte Carlo precision will require to
generate a larger number of events, possibly with cuts that select
only the kinematic bins affected by the largest MC uncertainties. If this
turned out to be computationally too expensive, it would be desirable to treat
this uncertainty as an additional theoretical uncertainty in the PDF
fit~\cite{Ball:2018lag}.

\subsubsection{Top-quark pair production.}
\label{sec:toppair}

\paragraph{Experimental measurements and process features.}
We select the single-differential distribution in either the transverse
momentum of the top quark, $p_\mathrm{T}^\mathrm{t}$, or the invariant mass of the top-quark
pair, $m_{\mathrm{t}\bar{\mathrm{t}}}$, measured by the ATLAS and CMS experiments at a centre-of-mass
energy of \SI{8}{\tera\electronvolt}~\cite{Aad:2015mbv,Khachatryan:2015oqa}. These measurements have
been extensively studied in the context of PDF fits in
refs.~\cite{Czakon:2016olj,Bailey:2019yze,Amoroso:2020lgh,Kadir:2020yml} (see
also refs.~\cite{Pagani:2016caq,Czakon:2017wor} for studies related to the photon density) and
included by default in the CT18~\cite{Hou:2019efy} analysis.
Because EW corrections are significantly smaller for distributions differential
in the rapidity of either the top quark or the top-quark
pair~\cite{Czakon:2017wor}, these distributions were preferred for inclusion
in the NNPDF3.1 analysis~\cite{Ball:2017nwa}. The process receives
pure QCD contributions at LO, $\mathcal{O}(\alphas^2)$, and
at NLO, $\mathcal{O}(\alphas^3)$. These orders make up our NLO QCD
computation. The NLO QCD+EW computation includes the $\mathcal{O}(\alphas^2)$
and $\mathcal{O}(\alphas^3)$ QCD contributions, the LO contribution
$\mathcal{O}(\alphas\alpha)$ and the NLO contribution
$\mathcal{O}(\alphas^2\alpha)$.
We do not consider the LO contribution $\mathcal{O}(\alpha^2)$ nor the
NLO contributions $\mathcal{O}(\alphas\alpha^2)$ and $\mathcal{O}(\alpha^3)$, which have been shown to be negligible~\cite{Czakon:2017wor,Frederix:2018nkq}.
EW corrections for this process
were computed in refs.~\cite{Kuhn:2006vh,Bernreuther:2010ny,Hollik:2011ps,Kuhn:2011ri,Bernreuther:2012sx,Pagani:2016caq,Denner:2016jyo,Czakon:2017wor,Czakon:2017lgo,Czakon:2017mmr,Gutschow:2018tuk,Frederix:2018nkq,Czakon:2019bcq,Czakon:2019txp}. The process receives contributions from
7 (37) parton luminosities at NLO QCD (NLO QCD+EW),
see appendix~\ref{app:lumis} for details.

\paragraph{Process-specific settings.}
We employ the following functional form for the renormalisation and factorisation scales;
$\mu_\mathrm{R}=\mu_\mathrm{F}=\sqrt{m_\mathrm{t}^2+(p_\mathrm{T}^\mathrm{t})^2}{\Big /}2$ for the distribution differential
in the transverse momentum of the top quark, and $\mu_\mathrm{R}=\mu_\mathrm{F}=H_\mathrm{T}/4$ for the
distribution differential in the invariant mass of the top-quark pair, where
$H_\mathrm{T}=\sqrt{m_\mathrm{t}^2+(p_\mathrm{T}^\mathrm{t})^2}+\sqrt{m_\mathrm{t}^2+(p_\mathrm{T}^{\bar{\mathrm{t}}})}$, with $m_\mathrm{t}$,
$p_\mathrm{T}^\mathrm{t}$ and $p_\mathrm{T}^{\bar{\mathrm{t}}}$ the mass of the top quark and the transverse momenta
of the top and antitop quarks, respectively. These choices were demonstrated
to maximise the convergence of the perturbative expansion~\cite{Czakon:2016dgf}. No cuts are imposed.

\paragraph{Numerical results.}

\begin{figure}[!t]
    \centering
    \includegraphics[width=0.5\textwidth]{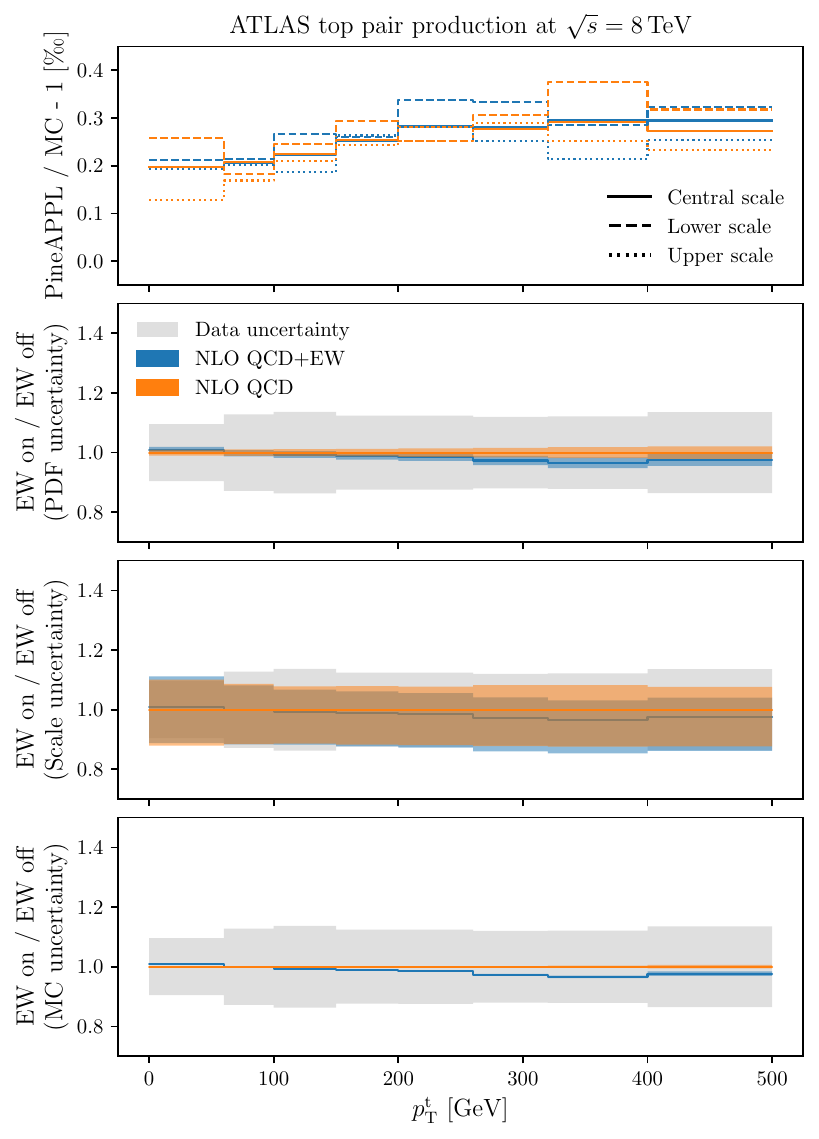}%
    \includegraphics[width=0.5\textwidth]{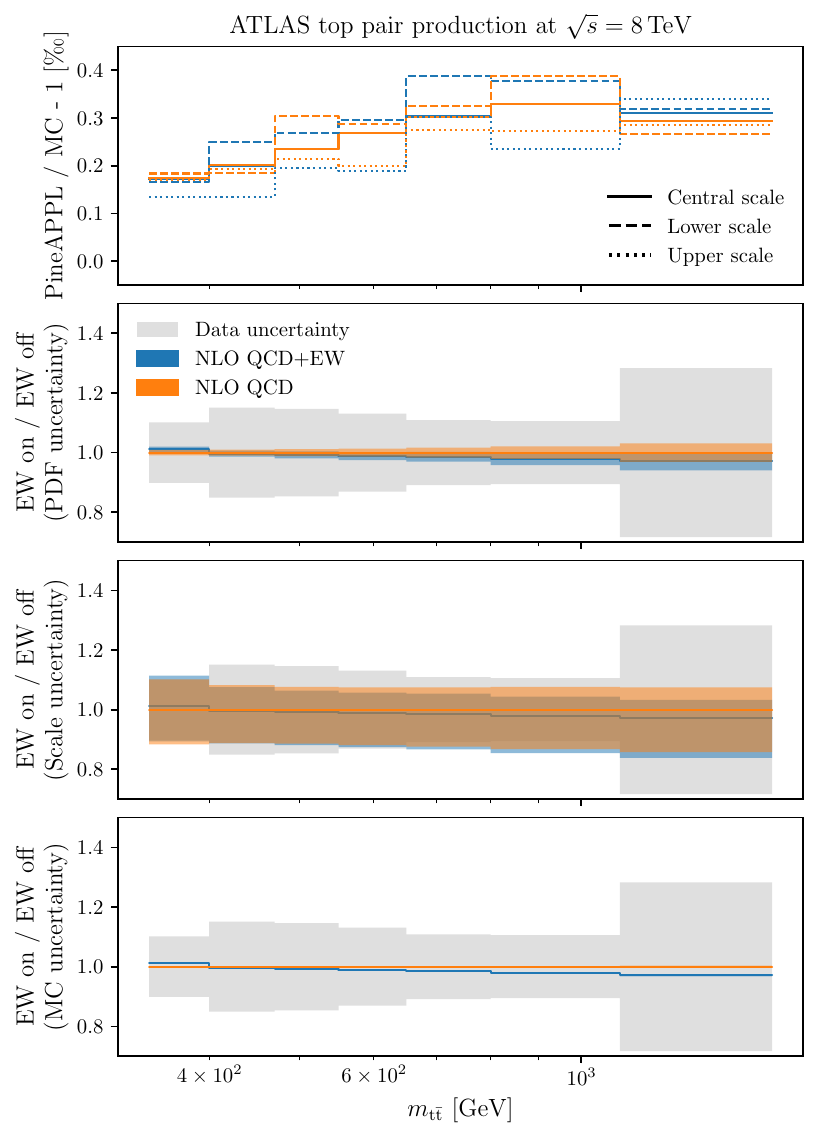}
    \caption{Same as figure~\ref{fig:atlaszhighmass49fb}, but for the ATLAS
      differential top-quark pair measurement at a centre-of-mass energy of
      \SI{8}{\tera\electronvolt}~\cite{Aad:2015mbv}. Displayed are the distributions in the
      transverse momentum of the top quark $p_\mathrm{T}^\mathrm{t}$ (left), and in the invariant
      mass of the top-quark pair $m_{\mathrm{t}\bar{\mathrm{t}}}$ (right).}
    \label{fig:atlastop}
\end{figure}

In figure~\ref{fig:atlastop} we report the
distributions differential in the transverse momentum of the top quark,
$p_\mathrm{T}^\mathrm{t}$, and in the invariant mass of the top-quark pair,
$m_{\mathrm{t}\bar{\mathrm{t}}}$. Analogous plots for the distributions differential
in the rapidity of either the top quark or the top-quark pair are collected in
appendix~\ref{app:add_plots}. From figure~\ref{fig:atlastop}, we immediately
validate the \textsc{PineAPPL} result: its relative difference with respect to
the MC result is at most as large as \SI{0.4}{\permille}, irrespective of the
accuracy of the theory, of the choice of scale and of the distribution
considered.

The process receives its leading contribution from the $\mathrm{gg}$ channel,
which varies between \SI{81}{\percent} and \SI{61}{\percent} (\SI{76}{\percent} and \SI{83}{\percent}) of the $p_\mathrm{T}^\mathrm{t}$
($m_{\mathrm{t}\bar{\mathrm{t}}}$) differential cross section as the value of the
transverse momentum of the top quark (the invariant mass of the top-quark pair)
increases; the largest PI contribution for this process comes from $\gamma\mathrm{g}$ scattering,
which accounts for about \SIrange{0.5}{1}{\percent} (\SIrange{0.5}{0.7}{\percent}) of the cross
section, and is almost entirely (\SI{90}{\percent}) due to the LO contribution at $\mathcal{O}(\alphas\alpha)$; the contribution from other PI parton luminosities is comparatively
negligible. Overall, the EW corrections suppress the $p_\mathrm{T}^\mathrm{t}$
($m_{\mathrm{t}\bar{\mathrm{t}}}$) distribution by about \SIrange{0.2}{3.5}{\percent}
(\SIrange{0.5}{0.2}{\percent}) for increasing values of $p_\mathrm{T}^\mathrm{t}$
($m_{\mathrm{t}\bar{\mathrm{t}}}$), except in the first bin of the
$p_\mathrm{T}^\mathrm{t}$ distribution, where they enhance the cross section by
about \SI{1}{\percent}. The size of these shifts, however, remains always significantly
smaller than the data uncertainty.\footnote{In
  figure~\ref{fig:atlastop} the data uncertainty corresponds to the ATLAS
  measurement~\cite{Aad:2015auj}. Similar considerations apply also for the CMS
  measurement~\cite{Khachatryan:2015oaa}.}
As a consequence, we anticipate that the more accurate NLO QCD+EW theory is
likely to be easily accommodated by the large data uncertainty, should the data
be fitted with the inclusion of EW corrections.

The size of the EW correction is comparable to the size of the PDF uncertainty,
except at large values of transverse momentum or invariant mass, where the
former becomes larger than the latter. This fact suggests that, once included in
a global fit, EW corrections can improve the accuracy of the PDFs. In comparison
to the scale uncertainty, the size of the EW corrections remains negligible:
despite the fact that the choice of factorisation and renormalisation scales
have been devised to optimise the convergence of the perturbative expansion,
NNLO QCD corrections remain large~\cite{Czakon:2013goa,Czakon:2014xsa,Czakon:2015owf,Czakon:2016ckf,Catani:2019iny,Behring:2019iiv,Catani:2019hip,Catani:2020tko,Czakon:2020qbd}, as expected in a process mostly initiated
by gluons. Their inclusion is therefore mandatory in a fit of PDFs. Finally,
the Monte Carlo statistical uncertainty remains negligible in comparison to the
data, PDF and scale uncertainties, and to the size of the EW correction. Our
conclusions are therefore not affected by Monte Carlo inefficiencies.

\subsubsection{Z-boson production with non-zero transverse momentum.}
\label{sec:Zpt}

\paragraph{Experimental measurements and process features.}
We select the single-differential transverse momentum distribution of the
sum of the two leptons (the \enquote{Z boson}), $p_\mathrm{T}^{\ell \bar{\ell}}$, measured by the CMS experiment at a
centre-of-mass energy of \SI{13}{\tera\electronvolt}~\cite{Sirunyan:2019bzr}.
This measurement, which has not been included in any PDF determination so far,
shows very low experimental uncertainties (at the percent level or below).
EW corrections are therefore expected to be essential to achieve a good
description of it, and to constrain accurately the PDFs, together with
NNLO QCD corrections, which are already well
known~\cite{Boughezal:2015ded,Boughezal:2016isb,Boughezal:2016yfp,Ridder:2015dxa,Ridder:2016nkl,Gehrmann-DeRidder:2017mvr,Bizon:2019zgf}.
Analogous measurements,
from the ATLAS~\cite{Aad:2015auj} and CMS~\cite{Khachatryan:2015oaa}
experiments at a centre-of-mass energy of \SI{8}{\tera\electronvolt},
were partly included (upon
selection of an appropriate kinematic cut that excluded bins with large EW
corrections) in a dedicated study~\cite{Boughezal:2017nla},
in the NNPDF3.1 PDF set~\cite{Ball:2017nwa} and in variants of
the CT18 PDF set~\cite{Hou:2019efy}. In the QCD computation, we consider a
single LO contribution $\mathcal{O}(\alphas\alpha^2)$ and a single NLO
contribution $\mathcal{O}(\alphas^2\alpha^2)$. In the NLO QCD+EW computation,
we supplement these with another LO and NLO, which are
$\mathcal{O}(\alpha^3)$ and $\mathcal{O}(\alphas\alpha^3)$;
contributions of the order $\mathcal{O}(\alpha^4)$ are not considered (see ref.~\cite{Denner:2019zfp}).
EW corrections for this process were computed in
refs.~\cite{Kuhn:2005az,Denner:2011vu,Hollik:2015pja,Kallweit:2015dum,Frederix:2018nkq}.
The process receives contributions from 101 (166) parton luminosities,
see appendix~\ref{app:lumis}.

\paragraph{Process-specific settings.}

We choose a dynamic value for the renormalisation and factorisation scales $\mu_\mathrm{R}=\mu_\mathrm{F}=\sqrt{M_{\ell \bar{\ell}}^2 + \left( p_\mathrm{T}^{\ell \bar{\ell}} \right)^2}$, used for example in ref.~\cite{Ridder:2016nkl}, where $M_{\ell \bar{\ell}}$ is the invariant mass of the lepton pair and $p_\mathrm{T}^{\ell \bar{\ell}}$ its transverse momentum.\footnote{In the last bin the dynamic scale is $\mu_\mathrm{R} = \mu_\mathrm{F} \approx p_\mathrm{T}^{\ell \bar{\ell}}$, and therefore the bin limits drive the scale beyond the limit $Q_\text{max}^2 = (\SI{1}{\tera\electronvolt})^2$ defined in sec.~\ref{sec:grid-representation}, which we increase for this process to $Q_\text{max}^2 = (\SI{10}{\tera\electronvolt})^2$, along with $N_\tau = 40$.}
Consistently with the
experimental analysis, we require $p_\mathrm{T}^\ell>\SI{25}{\giga\electronvolt}$,
$|\eta_\ell|<2.4$, $|M_{\ell\bar\ell} - M_\mathrm{Z}| < \SI{15}{\giga\electronvolt}$,
$|y_{\ell\bar\ell}|<2.4$ and $\SI{20}{\giga\electronvolt}<p_\mathrm{T}^{\ell\bar\ell}<\SI{1500}{\giga\electronvolt}$ for the
transverse momentum and pseudorapidity of each lepton, and for the invariant
mass, pseudorapidity and transverse momentum of the lepton pair.
We finally discard all bins with
$p_\mathrm{T}^{\ell \bar{\ell}}<\SI{20}{\giga\electronvolt}$ to avoid a kinematic
region where resummation effects are sizeable.

\begin{figure}[!t]
    \centering
    \includegraphics[width=0.5\textwidth]{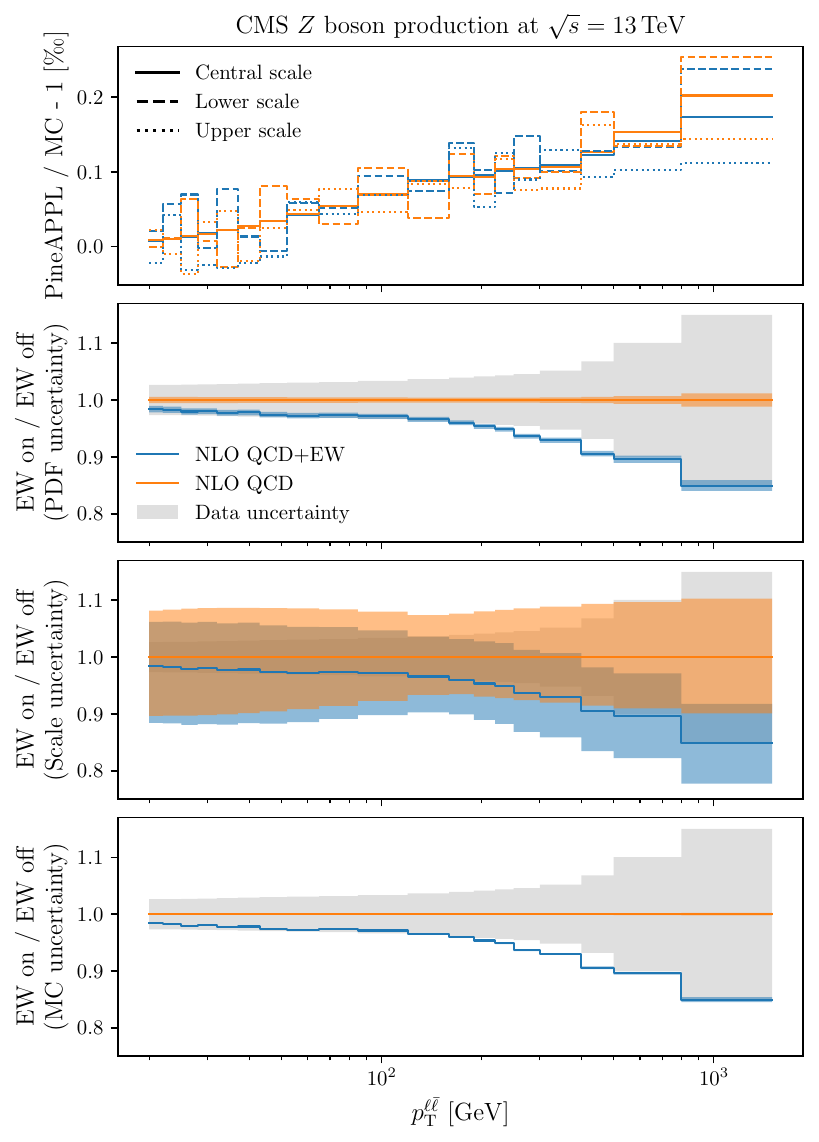}
    \caption{Same as figure~\ref{fig:atlaszhighmass49fb} but for the 
      CMS differential Z $p_\mathrm{T}$ measurement at a centre-of-mass energy of
      \SI{13}{\tera\electronvolt}~\cite{Sirunyan:2019bzr}.}
    \label{fig:cmsZ13TeV}
\end{figure}

\paragraph{Numerical results.} In figure~\ref{fig:cmsZ13TeV} we report the
distribution differential in the transverse momentum of the Z
boson. Also in this case, the \textsc{PineAPPL} result is well validated, as it
differs from the MC result by $\SI{0.5}{\permille}$ at most. The accuracy of the
theory or the choice of scale do not alter this conclusion.

The process receives its leading contribution from $\mathrm{qg}$- and
$\bar{\mathrm{q}}\mathrm{g}$-initiated channels, which account for about \SI{79}{\percent}
of the cross section, with some variations in the relative contributions from
individual quark and antiquark flavours as the transverse momentum of the
Z boson varies; the PI contribution always remain negligible.
Overall, the EW corrections suppress the theoretical predictions by about
\SIrange{2}{15}{\percent} as the transverse momentum of the Z boson increases.
The size of this shift is as large (or slightly larger) as the data uncertainty over
the entire range of $p_{\mathrm{T}}^{\ell\bar\ell}$. As a consequence,
we anticipate the inclusion of EW corrections to be relevant for an accurate
fit of this data.

The size of the EW correction is between four and fifteen times larger than the
size of the PDF uncertainty: as previously noted in the other cases, this fact
suggests that, once included in a PDF fit, EW corrections can improve the
accuracy of the PDFs. In comparison to the scale uncertainty, the size of the
EW corrections remains negligible at small values of $p_{\mathrm{T}}^{\ell\bar\ell}$,
roughly $p_{\mathrm{T}}^{\ell\bar\ell}\lesssim\SI{400}{\giga\electronvolt}$, while it
becomes larger than it in the two bins at the largest value of
$p_{\mathrm{T}}^{\ell\bar\ell}$. In this kinematic region, NLO EW corrections might
therefore become even more relevant than NNLO QCD corrections, and should
therefore be mandatorily included in a fit of PDFs to this data set.
Finally, the Monte Carlo uncertainty is well under control, as it remains
mostly negligible in comparison to the PDF, scale and data uncertainty, and to
the size of the EW correction.

\section{Subtraction of EW effects from data}
\label{sec:doublecounting}

The ability to perform theoretical calculations simultaneously accurate in both
the QCD and EW couplings is not sufficient to make a \emph{consistent}
comparison with experimental measurements. In this section
we formulate some
guidelines to facilitate this task. We focus on the problem of data with (partially) subtracted EW effects, which, if compared to theory predictions including them, leads to a double counting issue.
Our guidelines are intended to make the reader aware of an
emerging new issue, whose definitive solution remains however beyond the scope
of this work.

\begin{figure}[!t]
    \centering
    \begin{overpic}[width=0.6\textwidth, trim=0.cm 11cm 0.cm 10cm, clip=True]{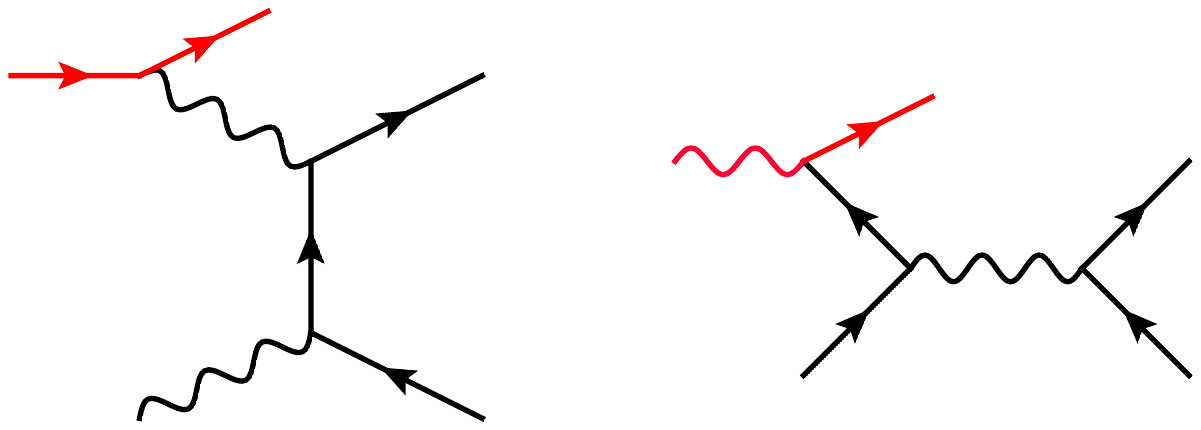}
        \put (5, 33) {\large $\mathrm{q}$}
        \put (20, 37) {\large $\mathrm{q}$}
        \put (38, 5) {\large $\bar{\ell}$}
        \put (38, 31) {\large $\ell$}
        \put (11, 6) {\large $\gamma$}
        \put (65, 9) {\large $\mathrm{q}$}
        \put (73, 30) {\large $\mathrm{q}$}
        \put (96, 25) {\large $\ell$}
        \put (96, 8) {\large $\bar{\ell}$}
        \put (55, 26) {\large $\gamma$}
    \end{overpic}
    \caption{\label{fig:dy-pi}
    Photon-induced (left) and quark-induced (right) contributions to the Drell-Yan process. In black, the LO process is shown.
    In red, the initial-state splitting leading to the real-emission $\mathrm{q} \gamma \to \ell \bar{\ell} \mathrm{q}$ is highlighted. Such a
    real emission enters in the NLO EW corrections.}
\end{figure}

A first example is the subtraction of (irreducible) background processes which must not be considered as such. A very blatant case
is neutral-current Drell--Yan, where the signal process is the production of an opposite-sign lepton pair, which starts
at $\mathcal O(\alpha^2)$. Because this process is usually thought
as a quark-initiated $s$-channel mechanism ($\mathrm{q} \bar{\mathrm{q}} \to \gamma^*/\mathrm{Z} \to \ell \bar{\ell}$), in many analyses the PI component,
$\gamma \gamma \to \ell \bar{\ell}$ in the $t$ channel, is considered a different process, and therefore as a background and subtracted.
The subtraction from the measured data is done by calculating the theoretical predictions of the double-photon initiated contribution, possibly including (ill-defined) higher-order
corrections. For example, in refs.~\cite{Aaboud:2017ffb,Aad:2016zzw} (a similar statement appears also in an older analysis~\cite{Aad:2013iua}), one reads:
\begin{quote}
The photon-induced process, $\gamma\gamma \to \ell \bar{\ell}$, is simulated at LO using Pythia 8
and the MRST2004qed PDF set~\cite{Martin:2004dh}. The expected yield for this process also accounts for 
NLO QED/EW corrections from references~\cite{Bardin:2012jk,Bondarenko:2013nu}, which decrease the yield by approximately \SI{30}{\percent}.
\end{quote}
Such a distinction, which is unphysical and incorrect in quantum mechanics, may be somehow justified at LO\@. Beyond this order, it is simply wrong.
Indeed, at $\mathcal O(\alpha^3)$, the reaction $\mathrm{q} \gamma \to \ell \bar{\ell} \mathrm{q}$ becomes possible, which
includes both kind of topologies discussed above, and needs both in order to yield an IR-finite result, see Fig.~\ref{fig:dy-pi} (as a consequence, one cannot speak of EW corrections to $\gamma \gamma \to \ell \bar{\ell}$). It may be useful for the reader to consider a QCD counterpart of this issue, 
e.g.\ the subtraction of the gluon-initiated contribution to top-pair production, in order to understand the incorrectness of this procedure.

A second example is related to removing EW effects from data. These can be either the full EW corrections
or just a part of them. In either case, a comparison between these data and a NLO-EW accurate simulation aimed at the extraction of some parameter would be meaningless, as some effects included in the latter
have been removed from the former. The typical example relevant for the LHC is the deconvolution of effects due to multiple-photon radiation
from light particles in the final state. This applies mostly
to processes such as neutral- or charged-current Drell--Yan, especially when electrons are considered. The problem lies in the fact that
 electrons, and to a lesser extent muons, tend to radiate collinear photons, which are not accounted
for in QCD-only matrix elements. Thus, leptons that are measured in the detector are less energetic, and this fact is compensated for
by inverting a photon shower. The resulting dataset is e.g.\ referred to as \emph{pre-FSR} with observables defined in terms of \emph{Born-level electrons} (see e.g.\ ref.~\cite{Aad:2015auj} for its definition).
These datasets are needed for and correctly used in QCD-only PDF determinations, since the EW corrections to some DY observables can be significant, and excluding them would therefore degrade the quality of the fit.
In fits including fixed-order EW corrections the problem with this definition, besides double counting, is that the first photon emission is included exactly at the matrix-element level. The inclusion of
subsequent emission would require the matching with the QED shower, which is not yet available for general processes.

It is interesting
to note that one can tune the QED parton shower to mimic NLO EW effects for specific processes and observables, so that a prediction only accurate at NLO
QCD displays a remarkable agreement with another at NLO QCD+EW when the photonic shower is included (see also the behaviour of predictions showered with
\textsc{Photos}~\cite{Barberio:1990ms,Barberio:1993qi,Golonka:2005pn} in ref.~\cite{CarloniCalame:2016ouw}).
However, this kind of agreement
always comes \emph{a posteriori}, and cannot be ensured in general.
The transverse momentum of the Z boson as shown in section~\ref{sec:Zpt} is such an example: the difference between the two datasets is rather small (less than \SI{6}{\percent}), but the NLO EW corrections are as large as \SI{25}{\percent}.
Furthermore, the deconvolution of QED effects in data introduces a dependence on the program (and possibly on 
the specific version) employed for the shower inversion. 
This fact is especially problematic if deconvolved datasets are the only ones which are published, since undoing the exact deconvolution can be very difficult, or practically impossible.

A more physical definition of leptonic observables would be one making use of either bare leptons (the leptons as they emerge after FSR)
or of dressed leptons (leptons and photons are clustered together and their momenta are combined, in analogy with jets in QCD). The problem with the former is that 
electrons are never measured as bare particles, because of the finite resolution 
of the electromagnetic calorimeter. For what concerns muons, while in principle the concept of a bare muon is physical, it should be kept in
mind that modern, general-purpose codes employed to
compute EW corrections treat leptons as massless, to ensure numerical stability of the matrix elements. In this case, using bare leptons is not collinear safe.
Dressed leptons avoid all these shortcomings, with the further advantage of being
inclusive on the effect of extra collinear emissions. This fact encourages to explore the possibility of employing a dressed-lepton
definition, regardless of the leptonic flavour. We acknowledge that this practice is already being followed in (some) experimental analyses:
indeed, to mention two examples discussed in this paper, in ref.~\cite{Aad:2015auj} data for dressed leptons are published, together with the Born-level
and bare ones, while refs.~\cite{Sirunyan:2019bzr} employs a dressed-lepton definition. We therefore recommend that these ways of presenting the experimental
data become standard in the future. 

\section{Conclusions and Outlook}
\label{sec:conclusion}

The systematic inclusion of EW corrections in accurate theoretical computations
for several LHC processes is becoming more and more important in order to
match the increasing precision of the data. In this paper we simplified the
computational aspect of this task, building upon the automation of QCD and EW
computations pioneered in recent years~\cite{Kallweit:2014xda,Biedermann:2017yoi,Frederix:2018nkq}.
Specifically we developed \textsc{PineAPPL}, a new library that stores perturbative calculations from an external Monte Carlo generator in a PDF-independent way using interpolation grids.
This offers the advantage of fast a posteriori convolutions with PDFs, for example to study the uncertainties coming from different PDF sets and/or the strong coupling $\alphas$, and to determine the PDFs themselves, a task for
which fast-interpolation grids are fundamental.
We tested \textsc{PineAPPL} together with \textsc{mg5\_aMC} and found a precision of \numrange{e-4}{e-5} relative to the MC result, which is excellent for all
foreseeable practical purposes.
Although we used \textsc{mg5\_aMC}, we note that \textsc{PineAPPL} is not tied in any way to a specific Monte Carlo generator, and can be easily interfaced with any of them.

We emphasise that a distinguishing feature of \textsc{PineAPPL} is the support for arbitrary coupling orders not only in the strong, but also in the electroweak coupling.
This enables us to generate, for the first time, NLO EW and NLO combined QCD--EW interpolation grids.
Using \textsc{mg5\_aMC} we calculated and showcased the impact of these corrections for specific measurements of some representative LHC processes: Drell--Yan lepton-pair production, top-pair production, and Z-boson production with non-zero transverse momentum. 

Finally, we discussed the issue of subtracting EW corrections in experimental data, which becomes important when theoretical predictions including EW corrections are compared to experimental data. In particular, with the
development of \textsc{PineAPPL}, all technical requirements are fulfilled for producing the first PDF fit of LHC data including EW and combined QCD--EW corrections.
This will have at least two advantages: in PDF fits phase-space regions are usually cut away if they exhibit large EW corrections; including them therefore increases the number of data points in a fit and therefore indirectly enlarges a PDF set's interpolation region.
Secondly, this makes it possible to use experimental data that are closer to the actual measurement, without the need to compensate for missing EW corrections.
We plan to address this task in a future work.

\vspace{0.5cm}
\hrule
\begin{center}
The \textsc{PineAPPL} library is available at \url{https://n3pdf.github.io/pineappl}.
\end{center}

\appendix

\acknowledgments
We would like to thank Valerio Bertone, Stefan Dittmaier, Stefano Forte, Zahari Kassabov, Davide Pagani, and Juan Rojo for a critical reading of the manuscript, and Rikkert Frederix for discussions about \textsc{aMCfast}.
We also acknowledge discussions with
Florencia Canelli, Stefano Camarda, Paolo Francavilla, Abideh Jafari, Andreas Jung, Elizaveta Shabalina, Wolfgang Wagner about the 
treatment of EW effects in experimental data.
S.C.\ and C.S.\ are supported by the European Research Council under the European Union's
Horizon 2020 research and innovation Programme (grant agreement no.\ 740006).
E.R.N.\ is supported by the European Commission through a Marie
Sk\l odowska-Curie Action (grant agreement no.\ 752748).
M.Z.\ thanks the Nikhef institute in Amsterdam,
where he was employed when this paper was started.

\section{Installation and usage of PineAPPL}
\label{app:pineappl}

\textsc{PineAPPL} currently consists of three parts: 1) the library itself, which is a dependency for the other parts, 2) the helper program \texttt{pineappl}, which allows one to read \textsc{PineAPPL} grids from the command line and make predictions with it (explained in section~\ref{app:pineappl-demo}), and finally 3) the C interface, which is intended to be used in Monte Carlo integrators to generate the grids.

\subsection{Demonstration of \texorpdfstring{\texttt{pineappl}}{pineappl}}
\label{app:pineappl-demo}

The program \texttt{pineappl} can be used to perform quick convolutions and other calculations with existing grids on the command line.
If started without any arguments, it prints its help and lists all supported subcommands:
\begin{verbatim}
 $ pineappl
pineappl 0.3.0
Read, write, and query PineAPPL grids

USAGE:
    pineappl <SUBCOMMAND>

FLAGS:
    -h, --help       Prints help information
    -V, --version    Prints version information

SUBCOMMANDS:
    channels           Shows the contribution for each partonic channel
    convolute          Convolutes a PineAPPL grid with a PDF set
    diff               Compares the contents of two grids with each other
    info               Shows information about the grid
    luminosity         Shows the luminosity function
    merge              Merges one or more PineAPPL grids together
    orders             Shows the predictions for all bins for each order
                       separately
    pdf_uncertainty    Calculates PDF uncertainties
\end{verbatim}

\paragraph{Convolutions.}
The most important subcommand is \texttt{convolute}, which performs a convolution of a single grid with a single or multiple PDF sets.
As an example we show the grid produced for the ATLAS Drell--Yan high-mass lepton-pair production from section~\ref{sec:dy-lepton-pair-production}, convoluted with \texttt{NNPDF31\_nlo\_as\_0118\_luxqed} as the main PDF set and with \texttt{CT18NLO} as a second PDF set.
\begin{verbatim}
 $ pineappl convolute ATLASZHIGHMASS49FB.pineappl \
 > NNPDF31_nlo_as_0118_luxqed CT18NLO
bin xmin xmax     diff        integ    neg unc pos unc       CT18NLO
---+----+----+------------+-----------+-------+-------+------------+------
  0  116  130 2.0630698e-1  2.888297e0  -2.08%   1.69% 2.0246802e-1 -1.86%
  1  130  150 9.1818985e-2  1.836379e0  -1.79%   1.86% 8.9766355e-2 -2.24%
  2  150  170 4.5306370e-2 9.061274e-1  -1.60%   1.98% 4.4115960e-2 -2.63%
  3  170  190 2.5894856e-2 5.178971e-1  -1.66%   2.06% 2.5138525e-2 -2.92%
  4  190  210 1.6075267e-2 3.215053e-1  -1.70%   2.10% 1.5566535e-2 -3.16%
  5  210  230 1.0526659e-2 2.105331e-1  -1.71%   2.12% 1.0173163e-2 -3.36%
  6  230  250 7.1928162e-3 1.438563e-1  -1.71%   2.13% 6.9403972e-3 -3.51%
  7  250  300 4.0776555e-3 2.038827e-1  -1.70%   2.44% 3.9255068e-3 -3.73%
  8  300  400 1.4775481e-3 1.477548e-1  -1.94%   2.87% 1.4182754e-3 -4.01%
  9  400  500 4.5473785e-4 4.547378e-2  -2.30%   3.19% 4.3525336e-4 -4.28%
 10  500  700 1.2164277e-4 2.432855e-2  -2.41%   3.12% 1.1612523e-4 -4.54%
 11  700 1000 1.9792340e-5 5.937701e-3  -2.05%   2.12% 1.8813113e-5 -4.95%
 12 1000 1500 2.0228761e-6 1.011438e-3  -1.29%   0.47% 1.9221978e-6 -4.98%
\end{verbatim}
The output shows all 13 bins with lower (\texttt{xmin}) and upper limit (\texttt{xmax}) of the invariant mass $M_{\ell \bar{\ell}}$ of the lepton pair, with the differential cross section $\mathrm{d} \sigma / \mathrm{d} M_{\ell \bar{\ell}}$ (\texttt{diff}), integrated cross section $(M_{\ell \bar{\ell}}^\mathrm{max} - M_{\ell \bar{\ell}}^\mathrm{min}) \mathrm{d} \sigma / \mathrm{d} M_{\ell \bar{\ell}}$ (\texttt{integ}), and the perturbative uncertainty estimated from a 7-point scale variation (envelope given by \texttt{neg unc} and \texttt{pos unc}).
The uncertainty estimation can alternatively use a 3- or a 9-point scale variation using the optional program switch \texttt{-{}-scales 3} or \texttt{-{}-scales 9}, respectively.
The (differential) results for the second PDF set (\texttt{CT18NLO}) is shown in absolute numbers and also as a percentage relative to the result of the first PDF set.

\paragraph{Perturbative orders.}
Often it is helpful to see the impact of the different perturbative orders to the cross section.
The subcommand \texttt{orders} shows this:
\begin{verbatim}
 $ pineappl orders ATLASZHIGHMASS49FB.pineappl \
 > NNPDF31_nlo_as_0118_luxqed
bin xmin xmax     diff     O(as^0 a^2) O(as^1 a^2) O(as^0 a^3)
---+----+----+------------+-----------+-----------+-----------
  0  116  130 2.0630698e-1     100.00%      15.97%      -5.26%
  1  130  150 9.1818985e-2     100.00%      18.07%      -4.29%
  2  150  170 4.5306370e-2     100.00%      19.66%      -3.64%
  3  170  190 2.5894856e-2     100.00%      20.69%      -3.21%
  4  190  210 1.6075267e-2     100.00%      21.26%      -2.91%
  5  210  230 1.0526659e-2     100.00%      21.48%      -2.85%
  6  230  250 7.1928162e-3     100.00%      21.60%      -2.62%
  7  250  300 4.0776555e-3     100.00%      21.36%      -2.75%
  8  300  400 1.4775481e-3     100.00%      20.32%      -3.11%
  9  400  500 4.5473785e-4     100.00%      17.83%      -3.65%
 10  500  700 1.2164277e-4     100.00%      14.04%      -4.68%
 11  700 1000 1.9792340e-5     100.00%       7.21%      -6.75%
 12 1000 1500 2.0228761e-6     100.00%      -3.05%      -9.99%
\end{verbatim}
The first four columns are the same as in \texttt{convolute}, and the remaining ones show all orders normalised to the sum of the leading orders, which in this case is only the $\mathcal{O} (\alpha^2)$.
Absolute numbers are shown if the switch \texttt{-{}-absolute} or \texttt{-a} is passed to the program.

\paragraph{Channels and Luminosity function.}
Sometimes it is useful to know which partons contribute the most and by how much.
This is what the subcommand \texttt{channels} shows:
\begin{verbatim}
 $ pineappl channels ATLASZHIGHMASS49FB.pineappl \
 > NNPDF31_nlo_as_0118_luxqed --limit 5
bin xmin xmax lumi  size
---+----+----+----+------+---+------+---+------+---+------+---+-----
  0  116  130  #15 27.42%  #0 27.42%  #5 24.52% #20 24.51% #30 1.30%
  1  130  150  #15 28.83%  #0 28.83%  #5 21.78% #20 21.78% #30 1.86%
  2  150  170  #15 29.87%  #0 29.82%  #5 19.63% #20 19.62% #30 2.32%
  3  170  190  #15 30.54%  #0 30.53% #20 18.17%  #5 18.13% #30 2.58%
  4  190  210   #0 31.06% #15 31.04%  #5 17.07% #20 17.04% #30 2.75%
  5  210  230   #0 31.46% #15 31.38% #20 16.27%  #5 16.24% #30 2.86%
  6  230  250   #0 31.73% #15 31.64% #20 15.66%  #5 15.62% #30 2.91%
  7  250  300   #0 32.07% #15 32.02%  #5 14.91% #20 14.86% #30 2.98%
  8  300  400   #0 32.62% #15 32.56%  #5 13.83% #20 13.81% #30 2.98%
  9  400  500  #15 33.14%  #0 33.12% #20 12.89%  #5 12.87% #30 2.93%
 10  500  700   #0 33.71% #15 33.31% #20 12.28%  #5 12.26% #30 2.93%
 11  700 1000  #15 34.08%  #0 33.71% #20 11.59%  #5 11.51% #30 3.06%
 12 1000 1500   #0 33.95% #15 33.88% #20 11.12%  #5 10.88% #30 3.56%
\end{verbatim}
The first three columns are known from \texttt{convolute}.
The next columns (the switch \texttt{-{}-limit 5} limits the output to five columns) show first the channel index and then the relative size of the corresponding contribution.
Since the contribution of a partonic channel can be negative, the columns are sorted ignoring the sign of the contribution.
The first line shows that for bin 0, i.e.\ for the range $\SI{116}{\giga\electronvolt} < M_{\ell \bar{\ell}} < \SI{130}{\giga\electronvolt}$, the cross section is dominated by partonic channel \texttt{\#15} (\SI{27.42}{\percent}), following by partonic channel \texttt{\#0} with same size, then channel \texttt{\#5}, etc.
The meaning of the channel numbers is given by using the subcommand \texttt{luminosity} (only an excerpt is shown):
\begin{verbatim}
id    entry
--+------------+------------
0  1 × ( 2, -2) 1 × ( 4, -4)
5  1 × ( 1, -1) 1 × ( 3, -3)
15 1 × (-4,  4) 1 × (-2,  2)
20 1 × (-3,  3) 1 × (-1,  1)
30 1 × (22, 22)
\end{verbatim}
This shows that channel \texttt{\#0} represents the up-type quark--anti-quark contributions (shown with PDG id 2 and 4 for up and charm quarks, which have the same matrix elements), channel \texttt{\#15} is the same channel with its initial states transposed, channels \texttt{\#5} and \texttt{\#20} are the down-type quark--anti-quark channels, and channel \texttt{\#30} is the photon--photon channel.
The size of the remaining channels is smaller than the photon--photon channel.
The factors \texttt{1} are not important here, but in general they can contain CKM values and charge factors that, if kept in the squared matrix elements, would not allow for sharing a single matrix element for different quark flavours and therefore slow down the calculation.
A complete list of all channels and of their contribution to the cross section for all of the processes discussed in section~\ref{sec:results} is collected in appendix~\ref{app:lumis}.

\subsection{Sample runcard for mg5\_aMC@NLO}
\label{app:sample-runcard}

The following run card was used to produce the results shown in section~\ref{sec:dy-lepton-pair-production}.
The only difference with respect to a standard \textsc{mg5\_aMC} run is the switch \texttt{set pineappl True}, which enables to fill a \textsc{PineAPPL} grid.
For a complete set of runcards and patches see \url{https://n3pdf.github.io/pineappl}.
\begin{verbatim}
set complex_mass_scheme True
import model loop_qcd_qed_sm_Gmu
define p = p b b~
define j = p
generate p p > e+ e- [QCD QED]
output @OUTPUT@
launch @OUTPUT@
fixed_order = ON
set mz @MZ@
set ymt @YMT@
set ebeam1 3500
set ebeam2 3500
set pdlabel lhapdf
set lhaid 324900
set fixed_ren_scale True
set fixed_fac_scale True
set mur_ref_fixed @MZ@
set muf_ref_fixed @MZ@
set reweight_scale True
set ptl = 25.0
set etal = 2.5
set mll = 116
#user_defined_cut set mmllmax = 1500.0
set req_acc_FO 0.0001
set pineappl True
done
quit
\end{verbatim}

\subsection{Example Monte Carlo program in C++}
\label{app:example-program}

\definecolor{mygreen}{rgb}{0,0.6,0}
\definecolor{mygray}{rgb}{0.5,0.5,0.5}
\definecolor{mymauve}{rgb}{0.58,0,0.82}

\lstset{
  belowcaptionskip=1\baselineskip,
  breaklines=true,
  frame=L,
  numbers=left,
  xleftmargin=\parindent,
  showstringspaces=false,
  basicstyle=\footnotesize\ttfamily,
  keywordstyle=\bfseries\color{green!40!black},
  commentstyle=\itshape\color{purple!40!black},
  identifierstyle=\color{blue},
  stringstyle=\color{orange},
}

The following listing shows how to setup \textsc{PineAPPL} using its C interface in a simple Monte Carlo integrator for calculating the double-photon contribution to Drell--Yan lepton-pair production at the LHC\@.
All \textsc{PineAPPL} functions have the prefix \texttt{pineappl\_}.
The full example together with a makefile can be found at \url{https://github.com/N3PDF/pineappl/tree/master/examples/capi-dy-aa}.
The documentation of the C API can be found at \url{https://docs.rs/pineappl_capi}.

\begin{lstlisting}[language=C++,mathescape=true]
#include <pineappl_capi.h>
#include <LHAPDF/LHAPDF.h>

#include <cmath>
#include <cstddef>
#include <cstdio>
#include <random>
#include <vector>

double int_photo(double s, double t, double u) {
    double alpha0 = 1.0 / 137.03599911;
    return alpha0 * alpha0 / 2.0 / s * (t / u + u / t);
}

struct Psp2to2 {
    double s;
    double t;
    double u;
    double x1;
    double x2;
    double jacobian;
};

Psp2to2 hadronic_pspgen(std::mt19937& rng, double mmin, double mmax) {
    using std::acos;
    using std::log;
    using std::pow;

    double smin = mmin * mmin;
    double smax = mmax * mmax;

    double r1 = std::generate_canonical<double, 53>(rng);
    double r2 = std::generate_canonical<double, 53>(rng);
    double r3 = std::generate_canonical<double, 53>(rng);

    double tau0 = smin / smax;
    double tau = pow(tau0, r1);
    double y = pow(tau, 1.0 - r2);
    double x1 = y;
    double x2 = tau / y;
    double s = tau * smax;

    double jacobian = tau * log(tau0) * log(tau0) * r1;

    // theta integration (in the CMS)
    double  cos_theta = 2.0 * r3 - 1.0;
    jacobian *= 2.0;

    double  t = -0.5 * s * (1.0 - cos_theta);
    double  u = -0.5 * s * (1.0 + cos_theta);

    // phi integration
    jacobian *= 2.0 * acos(-1.0);

    return { s, t, u, x1, x2, jacobian };
}

void fill_grid(pineappl_grid* grid, std::size_t calls) {
    using std::acosh;
    using std::fabs;
    using std::log;
    using std::sqrt;

    auto rng = std::mt19937();

    // in GeV^2 pbarn
    double  hbarc2 = 389379372.1;

    for (std::size_t i = 0; i != calls; ++i) {
        // generate a phase-space point
        auto tmp = hadronic_pspgen(rng, 10.0, 7000.0);
        auto s = tmp.s;
        auto t = tmp.t;
        auto u = tmp.u;
        auto x1 = tmp.x1;
        auto x2 = tmp.x2;
        auto jacobian = tmp.jacobian;

        double ptl = sqrt((t * u / s));
        double mll = sqrt(s);
        double yll = 0.5 * log(x1 / x2);
        double ylp = fabs(yll + acosh(0.5 * mll / ptl));
        double ylm = fabs(yll - acosh(0.5 * mll / ptl));

        jacobian *= hbarc2 / calls;

        // cuts for LO for the invariant-mass slice containing the
        // Z-peak from CMSDY2D11
        if ((ptl < 14.0) || (fabs(yll) > 2.4) || (ylp > 2.4)
            || (ylm > 2.4) || (mll < 60.0) || (mll > 120.0))
        {
            continue;
        }

        auto weight = jacobian * int_photo(s, u, t);
        double  q2 = 90.0 * 90.0;

        pineappl_grid_fill(grid, x1, x2, q2, 0, fabs(yll), 0, weight);
    }
}

int main() {
    // create a new luminosity function for the $\gamma\gamma$ initial state
    auto* lumi = pineappl_lumi_new();
    int32_t pdg_ids[] = { 22, 22 };
    double ckm_factors[] = { 1.0 };
    pineappl_lumi_add(lumi, 1, pdg_ids, ckm_factors);

    // only LO $\alpha_\mathrm{s}^0 \alpha^2 \log^0(\xi_\mathrm{R}) \log^0(\xi_\mathrm{F})$
    uint32_t orders[] = { 0, 2, 0, 0 };

    // we bin in rapidity from 0 to 2.4 in steps of 0.1
    double bins[] = {
        0.0,
        0.1, 0.2, 0.3, 0.4, 0.5, 0.6, 0.7, 0.8, 0.9, 1.0, 1.1, 1.2,
        1.3, 1.4, 1.5, 1.6, 1.7, 1.8, 1.9, 2.0, 2.1, 2.2, 2.3, 2.4
    };

    // create the PineAPPL grid with default interpolation and binning parameters
    auto* keyval = pineappl_keyval_new();
    auto* grid = pineappl_grid_new(lumi, 1, orders, 24, bins, keyval);

    // now we no longer need `keyval` and `lumi`
    pineappl_keyval_delete(keyval);
    pineappl_lumi_delete(lumi);

    // fill the grid with phase-space points
    fill_grid(grid, 10000000);

    // perform a convolution of the grid with PDFs
    auto* pdf = LHAPDF::mkPDF("NNPDF31_nlo_as_0118_luxqed", 0);
    auto xfx = [](int32_t id, double x, double q2, void* pdf) {
        return static_cast <LHAPDF::PDF*> (pdf)->xfxQ2(id, x, q2);
    };
    auto alphas = [](double q2, void* pdf) {
        return static_cast <LHAPDF::PDF*> (pdf)->alphasQ2(q2);
    };

    std::vector<double> dxsec(24);
    pineappl_grid_convolute(grid, xfx, xfx, alphas, pdf, nullptr,
        nullptr, 1.0, 1.0, dxsec.data());

    // print the results
    for (std::size_t i = 0; i != 24; ++i) {
        std::printf("%.1f %.1f %.3e\n", bins[i], bins[i + 1], dxsec[i]);
    }

    // write the grid to disk
    pineappl_grid_write(grid, "DY-LO-AA.pineappl");

    // destroy the object
    pineappl_grid_delete(grid);
}
\end{lstlisting}

\subsection{Installation}
\label{app:installation}

Updated installation instructions are kept in the file \texttt{README.md} in \textsc{PineAPPL}'s repository at \url{https://github.com/N3PDF/pineappl/} and on its homepage \url{https://n3pdf.github.io/pineappl/}.

\subsubsection*{Installation of Rust}

\lstset{
  basicstyle=\ttfamily\small,
  showstringspaces=false,
  commentstyle=\color{red},
  keywordstyle=\color{blue},
}

All parts are written in Rust: a Rust compiler and related tools are needed.
On operating systems with a \texttt{bash} shell (such as Linux or MacOS) the installation is as simple as
\begin{verbatim}
 $ curl --proto '=https' --tlsv1.2 -sSf https://sh.rustup.rs | sh
\end{verbatim}
which downloads the compiler \texttt{rustc}, the package manager \texttt{cargo}, and a few other helpful tools.
When the installation has completed make sure to read and follow the instructions printed on screen.
See also \url{https://www.rust-lang.org/tools/install} for more details and for installation instructions for other operating systems.

\subsubsection*{Installation of the command-line program \texorpdfstring{\texttt{pineappl}}{pineappl}}

The command-line program \texttt{pineappl} is compiled and installed using
\begin{verbatim}
 $ cargo install pineappl_cli
\end{verbatim}
This program also needs \texttt{LHAPDF} \cite{Buckley:2014ana} installed; make sure that the environment variables \texttt{PATH}, \texttt{LD\_LIBRARY\_PATH}, and \texttt{PKG\_CONFIG\_PATH} are properly set.
For usage instructions simply type \texttt{pineappl} in your shell and read the help message.

\subsubsection*{Installation of the C-language interface (optional)}

For the C interface you need to first install \texttt{cargo-c},
\begin{verbatim}
 $ cargo install cargo-c
\end{verbatim}
and then download the \textsc{PineAPPL} repository, compile and finally install into it into a directory \texttt{\$prefix} as follows:
\begin{verbatim}
 $ git clone https://github.com/N3PDF/pineappl/
 $ cd pineappl_capi/
 $ cargo cinstall --release --prefix=DIRECTORY
\end{verbatim}
The last line will install the C header \texttt{pineappl\_capi.h}, the library, and a pkg-config\footnote{A standard way on Linux to express how dependencies are compiled/linked against, see \url{https://www.freedesktop.org/wiki/Software/pkg-config/}} file (\texttt{pineappl\_capi.pc}) into the directory specified as \texttt{DIRECTORY}.
Make sure that the environment variables \texttt{PATH}, \texttt{LD\_LIBRARY\_PATH}, and \texttt{PKG\_CONFIG\_PATH} are properly set.
The latter is needed for \texttt{pkg-config -{}-cflags -{}-libs pineappl\_capi} to work, which prints the necessary compiler/linker flags.

After being installed, one can compile and link against the library.
See appendix~\ref{app:example-program} for an example.

\section{Parton Luminosities}
\label{app:lumis}

\begin{table}[!t]
  \centering
  \tiny
  \begin{tabularx}{\textwidth}{rXXX@{}S@{}S@{}S@{}S@{}S@{}S@{}S@{}S@{}S@{}S@{}S@{}S@{}S}
  \toprule
    &                  &                  &                 &    1  &    2  &     3 &     4 &     5 &     6 &     7 &     8 &     9 &    10 &    11 &    12 &    13 \\
  \midrule
  1 & $u,\bar u$       & $ c,\bar c$      &                 & 27.42 & 28.83 & 29.82 & 30.53 & 31.06 & 31.46 & 31.73 & 32.07 & 32.62 & 33.12 & 33.71 & 33.71 & 33.95 \\
  2 & $\gamma,\bar c$  & $\gamma, \bar u$ &                 & -0.03 & -0.03 & -0.03 & -0.03 & -0.02 & -0.01 & -0.01 & -0.00 &  0.01 &  0.01 &  0.02 &  0.02 &  0.02 \\
  3 & $g,\bar c$       & $g,\bar u$       &                 & -0.77 & -0.58 & -0.39 & -0.27 & -0.17 & -0.10 & -0.05 &  0.02 &  0.08 &  0.12 &  0.13 &  0.10 &  0.09 \\
  4 & $u,\gamma$       & $c,\gamma$       &                 & -0.09 & -0.10 & -0.09 & -0.07 & -0.05 & -0.02 &  0.01 &  0.05 &  0.11 &  0.19 &  0.21 &  0.30 &  0.40 \\
  5 & $u,g$            & $c,g$            &                 & -0.76 & -0.42 & -0.08 &  0.19 &  0.42 &  0.59 &  0.76 &  0.93 &  1.22 &  1.50 &  1.68 &  1.88 &  2.15 \\
  6 & $d,\bar d$       & $s,\bar s$       &                 & 24.52 & 21.78 & 19.63 & 18.13 & 17.07 & 16.24 & 15.62 & 14.91 & 13.83 & 12.87 & 12.26 & 11.51 & 10.88 \\
  7 & $\gamma, \bar s$ & $\gamma, \bar d$ &                 &  0.00 &  0.00 &  0.00 &  0.00 &  0.00 &  0.00 &  0.00 &  0.00 &  0.00 &  0.00 &  0.00 &  0.00 &  0.00 \\
  8 & $g,\bar s$       & $g,\bar d$       &                 & -0.84 & -0.55 & -0.34 & -0.21 & -0.13 & -0.06 & -0.02 &  0.02 &  0.08 &  0.10 &  0.10 &  0.09 &  0.07 \\
  9 & $d,\gamma$       & $s,\gamma$       &                 &  0.00 &  0.00 &  0.00 &  0.00 &  0.01 &  0.01 &  0.01 &  0.01 &  0.01 &  0.01 &  0.01 &  0.01 &  0.01 \\
 10 & $d,g$            & $s,g$            &                 & -0.89 & -0.53 & -0.27 & -0.10 &  0.01 &  0.10 &  0.17 &  0.23 &  0.33 &  0.39 &  0.42 &  0.45 &  0.44 \\
 11 & $b,\bar b$       &                  &                 &  1.15 &  0.92 &  0.74 &  0.60 &  0.51 &  0.43 &  0.39 &  0.31 &  0.22 &  0.14 &  0.09 &  0.05 &  0.02 \\
 12 & $\gamma,\bar b$  &                  &                 &  0.00 &  0.00 &  0.00 &  0.00 &  0.00 &  0.00 &  0.00 &  0.00 &  0.00 &  0.00 &  0.00 &  0.00 &  0.00 \\
 13 & $g,\bar b$       &                  &                 & -0.17 & -0.11 & -0.07 & -0.05 & -0.03 & -0.02 & -0.01 & -0.00 &  0.00 &  0.01 &  0.01 &  0.01 &  0.01 \\
 14 & $b,\gamma$       &                  &                 &  0.00 &  0.00 &  0.00 &  0.00 &  0.00 &  0.00 &  0.00 &  0.00 &  0.00 &  0.00 &  0.00 &  0.00 &  0.00 \\
 15 & $b,g$            &                  &                 & -0.17 & -0.11 & -0.07 & -0.04 & -0.03 & -0.02 & -0.01 & -0.00 &  0.00 &  0.01 &  0.01 &  0.01 &  0.00 \\
 16 & $\bar c,c$       & $\bar u,u$       &                 & 27.42 & 28.83 & 29.87 & 30.54 & 31.04 & 31.38 & 31.64 & 32.02 & 32.56 & 33.14 & 33.31 & 34.08 & 33.88 \\
 17 & $\gamma, u$      & $\gamma, c$      &                 & -0.09 & -0.10 & -0.09 & -0.07 & -0.04 & -0.03 & -0.01 &  0.04 &  0.11 &  0.18 &  0.23 &  0.31 &  0.44 \\
 18 & $g,u$            & $g,c$            &                 & -0.77 & -0.42 & -0.08 &  0.19 &  0.41 &  0.59 &  0.73 &  0.93 &  1.23 &  1.48 &  1.68 &  1.90 &  2.12 \\
 19 & $\bar c,\gamma$  & $\bar u,\gamma$  &                 & -0.03 & -0.03 & -0.03 & -0.03 & -0.02 & -0.01 & -0.01 & -0.00 &  0.01 &  0.01 &  0.02 &  0.02 &  0.02 \\
 20 & $\bar c, g$      & $\bar u, g$      &                 & -0.76 & -0.57 & -0.40 & -0.27 & -0.17 & -0.10 & -0.04 &  0.02 &  0.08 &  0.12 &  0.13 &  0.11 &  0.09 \\
 21 & $\bar s, s$      & $\bar d, d$      &                 & 24.51 & 21.78 & 19.62 & 18.17 & 17.04 & 16.27 & 15.66 & 14.86 & 13.81 & 12.89 & 12.28 & 11.59 & 11.12 \\
 22 & $\gamma, d$      & $\gamma, s$      &                 &  0.00 &  0.00 &  0.00 &  0.01 &  0.01 &  0.01 &  0.01 &  0.01 &  0.01 &  0.01 &  0.01 &  0.01 &  0.01 \\
 23 & $g,d$            & $g,s$            &                 & -0.88 & -0.53 & -0.26 & -0.10 &  0.01 &  0.10 &  0.15 &  0.24 &  0.32 &  0.39 &  0.43 &  0.44 &  0.43 \\
 24 & $\bar s,\gamma$  & $\bar d,\gamma$  &                 &  0.00 &  0.00 &  0.00 &  0.00 &  0.00 &  0.00 &  0.00 &  0.00 &  0.00 &  0.00 &  0.00 &  0.00 &  0.00 \\
 25 & $\bar s,g$       & $\bar d, g$      &                 & -0.84 & -0.55 & -0.34 & -0.21 & -0.12 & -0.06 & -0.02 &  0.03 &  0.07 &  0.10 &  0.10 &  0.09 &  0.07 \\
 26 & $\bar b,b$       &                  &                 &  1.15 &  0.92 &  0.73 &  0.60 &  0.51 &  0.44 &  0.38 &  0.31 &  0.22 &  0.14 &  0.09 &  0.05 &  0.02 \\
 27 & $\gamma, b$      &                  &                 &  0.00 &  0.00 &  0.00 &  0.00 &  0.00 &  0.00 &  0.00 &  0.00 &  0.00 &  0.00 &  0.00 &  0.00 &  0.00 \\
 28 & $g,b$            &                  &                 & -0.17 & -0.11 & -0.07 & -0.04 & -0.03 & -0.02 & -0.01 & -0.00 &  0.00 &  0.01 &  0.01 &  0.01 &  0.01 \\
 29 & $\bar b,\gamma$  &                  &                 &  0.00 &  0.00 &  0.00 &  0.00 &  0.00 &  0.00 &  0.00 &  0.00 &  0.00 &  0.00 &  0.00 &  0.00 &  0.00 \\
 30 & $\bar b,g$       &                  &                 & -0.17 & -0.11 & -0.07 & -0.04 & -0.03 & -0.02 & -0.01 & -0.00 &  0.00 &  0.01 &  0.01 &  0.01 &  0.01 \\
 31 & $\gamma,\gamma$  &                  &                 &  1.30 &  1.86 &  2.32 &  2.58 &  2.75 &  2.86 &  2.91 &  2.98 &  2.98 &  2.93 &  2.93 &  3.06 &  3.56 \\
 32 & $\bar b,\gamma$  & $\bar s,\gamma$  & $\bar d,\gamma$ & -0.01 & -0.01 & -0.01 & -0.01 & -0.00 & -0.00 &  0.00 &  0.00 &  0.01 &  0.01 &  0.01 &  0.01 &  0.01 \\
 33 & $d,\gamma$       & $s,\gamma$       & $b,\gamma$      & -0.02 & -0.02 & -0.02 & -0.01 & -0.01 & -0.00 &  0.01 &  0.01 &  0.03 &  0.04 &  0.05 &  0.07 &  0.08 \\
 34 & $\gamma,\bar b$  & $\gamma,\bar s$  & $\gamma,\bar d$ & -0.01 & -0.01 & -0.01 & -0.01 & -0.01 & -0.00 &  0.00 &  0.00 &  0.01 &  0.01 &  0.01 &  0.02 &  0.01 \\
 35 & $\gamma,d$       & $\gamma,s$       & $\gamma, b$     & -0.02 & -0.02 & -0.02 & -0.01 & -0.01 & -0.00 &  0.01 &  0.02 &  0.03 &  0.03 &  0.05 &  0.07 &  0.08 \\
 \bottomrule
\end{tabularx}
\\
  \caption{The 35 parton luminosities contributing to the predictions, accurate
    to NLO QCD+EW, of the Drell--Yan lepton-pair production measured by the
    ATLAS experiment at a centre-of-mass energy of
    \SI{7}{\tera\electronvolt}~\cite{Aad:2013iua}. The contributions
    are reported, in percentage, for each of the 13 bins. The bin edges
    are in the invariant mass of the lepton pair, $M_{\ell\bar\ell}$, as follows:
    116, 130, 150, 170, 190,
    210, 230, 250, 300, 400, 500, 700, 1000, 1500 GeV.}
  \label{tab:B1}
\end{table}

\begin{table}[!p]
  \centering
  \tiny
  \begin{tabularx}{\textwidth}{rXXX@{}S@{}S@{}S@{}S@{}S@{}S@{}S@{}S@{}S@{}S@{}S@{}S}
  \toprule
    &                  &                  &                 &     1 &     2 &     3 &     4 &     5 &     6 &     7 &     8 &     9 &    10 &    11 &    12 \\
  \midrule
  1 & $u,\bar u$       & $ c,\bar c$      &                 & 16.87 & 17.00 & 16.94 & 17.06 & 16.99 & 17.02 & 17.03 & 17.19 & 17.50 & 17.21 & 17.34 & 17.44 \\
  2 & $\gamma,\bar c$  & $\gamma, \bar u$ &                 &  0.22 &  0.23 &  0.23 &  0.22 &  0.23 &  0.23 &  0.22 &  0.24 &  0.23 &  0.23 &  0.22 &  0.23 \\
  3 & $g,\bar c$       & $g,\bar u$       &                 &  8.98 &  8.93 &  8.96 &  8.84 &  8.91 &  8.76 &  8.76 &  8.63 &  8.51 &  8.38 &  8.44 &  8.20 \\
  4 & $u,\gamma$       & $c,\gamma$       &                 &  0.56 &  0.56 &  0.56 &  0.57 &  0.54 &  0.55 &  0.58 &  0.56 &  0.55 &  0.54 &  0.52 &  0.55 \\
  5 & $u,g$            & $c,g$            &                 & 10.87 & 10.88 & 10.89 & 10.89 & 10.96 & 10.86 & 11.14 & 11.06 & 10.94 & 11.31 & 11.31 & 11.44 \\
  6 & $d,\bar d$       & $s,\bar s$       &                 &  5.12 &  5.21 &  5.15 &  5.12 &  5.23 &  5.11 &  5.17 &  5.13 &  5.11 &  5.15 &  5.07 &  5.11 \\
  7 & $\gamma, \bar s$ & $\gamma, \bar d$ &                 &  0.00 &  0.01 &  0.01 &  0.00 &  0.00 &  0.01 &  0.01 &  0.00 &  0.00 &  0.00 &  0.01 &  0.00 \\
  8 & $g,\bar s$       & $g,\bar d$       &                 &  2.72 &  2.69 &  2.71 &  2.66 &  2.71 &  2.63 &  2.64 &  2.64 &  2.59 &  2.57 &  2.58 &  2.53 \\
  9 & $d,\gamma$       & $s,\gamma$       &                 &  0.01 &  0.00 &  0.01 &  0.01 &  0.00 &  0.00 &  0.01 &  0.00 &  0.01 &  0.01 &  0.01 &  0.01 \\
 10 & $d,g$            & $s,g$            &                 &  2.91 &  2.96 &  2.94 &  3.00 &  2.97 &  2.94 &  2.92 &  2.92 &  2.98 &  2.93 &  2.93 &  2.95 \\
 11 & $b,\bar b$       &                  &                 &  0.29 &  0.32 &  0.32 &  0.30 &  0.29 &  0.30 &  0.30 &  0.30 &  0.30 &  0.30 &  0.28 &  0.28 \\
 12 & $\gamma,\bar b$  &                  &                 &  0.00 &  0.00 &  0.00 &  0.00 &  0.00 &  0.00 &  0.00 &  0.00 &  0.00 &  0.00 &  0.00 &  0.00 \\
 13 & $g,\bar b$       &                  &                 &  0.48 &  0.45 &  0.44 &  0.47 &  0.45 &  0.46 &  0.44 &  0.44 &  0.43 &  0.43 &  0.43 &  0.42 \\
 14 & $b,\gamma$       &                  &                 &  0.00 &  0.00 &  0.00 &  0.00 &  0.00 &  0.00 &  0.00 &  0.00 &  0.00 &  0.00 &  0.00 &  0.00 \\
 15 & $b,g$            &                  &                 &  0.47 &  0.48 &  0.44 &  0.48 &  0.48 &  0.45 &  0.47 &  0.44 &  0.42 &  0.43 &  0.43 &  0.40 \\
 16 & $\bar c,c$       & $\bar u,u$       &                 & 17.00 & 16.92 & 16.99 & 17.05 & 17.01 & 17.18 & 16.99 & 17.20 & 17.16 & 17.27 & 17.49 & 17.38 \\
 17 & $\gamma, u$      & $\gamma, c$      &                 &  0.58 &  0.57 &  0.57 &  0.57 &  0.54 &  0.56 &  0.54 &  0.55 &  0.57 &  0.55 &  0.56 &  0.58 \\
 18 & $g,u$            & $g,c$            &                 & 10.82 & 10.74 & 10.86 & 10.74 & 10.81 & 10.92 & 11.07 & 10.96 & 11.23 & 11.33 & 11.28 & 11.55 \\
 19 & $\bar c,\gamma$  & $\bar u,\gamma$  &                 &  0.22 &  0.22 &  0.23 &  0.22 &  0.23 &  0.23 &  0.22 &  0.23 &  0.22 &  0.22 &  0.24 &  0.24 \\
 20 & $\bar c, g$      & $\bar u, g$      &                 &  8.96 &  8.95 &  8.92 &  8.93 &  8.87 &  8.88 &  8.69 &  8.69 &  8.59 &  8.42 &  8.34 &  8.27 \\
 21 & $\bar s, s$      & $\bar d, d$      &                 &  5.12 &  5.10 &  5.06 &  5.11 &  5.12 &  5.18 &  5.10 &  5.19 &  5.14 &  5.11 &  5.12 &  5.09 \\
 22 & $\gamma, d$      & $\gamma, s$      &                 &  0.00 &  0.01 &  0.00 &  0.01 &  0.00 &  0.01 &  0.00 &  0.01 &  0.00 &  0.00 &  0.01 &  0.00 \\
 23 & $g,d$            & $g,s$            &                 &  2.94 &  2.93 &  2.95 &  2.98 &  2.92 &  2.92 &  2.94 &  2.91 &  2.92 &  2.97 &  2.91 &  2.93 \\
 24 & $\bar s,\gamma$  & $\bar d,\gamma$  &                 &  0.01 &  0.01 &  0.00 &  0.01 &  0.00 &  0.00 &  0.01 &  0.00 &  0.01 &  0.01 &  0.01 &  0.01 \\
 25 & $\bar s,g$       & $\bar d, g$      &                 &  2.73 &  2.72 &  2.71 &  2.71 &  2.67 &  2.69 &  2.70 &  2.63 &  2.58 &  2.62 &  2.52 &  2.49 \\
 26 & $\bar b,b$       &                  &                 &  0.29 &  0.31 &  0.31 &  0.29 &  0.29 &  0.31 &  0.28 &  0.30 &  0.28 &  0.29 &  0.28 &  0.26 \\
 27 & $\gamma, b$      &                  &                 &  0.00 &  0.00 &  0.00 &  0.00 &  0.00 &  0.00 &  0.00 &  0.00 &  0.00 &  0.00 &  0.00 &  0.00 \\
 28 & $g,b$            &                  &                 &  0.46 &  0.44 &  0.45 &  0.44 &  0.48 &  0.46 &  0.47 &  0.44 &  0.45 &  0.45 &  0.40 &  0.39 \\
 29 & $\bar b,\gamma$  &                  &                 &  0.00 &  0.00 &  0.00 &  0.00 &  0.00 &  0.00 &  0.00 &  0.00 &  0.00 &  0.00 &  0.00 &  0.00 \\
 30 & $\bar b,g$       &                  &                 &  0.46 &  0.48 &  0.47 &  0.45 &  0.42 &  0.45 &  0.43 &  0.44 &  0.42 &  0.44 &  0.42 &  0.42 \\
 31 & $\gamma,\gamma$  &                  &                 &  0.49 &  0.48 &  0.48 &  0.48 &  0.46 &  0.49 &  0.49 &  0.48 &  0.45 &  0.46 &  0.45 &  0.44 \\
 32 & $\bar b,\gamma$  & $\bar s,\gamma$  & $\bar d,\gamma$ &  0.08 &  0.07 &  0.08 &  0.08 &  0.07 &  0.08 &  0.08 &  0.08 &  0.08 &  0.07 &  0.08 &  0.07 \\
 33 & $d,\gamma$       & $s,\gamma$       & $b,\gamma$      &  0.13 &  0.12 &  0.12 &  0.11 &  0.12 &  0.12 &  0.12 &  0.12 &  0.12 &  0.12 &  0.12 &  0.13 \\
 34 & $\gamma,\bar b$  & $\gamma,\bar s$  & $\gamma,\bar d$ &  0.07 &  0.08 &  0.07 &  0.07 &  0.08 &  0.08 &  0.08 &  0.07 &  0.08 &  0.07 &  0.08 &  0.08 \\
 35 & $\gamma,d$       & $\gamma,s$       & $\gamma, b$     &  0.12 &  0.12 &  0.12 &  0.12 &  0.11 &  0.11 &  0.12 &  0.12 &  0.11 &  0.11 &  0.12 &  0.12 \\
 \midrule
    &                  &                  &                 &    13 &    14 &    15 &    16 &    17 &    18 &    19 &    20 &    21 &    22 &    23 &    24 \\
 \midrule
  1 & $u,\bar u$       & $ c,\bar c$      &                 & 17.52 & 17.69 & 17.89 & 17.93 & 18.41 & 18.50 & 19.19 & 19.83 & 20.68 & 21.20 & 22.55 & 22.96 \\
  2 & $\gamma,\bar c$  & $\gamma, \bar u$ &                 &  0.25 &  0.24 &  0.22 &  0.24 &  0.21 &  0.23 &  0.20 &  0.23 &  0.21 &  0.17 &  0.15 &  0.15 \\
  3 & $g,\bar c$       & $g,\bar u$       &                 &  8.07 &  8.02 &  7.66 &  7.55 &  7.26 &  7.08 &  6.61 &  6.25 &  5.81 &  5.60 &  4.91 &  4.89 \\
  4 & $u,\gamma$       & $c,\gamma$       &                 &  0.55 &  0.56 &  0.54 &  0.53 &  0.49 &  0.50 &  0.49 &  0.41 &  0.39 &  0.40 &  0.39 &  0.59 \\
  5 & $u,g$            & $c,g$            &                 & 11.56 & 11.75 & 11.67 & 11.91 & 11.96 & 12.01 & 12.01 & 11.97 & 11.80 & 11.68 & 11.28 & 12.60 \\
  6 & $d,\bar d$       & $s,\bar s$       &                 &  5.09 &  5.05 &  5.06 &  5.11 &  5.16 &  5.03 &  5.18 &  5.34 &  5.45 &  5.41 &  5.47 &  5.61 \\
  7 & $\gamma, \bar s$ & $\gamma, \bar d$ &                 &  0.01 &  0.00 &  0.00 &  0.00 &  0.01 &  0.00 &  0.01 &  0.00 &  0.00 &  0.00 &  0.00 &  0.00 \\
  8 & $g,\bar s$       & $g,\bar d$       &                 &  2.44 &  2.45 &  2.41 &  2.33 &  2.28 &  2.16 &  2.13 &  2.02 &  1.86 &  1.82 &  1.63 &  1.41 \\
  9 & $d,\gamma$       & $s,\gamma$       &                 &  0.01 &  0.00 &  0.01 &  0.00 &  0.00 &  0.00 &  0.00 &  0.01 &  0.00 &  0.01 &  0.01 &  0.00 \\
 10 & $d,g$            & $s,g$            &                 &  2.91 &  2.94 &  2.93 &  2.87 &  2.82 &  2.87 &  2.81 &  2.76 &  2.66 &  2.49 &  2.45 &  2.09 \\
 11 & $b,\bar b$       &                  &                 &  0.27 &  0.26 &  0.25 &  0.25 &  0.23 &  0.25 &  0.24 &  0.26 &  0.20 &  0.26 &  0.19 &  0.16 \\
 12 & $\gamma,\bar b$  &                  &                 &  0.00 &  0.00 &  0.00 &  0.00 &  0.00 &  0.00 &  0.00 &  0.00 &  0.00 &  0.00 &  0.00 &  0.01 \\
 13 & $g,\bar b$       &                  &                 &  0.41 &  0.37 &  0.39 &  0.36 &  0.36 &  0.32 &  0.31 &  0.28 &  0.30 &  0.27 &  0.16 &  0.22 \\
 14 & $b,\gamma$       &                  &                 &  0.00 &  0.00 &  0.00 &  0.00 &  0.00 &  0.00 &  0.00 &  0.00 &  0.00 &  0.00 &  0.00 &  0.00 \\
 15 & $b,g$            &                  &                 &  0.40 &  0.38 &  0.36 &  0.36 &  0.34 &  0.33 &  0.29 &  0.28 &  0.27 &  0.27 &  0.20 &  0.23 \\
 16 & $\bar c,c$       & $\bar u,u$       &                 & 17.71 & 17.50 & 18.22 & 18.23 & 18.33 & 18.76 & 19.13 & 19.71 & 20.69 & 21.12 & 21.51 & 21.39 \\
 17 & $\gamma, u$      & $\gamma, c$      &                 &  0.54 &  0.55 &  0.56 &  0.53 &  0.54 &  0.51 &  0.49 &  0.44 &  0.55 &  0.41 &  0.25 &  0.43 \\
 18 & $g,u$            & $g,c$            &                 & 11.47 & 11.74 & 11.73 & 11.98 & 12.01 & 12.13 & 12.16 & 11.89 & 11.61 & 11.72 & 11.91 & 12.47 \\
 19 & $\bar c,\gamma$  & $\bar u,\gamma$  &                 &  0.23 &  0.22 &  0.23 &  0.23 &  0.22 &  0.23 &  0.20 &  0.20 &  0.21 &  0.17 &  0.13 &  0.19 \\
 20 & $\bar c, g$      & $\bar u, g$      &                 &  8.05 &  7.97 &  7.64 &  7.48 &  7.32 &  7.11 &  6.79 &  6.25 &  5.81 &  5.44 &  5.06 &  4.10 \\
 21 & $\bar s, s$      & $\bar d, d$      &                 &  5.16 &  5.08 &  5.07 &  5.07 &  5.07 &  5.13 &  5.15 &  5.31 &  5.35 &  5.75 &  5.94 &  5.34 \\
 22 & $\gamma, d$      & $\gamma, s$      &                 &  0.01 &  0.00 &  0.00 &  0.01 &  0.01 &  0.01 &  0.00 &  0.00 &  0.00 &  0.00 &  0.00 &  0.01 \\
 23 & $g,d$            & $g,s$            &                 &  2.90 &  2.94 &  2.93 &  2.86 &  2.87 &  2.84 &  2.78 &  2.80 &  2.67 &  2.56 &  2.31 &  2.29 \\
 24 & $\bar s,\gamma$  & $\bar d,\gamma$  &                 &  0.00 &  0.00 &  0.00 &  0.00 &  0.01 &  0.00 &  0.00 &  0.00 &  0.00 &  0.01 &  0.00 &  0.01 \\
 25 & $\bar s,g$       & $\bar d, g$      &                 &  2.51 &  2.42 &  2.38 &  2.34 &  2.24 &  2.23 &  2.10 &  2.15 &  1.87 &  1.69 &  1.86 &  1.58 \\
 26 & $\bar b,b$       &                  &                 &  0.27 &  0.26 &  0.27 &  0.24 &  0.24 &  0.24 &  0.23 &  0.21 &  0.21 &  0.21 &  0.24 &  0.21 \\
 27 & $\gamma, b$      &                  &                 &  0.00 &  0.00 &  0.00 &  0.00 &  0.00 &  0.00 &  0.00 &  0.00 &  0.00 &  0.00 &  0.00 &  0.00 \\
 28 & $g,b$            &                  &                 &  0.43 &  0.37 &  0.38 &  0.38 &  0.34 &  0.34 &  0.34 &  0.27 &  0.27 &  0.26 &  0.30 &  0.22 \\
 29 & $\bar b,\gamma$  &                  &                 &  0.00 &  0.00 &  0.00 &  0.00 &  0.00 &  0.00 &  0.00 &  0.00 &  0.00 &  0.00 &  0.00 &  0.00 \\
 30 & $\bar b,g$       &                  &                 &  0.40 &  0.38 &  0.38 &  0.37 &  0.34 &  0.33 &  0.30 &  0.31 &  0.27 &  0.24 &  0.16 &  0.19 \\
 31 & $\gamma,\gamma$  &                  &                 &  0.46 &  0.46 &  0.44 &  0.46 &  0.56 &  0.47 &  0.50 &  0.52 &  0.55 &  0.55 &  0.59 &  0.50 \\
 32 & $\bar b,\gamma$  & $\bar s,\gamma$  & $\bar d,\gamma$ &  0.08 &  0.08 &  0.08 &  0.08 &  0.07 &  0.07 &  0.07 &  0.07 &  0.05 &  0.06 &  0.10 & -0.00 \\
 33 & $d,\gamma$       & $s,\gamma$       & $b,\gamma$      &  0.12 &  0.12 &  0.11 &  0.11 &  0.10 &  0.10 &  0.11 &  0.10 &  0.09 &  0.11 &  0.06 &  0.05 \\
 34 & $\gamma,\bar b$  & $\gamma,\bar s$  & $\gamma,\bar d$ &  0.08 &  0.08 &  0.09 &  0.07 &  0.08 &  0.08 &  0.07 &  0.07 &  0.06 &  0.05 &  0.06 &  0.06 \\
 35 & $\gamma,d$       & $\gamma,s$       & $\gamma, b$     &  0.10 &  0.12 &  0.10 &  0.11 &  0.12 &  0.10 &  0.11 &  0.07 &  0.09 &  0.07 &  0.09 &  0.04 \\
 \bottomrule
\end{tabularx}
\\
  \caption{The 35 parton luminosities contributing to the predictions, accurate
    to NLO QCD+EW, of the double differential Drell--Yan lepton pair production
    measured by the CMS experiment at a centre-of-mass energy of
    \SI{7}{\tera\electronvolt}~\cite{Chatrchyan:2013tia}. The invariant mass
    bin is $\SI{20}{\giga\electronvolt}<M_{\ell\bar\ell}<\SI{30}{\giga\electronvolt}$; bins 1--24 are in the rapidity of the lepton pair,
  $0.0<y_{\ell\bar\ell}<2.4$, and each bin has a width of 0.1.}
  \label{tab:B2}
\end{table}

\begin{table}[!p]
  \centering
  \tiny
\begin{tabularx}{\textwidth}{rXXX@{}S@{}S@{}S@{}S@{}S@{}S@{}S@{}S@{}S@{}S@{}S@{}S}
 \toprule
    &                  &                  &                 &    25 &    26 &    27 &    28 &    29 &    30 &    31 &    32 &    33 &    34 &    35 &    36 \\
 \midrule
  1 & $u,\bar u$       & $ c,\bar c$      &                 & 35.63 & 35.65 & 35.58 & 35.60 & 35.68 & 35.68 & 35.77 & 35.85 & 35.81 & 36.05 & 35.97 & 36.15 \\
  2 & $\gamma,\bar c$  & $\gamma, \bar u$ &                 & -0.01 & -0.02 & -0.02 & -0.02 & -0.02 & -0.02 & -0.01 & -0.02 & -0.01 & -0.02 & -0.01 & -0.02 \\
  3 & $g,\bar c$       & $g,\bar u$       &                 &  0.03 &  0.03 &  0.07 &  0.05 &  0.03 &  0.02 & -0.03 & -0.05 & -0.03 & -0.07 & -0.10 & -0.15 \\
  4 & $u,\gamma$       & $c,\gamma$       &                 &  0.03 &  0.02 &  0.03 &  0.04 &  0.04 &  0.03 &  0.02 &  0.01 &  0.02 &  0.01 &  0.01 &  0.01 \\
  5 & $u,g$            & $c,g$            &                 &  0.51 &  0.50 &  0.54 &  0.53 &  0.56 &  0.60 &  0.60 &  0.61 &  0.66 &  0.66 &  0.67 &  0.69 \\
  6 & $d,\bar d$       & $s,\bar s$       &                 & 11.37 & 11.33 & 11.40 & 11.33 & 11.33 & 11.32 & 11.29 & 11.25 & 11.22 & 11.15 & 11.09 & 11.11 \\
  7 & $\gamma, \bar s$ & $\gamma, \bar d$ &                 &  0.01 &  0.01 &  0.01 &  0.01 &  0.01 &  0.01 &  0.01 &  0.01 &  0.01 &  0.01 &  0.01 &  0.01 \\
\bottomrule
\end{tabularx}
\\
  \caption{Same as table~\ref{tab:B2}, but for the invariant mass bin
    $\SI{30}{\giga\electronvolt}<M_{\ell\bar\ell}<\SI{45}{\giga\electronvolt}$.}
  \label{tab:B3}
\end{table}

\begin{table}[!p]
  \centering
  \tiny
  \begin{tabularx}{\textwidth}{rXXX@{}S@{}S@{}S@{}S@{}S@{}S@{}S@{}S@{}S@{}S@{}S@{}S}
  \toprule
    &                  &                  &                 &    49 &    50 &    51 &    52 &    53 &    54 &    55 &    56 &    57 &    58 &    59 &    60 \\
  \midrule
  1 & $u,\bar u$       & $ c,\bar c$      &                 & 35.55 & 35.51 & 35.54 & 35.66 & 35.51 & 35.53 & 35.70 & 35.67 & 36.27 & 36.28 & 36.60 & 37.17 \\
  2 & $\gamma,\bar c$  & $\gamma, \bar u$ &                 & -0.08 & -0.08 & -0.07 & -0.09 & -0.08 & -0.07 & -0.09 & -0.09 & -0.08 & -0.10 & -0.10 & -0.13 \\
  3 & $g,\bar c$       & $g,\bar u$       &                 & -1.29 & -1.29 & -1.34 & -1.35 & -1.30 & -1.32 & -1.32 & -1.35 & -1.40 & -1.43 & -1.56 & -1.67 \\
  4 & $u,\gamma$       & $c,\gamma$       &                 & -0.13 & -0.13 & -0.11 & -0.09 & -0.14 & -0.10 & -0.13 & -0.13 & -0.16 & -0.16 & -0.22 & -0.25 \\
  5 & $u,g$            & $c,g$            &                 & -1.10 & -1.09 & -1.08 & -1.07 & -1.08 & -1.04 & -1.11 & -1.03 & -1.13 & -1.15 & -1.31 & -1.47 \\
  6 & $d,\bar d$       & $s,\bar s$       &                 & 15.51 & 15.31 & 15.25 & 15.23 & 15.35 & 15.22 & 15.13 & 15.08 & 15.03 & 14.85 & 15.15 & 15.04 \\
  7 & $\gamma, \bar s$ & $\gamma, \bar d$ &                 &  0.01 &  0.01 &  0.01 &  0.01 &  0.01 &  0.01 &  0.01 &  0.00 &  0.01 &  0.01 &  0.00 &  0.01 \\
\bottomrule
\end{tabularx}
\\
    \caption{Same as table~\ref{tab:B2}, but for the invariant mass bin
    $\SI{45}{\giga\electronvolt}<M_{\ell\bar\ell}<\SI{60}{\giga\electronvolt}$.}
  \label{tab:B4}
\end{table}

\begin{table}[!p]
  \centering
  \tiny
  \begin{tabularx}{\textwidth}{rXXX@{}S@{}S@{}S@{}S@{}S@{}S@{}S@{}S@{}S@{}S@{}S@{}S}
 \toprule
    &                  &                  &                 &    73 &    74 &    75 &    76 &    77 &    78 &    79 &    80 &    81 &    82 &    83 &    84 \\
 \midrule
  1 & $u,\bar u$       & $ c,\bar c$      &                 & 22.04 & 22.07 & 22.16 & 22.20 & 22.41 & 22.54 & 22.70 & 22.87 & 22.95 & 23.15 & 23.28 & 23.45 \\
  2 & $\gamma,\bar c$  & $\gamma, \bar u$ &                 & -0.00 & -0.01 & -0.01 & -0.01 & -0.01 & -0.01 & -0.01 & -0.01 & -0.01 & -0.01 & -0.01 & -0.01 \\
  3 & $g,\bar c$       & $g,\bar u$       &                 & -0.95 & -0.96 & -0.96 & -0.98 & -1.00 & -1.02 & -1.02 & -1.01 & -0.98 & -0.96 & -0.93 & -0.91 \\
  4 & $u,\gamma$       & $c,\gamma$       &                 & -0.01 & -0.01 & -0.01 & -0.01 & -0.01 & -0.01 & -0.01 & -0.01 & -0.01 & -0.01 & -0.01 & -0.01 \\
  5 & $u,g$            & $c,g$            &                 & -0.89 & -0.90 & -0.91 & -0.93 & -0.97 & -1.02 & -1.01 & -1.00 & -0.99 & -0.98 & -0.94 & -0.90 \\
  6 & $d,\bar d$       & $s,\bar s$       &                 & 30.85 & 30.88 & 30.83 & 30.87 & 30.86 & 30.94 & 30.88 & 30.65 & 30.47 & 30.25 & 30.02 & 29.76 \\
  7 & $\gamma, \bar s$ & $\gamma, \bar d$ &                 & -0.00 & -0.00 & -0.00 & -0.00 & -0.00 & -0.00 & -0.00 & -0.00 & -0.00 & -0.00 & -0.00 & -0.00 \\
\bottomrule
\end{tabularx}
\\
    \caption{Same as table~\ref{tab:B2}, but for the invariant mass bin
    $\SI{60}{\giga\electronvolt}<M_{\ell\bar\ell}<\SI{120}{\giga\electronvolt}$.}
  \label{tab:B5}
\end{table}

\begin{table}[!p]
  \centering
  \tiny
  \begin{tabularx}{\textwidth}{rXXX@{}S@{}S@{}S@{}S@{}S@{}S@{}S@{}S@{}S@{}S@{}S@{}S}
 \toprule
    &                  &                  &                 &    97 &    98 &    99 &   100 &   101 &   102 &   103 &   104 &   105 &   106 &   107 &   108 \\
 \midrule
  1 & $u,\bar u$       & $ c,\bar c$      &                 & 28.42 & 28.23 & 27.88 & 28.64 & 28.74 & 29.04 & 28.88 & 29.27 & 29.40 & 29.79 & 29.95 & 30.24 \\
  2 & $\gamma,\bar c$  & $\gamma, \bar u$ &                 & -0.11 & -0.06 & -0.11 & -0.07 & -0.06 & -0.07 & -0.07 & -0.06 & -0.06 & -0.04 & -0.03 & -0.05 \\
  3 & $g,\bar c$       & $g,\bar u$       &                 & -1.26 & -1.22 & -1.26 & -1.20 & -1.21 & -1.12 & -1.14 & -1.14 & -1.04 & -1.07 & -0.98 & -0.96 \\
  4 & $u,\gamma$       & $c,\gamma$       &                 & -0.27 & -0.19 & -0.24 & -0.25 & -0.19 & -0.16 & -0.16 & -0.13 & -0.14 & -0.13 & -0.11 & -0.09 \\
  5 & $u,g$            & $c,g$            &                 & -1.40 & -1.38 & -1.35 & -1.34 & -1.39 & -1.31 & -1.45 & -1.47 & -1.33 & -1.41 & -1.38 & -1.43 \\
  6 & $d,\bar d$       & $s,\bar s$       &                 & 24.28 & 24.35 & 24.49 & 24.18 & 23.96 & 23.93 & 23.73 & 23.71 & 23.68 & 23.03 & 22.98 & 22.78 \\
  7 & $\gamma, \bar s$ & $\gamma, \bar d$ &                 &  0.00 &  0.00 &  0.00 &  0.00 &  0.00 &  0.00 &  0.00 &  0.00 &  0.00 &  0.00 &  0.00 &  0.00 \\
\bottomrule
\end{tabularx}
\\
    \caption{Same as table~\ref{tab:B2}, but for the invariant mass bin
    $\SI{120}{\giga\electronvolt}<M_{\ell\bar\ell}<\SI{200}{\giga\electronvolt}$.}
  \label{tab:B6}
\end{table}

\begin{table}[!t]
  \centering
  \tiny
  \begin{tabularx}{\textwidth}{rXXX@{}S@{}S@{}S@{}S@{}S@{}S@{}S@{}S@{}S@{}S@{}S@{}S}
 \toprule
    &                  &                  &                 &   121 &   122 &   123 &   124 &   125 &   126 &   127 &   128 &   129 &   130 &   131 &   132 \\
 \midrule
  1 & $u,\bar u$       & $ c,\bar c$      &                 & 34.15 & 35.16 & 34.90 & 35.92 & 36.34 & 38.15 & 37.57 & 40.33 & 44.41 & 41.92 & 47.45 & 14.53 \\
  2 & $\gamma,\bar c$  & $\gamma, \bar u$ &                 & -0.19 & -0.13 & -0.17 & -0.13 & -0.13 & -0.13 & -0.11 & -0.10 & -0.09 & -0.07 & -0.03 & -0.02 \\
  3 & $g,\bar c$       & $g,\bar u$       &                 & -1.33 & -1.31 & -1.26 & -1.16 & -1.05 & -0.89 & -0.87 & -0.81 & -0.54 & -0.43 & -0.40 & -0.56 \\
  4 & $u,\gamma$       & $c,\gamma$       &                 & -0.88 & -0.71 & -0.60 & -0.48 & -0.43 & -0.36 & -0.38 & -0.21 & -0.19 & -0.17 & -0.25 & -0.14 \\
  5 & $u,g$            & $c,g$            &                 & -2.42 & -2.44 & -2.55 & -2.69 & -2.95 & -3.26 & -3.79 & -4.03 & -4.77 & -4.50 & -6.74 &-12.60 \\
  6 & $d,\bar d$       & $s,\bar s$       &                 & 19.16 & 18.52 & 19.19 & 18.54 & 17.96 & 17.04 & 17.41 & 17.52 & 15.05 & 11.73 & 12.88 & 20.30 \\
  7 & $\gamma, \bar s$ & $\gamma, \bar d$ &                 &  0.01 &  0.01 &  0.01 &  0.00 &  0.00 &  0.00 &  0.00 &  0.00 &  0.00 &  0.00 &  0.00 &  0.00 \\
\bottomrule
\end{tabularx}
\\
    \caption{Same as table~\ref{tab:B2}, but for the invariant mass bin
      $\SI{200}{\giga\electronvolt}<M_{\ell\bar\ell}<\SI{1500}{\giga\electronvolt}$. Each bin in the rapidity of the lepton pair has a width of 0.2.}
  \label{tab:B7}
\end{table}

\begin{table}[!p]
  \centering
  \tiny
  \begin{tabularx}{\textwidth}{rXXX@{}S@{}S@{}S@{}S@{}S@{}S@{}S@{}S@{}S@{}S@{}S@{}S@{}S}
  \toprule
     &                 &                 &                 &     1 &     2 &     3 &     4 &     5 &     6 &     7 &     8 &     1 &     2 &     3 &     4 &    5  \\
  \midrule
   1 & $g,g$           &                 &                 & 81.13 & 81.53 & 81.12 & 79.26 & 76.04 & 71.78 & 67.44 & 61.14 & 82.86 & 82.18 & 80.60 & 78.33 & 74.18 \\
   2 & $\bar s,g$      & $\bar d,g$      &                 &  0.10 &  0.12 &  0.17 &  0.20 &  0.29 &  0.40 &  0.34 &  0.44 &  0.17 &  0.15 &  0.10 &  0.07 &  0.04 \\
   3 & $d,g$           & $s,g$           &                 &  0.57 &  0.52 &  0.64 &  0.78 &  1.07 &  1.37 &  1.57 &  1.91 &  0.74 &  0.73 &  0.71 &  0.64 &  0.43 \\
   4 & $\bar b,g$      &                 &                 &  0.01 &  0.01 &  0.01 &  0.02 &  0.01 &  0.02 &  0.03 &  0.03 &  0.01 &  0.00 &  0.00 &  0.00 &  0.00 \\
   5 & $b,g$           &                 &                 &  0.01 &  0.01 &  0.01 &  0.01 &  0.03 &  0.02 &  0.05 &  0.02 &  0.01 &  0.01 &  0.01 &  0.01 &  0.00 \\
   6 & $\bar c,g$      & $\bar u,g$      &                 &  0.10 &  0.07 &  0.11 &  0.16 &  0.24 &  0.26 &  0.29 &  0.28 &  0.12 &  0.10 &  0.08 &  0.05 &  0.01 \\
   7 & $u,g$           & $b,g$           &                 &  1.21 &  1.21 &  1.35 &  1.75 &  2.33 &  3.02 &  3.71 &  4.52 &  1.59 &  1.67 &  1.82 &  1.81 &  1.60 \\
\bottomrule
\end{tabularx}
\\
  \begin{tabularx}{\textwidth}{rXXX@{}S@{}S@{}S@{}S@{}S@{}S@{}S@{}S@{}S@{}S@{}S@{}S}
  \toprule
     &                 &                 &                 &    1  &     2 &     3 &     4 &     5 &     6 &    7  &     1 &     2 &     3 &     4 &     5 \\
  \midrule
   1 & $g,g$           &                 &                 & 75.95 & 79.83 & 82.25 & 83.33 & 83.66 & 83.57 & 82.79 & 84.10 & 83.37 & 81.85 & 78.93 & 69.48 \\ 
   2 & $\bar s,g$      & $\bar d,g$      &                 &  0.21 &  0.16 &  0.14 &  0.13 &  0.13 &  0.12 &  0.14 &  0.17 &  0.14 &  0.11 &  0.06 &  0.02 \\
   3 & $d,g$           & $s,g$           &                 &  0.73 &  0.63 &  0.60 &  0.62 &  0.66 &  0.74 &  0.81 &  0.77 &  0.78 &  0.73 &  0.65 &  0.32 \\
   4 & $\bar b,g$      &                 &                 &  0.02 &  0.01 &  0.01 &  0.01 &  0.01 &  0.01 &  0.01 &  0.01 &  0.01 &  0.01 &  0.00 &  0.00 \\
   5 & $b,g$           &                 &                 &  0.02 &  0.01 &  0.01 &  0.01 &  0.01 &  0.01 &  0.01 &  0.01 &  0.01 &  0.00 &  0.00 &  0.00 \\
   6 & $\bar c,g$      & $\bar u,g$      &                 &  0.15 &  0.11 &  0.10 &  0.09 &  0.09 &  0.09 &  0.09 &  0.13 &  0.10 &  0.07 &  0.05 &  0.00 \\ 
   7 & $u,g$           & $b,g$           &                 &  1.51 &  1.36 &  1.33 &  1.42 &  1.57 &  1.85 &  2.26 &  1.64 &  1.80 &  1.91 &  1.91 &  1.45 \\
\bottomrule
\end{tabularx}
\\
  \caption{The 37 parton luminosities contributing to the predictions,
    accurate to NLO QCD+EW, to the top-quark pair distributions measured by the
    ATLAS and CMS experiments at a centre-of-mass-energy of
    $\SI{8}{\tera\electronvolt}$~\cite{Aad:2015mbv,Khachatryan:2015oqa}.
    From left to right, top to bottom,
    predictions are reported for the eight bins of transverse momentum of the
    top quark, for the five bins of top-quark rapidity, for the seven bins of
    top-quark pair invariant mass, and for the five bins of top-quark pair
  rapidity, see text for details.}
  \label{tab:B8}
\end{table}

\begin{table}[!p]
  \centering
  \tiny
  \begin{tabularx}{\textwidth}{rXXX@{}S@{}S@{}S@{}S@{}S@{}S@{}S@{}S@{}S@{}S}
  \toprule
  \multicolumn{4}{l}{Table~\ref{tab:B10} continues into next page} &    1  &     2 &     3 &     4 &     5 &     6 &    7  &     8 &     9 &    10 \\
  \midrule
  1 & $g,  u$          &                   &                  &  5.86 &  6.07 &  6.32 &  6.60 &  6.92 &  7.28 &  7.67 &  8.12 &  8.72 &  9.64 \\
  2 & $\bar d, u$      &                   &                  & -0.07 & -0.07 & -0.06 & -0.03 & -0.03 & -0.01 & -0.00 &  0.03 &  0.06 &  0.10 \\
  3 & $d, u$           &                   &                  & -0.09 & -0.07 & -0.05 & -0.03 & -0.01 &  0.02 &  0.07 &  0.12 &  0.19 &  0.30 \\
  4 & $\bar b, u$      & $\bar s, u$       &                  & -0.01 & -0.01 & -0.00 & -0.00 &  0.01 &  0.01 &  0.03 &  0.03 &  0.05 &  0.06 \\
  5 & $s, u$           & $b, u$            &                  & -0.01 & -0.01 & -0.00 &  0.00 &  0.01 &  0.02 &  0.03 &  0.04 &  0.06 &  0.09 \\
  6 & $\bar u, u$      &                   &                  & -0.05 & -0.04 & -0.04 & -0.03 & -0.02 & -0.01 & -0.01 &  0.01 &  0.03 &  0.05 \\
  7 & $u, u$           &                   &                  & -0.13 & -0.10 & -0.10 & -0.08 & -0.06 & -0.03 &  0.00 &  0.05 &  0.12 &  0.18 \\
  8 & $\bar c, u$      &                   &                  & -0.03 & -0.03 & -0.03 & -0.02 & -0.02 & -0.01 & -0.00 &  0.01 &  0.02 &  0.03 \\
  9 & $c, u$           &                   &                  & -0.04 & -0.03 & -0.03 & -0.03 & -0.02 & -0.01 & -0.01 &  0.00 &  0.01 &  0.02 \\
 10 & $g, \gamma$      &                   &                  & -0.00 & -0.01 & -0.01 & -0.00 & -0.00 & -0.00 & -0.00 & -0.00 & -0.00 & -0.00 \\
 11 & $g, g$           &                   &                  & -7.63 & -7.24 & -6.84 & -6.43 & -5.98 & -5.46 & -4.87 & -4.13 & -3.29 & -2.45 \\
 12 & $g, c$           &                   &                  &  1.95 &  2.02 &  2.01 &  2.06 &  2.06 &  2.05 &  2.05 &  1.99 &  1.91 &  1.75 \\
 13 & $\bar b, c$      & $\bar d, c$       &                  & -0.01 & -0.01 & -0.01 & -0.01 & -0.00 & -0.00 & -0.00 &  0.00 &  0.00 &  0.01 \\
 14 & $d, c$           & $b, c$            &                  & -0.02 & -0.01 & -0.01 & -0.01 & -0.01 & -0.00 &  0.00 &  0.01 &  0.01 &  0.02 \\
 15 & $\bar s, c$      &                   &                  & -0.01 & -0.01 & -0.01 & -0.01 & -0.01 & -0.00 & -0.00 &  0.00 &  0.00 &  0.00 \\
 16 & $s, c$           &                   &                  & -0.01 & -0.01 & -0.01 & -0.00 & -0.00 & -0.00 &  0.00 &  0.01 &  0.01 &  0.01 \\
 17 & $\bar u, c$      &                   &                  & -0.01 & -0.01 & -0.01 & -0.01 & -0.01 & -0.00 & -0.00 &  0.00 &  0.00 &  0.01 \\
 18 & $u, c$           &                   &                  & -0.04 & -0.03 & -0.03 & -0.02 & -0.02 & -0.01 & -0.01 &  0.00 &  0.01 &  0.02 \\
 19 & $\bar c, c$      &                   &                  & -0.00 & -0.00 & -0.00 & -0.00 & -0.00 & -0.00 & -0.00 & -0.00 &  0.00 &  0.00 \\
 20 & $c, c$           &                   &                  & -0.01 & -0.01 & -0.01 & -0.01 & -0.00 & -0.00 & -0.00 & -0.00 & -0.00 &  0.00 \\
 21 & $g, d$           &                   &                  &  6.23 &  6.40 &  6.64 &  6.85 &  7.13 &  7.38 &  7.67 &  8.00 &  8.31 &  8.81 \\
 22 & $\bar d, d$      &                   &                  & -0.04 & -0.03 & -0.03 & -0.02 & -0.02 & -0.01 &  0.00 &  0.01 &  0.03 &  0.06 \\
 23 & $d, d$           &                   &                  & -0.09 & -0.09 & -0.07 & -0.06 & -0.05 & -0.04 & -0.02 &  0.01 &  0.05 &  0.08 \\
 24 & $\bar b, d$      & $\bar s, d$       &                  & -0.02 & -0.01 & -0.02 & -0.01 & -0.01 &  0.00 &  0.01 &  0.01 &  0.02 &  0.04 \\
 25 & $s, d$           & $b, d$            &                  & -0.02 & -0.02 & -0.02 & -0.01 & -0.01 & -0.01 &  0.00 &  0.00 &  0.01 &  0.02 \\
 26 & $\bar u, d$      &                   &                  & -0.04 & -0.04 & -0.04 & -0.03 & -0.02 & -0.01 & -0.00 &  0.01 &  0.03 &  0.05 \\
 27 & $u, d$           &                   &                  & -0.09 & -0.08 & -0.05 & -0.03 & -0.01 &  0.02 &  0.06 &  0.11 &  0.20 &  0.30 \\
 28 & $\bar c, d$      &                   &                  & -0.01 & -0.01 & -0.01 & -0.01 & -0.00 & -0.00 & -0.00 &  0.01 &  0.01 &  0.01 \\
 29 & $c, d$           &                   &                  & -0.01 & -0.01 & -0.01 & -0.01 & -0.00 &  0.00 &  0.00 &  0.01 &  0.01 &  0.02 \\
 30 & $g, s$           &                   &                  &  3.62 &  3.70 &  3.74 &  3.81 &  3.84 &  3.87 &  3.88 &  3.84 &  3.74 &  3.53 \\
 31 & $\bar b, s$      & $\bar d, s$       &                  & -0.02 & -0.01 & -0.01 & -0.01 & -0.01 & -0.01 & -0.00 &  0.00 &  0.01 &  0.01 \\
 32 & $d, s$           & $b, s$            &                  & -0.03 & -0.03 & -0.03 & -0.02 & -0.02 & -0.02 & -0.01 & -0.00 &  0.00 &  0.01 \\
 33 & $\bar s, s$      &                   &                  & -0.01 & -0.01 & -0.01 & -0.01 & -0.01 & -0.00 & -0.00 &  0.00 &  0.00 &  0.01 \\
 34 & $s, s$           &                   &                  & -0.02 & -0.02 & -0.02 & -0.02 & -0.01 & -0.01 & -0.01 & -0.00 &  0.00 &  0.00 \\
 35 & $\bar u, s$      &                   &                  & -0.01 & -0.01 & -0.01 & -0.01 & -0.01 & -0.01 & -0.00 &  0.00 &  0.00 &  0.01 \\
 36 & $u, s$           &                   &                  & -0.05 & -0.04 & -0.04 & -0.03 & -0.02 & -0.01 & -0.01 &  0.01 &  0.02 &  0.04 \\
 37 & $\bar c, s$      &                   &                  & -0.01 & -0.01 & -0.01 & -0.01 & -0.01 & -0.00 & -0.00 &  0.00 &  0.00 &  0.01 \\
 38 & $c, s$           &                   &                  & -0.01 & -0.01 & -0.01 & -0.01 & -0.00 & -0.00 &  0.00 &  0.00 &  0.01 &  0.01 \\
 39 & $g,  b$          &                   &                  &  1.57 &  1.63 &  1.68 &  1.73 &  1.76 &  1.79 &  1.80 &  1.78 &  1.73 &  1.62 \\
 40 & $\bar s, b$      & $\bar d, b$       &                  & -0.01 & -0.01 & -0.01 & -0.01 & -0.01 & -0.00 & -0.00 &  0.00 &  0.00 &  0.01 \\
 41 & $d, b$           & $s, b$            &                  & -0.02 & -0.02 & -0.02 & -0.01 & -0.01 & -0.01 & -0.01 & -0.01 & -0.00 &  0.00 \\
 42 & $\bar b, b$      &                   &                  &  0.87 &  0.80 &  0.71 &  0.63 &  0.54 &  0.46 &  0.38 &  0.30 &  0.23 &  0.17 \\
 43 & $b, b$           &                   &                  & -0.00 & -0.00 & -0.00 & -0.00 & -0.00 & -0.00 & -0.00 & -0.00 & -0.00 &  0.00 \\
 44 & $\bar c, b$      & $\bar u, b$       &                  & -0.01 & -0.01 & -0.01 & -0.01 & -0.00 & -0.00 & -0.00 & -0.00 &  0.00 &  0.00 \\
 45 & $u, b$           & $c, b$            &                  & -0.02 & -0.03 & -0.02 & -0.02 & -0.01 & -0.01 & -0.00 &  0.00 &  0.01 &  0.02 \\
 46 & $g, \bar u$      &                   &                  &  3.48 &  3.53 &  3.61 &  3.66 &  3.72 &  3.76 &  3.79 &  3.77 &  3.72 &  3.54 \\
 47 & $\bar d, \bar u$ &                   &                  & -0.02 & -0.02 & -0.02 & -0.01 & -0.01 & -0.00 &  0.01 &  0.02 &  0.03 &  0.04 \\
 48 & $d, \bar u$      &                   &                  & -0.05 & -0.04 & -0.03 & -0.03 & -0.02 & -0.01 & -0.00 &  0.01 &  0.03 &  0.05 \\
 49 & $\bar b, \bar u$ & $\bar s, \bar u$  &                  & -0.01 & -0.01 & -0.01 & -0.01 & -0.00 & -0.00 &  0.00 &  0.01 &  0.01 &  0.02 \\
 50 & $s, \bar u$      & $b, \bar u$       &                  & -0.01 & -0.01 & -0.01 & -0.01 & -0.00 & -0.00 & -0.00 &  0.00 &  0.01 &  0.01 \\
 51 & $\bar u, \bar u$ &                   &                  & -0.02 & -0.02 & -0.02 & -0.02 & -0.01 & -0.01 & -0.01 & -0.00 &  0.00 &  0.01 \\
 52 & $u, \bar u$                          &                  & -0.05 & -0.04 & -0.04 & -0.03 & -0.02 & -0.01 & -0.01 &  0.00 &  0.03 &  0.05 \\
 53 & $\bar c, \bar u$ &                   &                  & -0.01 & -0.01 & -0.01 & -0.01 & -0.01 & -0.01 & -0.00 & -0.00 &  0.00 &  0.00 \\
 54 & $c, \bar u$      &                   &                  & -0.01 & -0.01 & -0.01 & -0.01 & -0.01 & -0.00 & -0.00 & -0.00 &  0.00 &  0.00 \\
 55 & $g, \bar c$      &                   &                  &  1.99 &  2.03 &  2.03 &  2.05 &  2.07 &  2.05 &  2.05 &  2.00 &  1.91 &  1.76 \\
 56 & $\bar b, \bar c$ & $\bar d, \bar c$  &                  & -0.01 & -0.01 & -0.01 & -0.00 & -0.00 & -0.00 &  0.00 &  0.00 &  0.01 &  0.01 \\
 57 & $d, \bar c$      & $b, \bar c$       &                  & -0.02 & -0.02 & -0.02 & -0.01 & -0.01 & -0.01 & -0.00 &  0.00 &  0.01 &  0.01 \\
 58 & $\bar s, \bar c$ &                   &                  & -0.01 & -0.01 & -0.00 & -0.01 & -0.00 & -0.00 &  0.00 &  0.01 &  0.01 &  0.01 \\
 59 & $s, \bar c$      &                   &                  & -0.01 & -0.01 & -0.01 & -0.01 & -0.01 & -0.00 & -0.00 &  0.00 &  0.00 &  0.01 \\
 60 & $\bar u, \bar c$ &                   &                  & -0.01 & -0.01 & -0.01 & -0.01 & -0.01 & -0.01 & -0.00 & -0.00 &  0.00 &  0.00 \\
 61 & $u, \bar c$      &                   &                  & -0.04 & -0.03 & -0.02 & -0.02 & -0.02 & -0.01 &  0.01 &  0.01 &  0.02 &  0.03 \\
 62 & $\bar c, \bar c$ &                   &                  & -0.01 & -0.01 & -0.00 & -0.01 & -0.00 & -0.00 & -0.00 & -0.00 & -0.00 &  0.00 \\
 63 & $c, \bar c$      &                   &                  & -0.00 & -0.00 & -0.00 & -0.00 & -0.00 & -0.00 & -0.00 & -0.00 &  0.00 &  0.00 \\
 64 & $g, \bar d$      &                   &                  &  4.65 &  4.74 &  4.85 &  4.95 &  5.02 &  5.12 &  5.19 &  5.21 &  5.15 &  5.04 \\
 65 & $\bar d, \bar d$ &                   &                  & -0.04 & -0.04 & -0.03 & -0.03 & -0.02 & -0.02 & -0.01 & -0.00 &  0.00 &  0.01 \\
 66 & $d, \bar d$      &                   &                  & -0.04 & -0.04 & -0.03 & -0.03 & -0.02 & -0.01 &  0.00 &  0.01 &  0.03 &  0.05 \\
 67 & $\bar b, \bar d$ & $\bar s, \bar d$  &                  & -0.02 & -0.02 & -0.01 & -0.01 & -0.01 & -0.01 & -0.01 & -0.00 &  0.00 &  0.00 \\
 68 & $s, \bar d$      & $b, \bar d$       &                  & -0.02 & -0.01 & -0.01 & -0.01 & -0.01 & -0.00 & -0.00 &  0.00 &  0.01 &  0.01 \\
 69 & $\bar u, \bar d$ &                   &                  & -0.02 & -0.02 & -0.02 & -0.01 & -0.01 & -0.00 &  0.01 &  0.02 &  0.03 &  0.04 \\
 70 & $u, \bar d$      &                   &                  & -0.08 & -0.07 & -0.05 & -0.04 & -0.03 & -0.01 & -0.00 &  0.03 &  0.06 &  0.11 \\
 71 & $\bar c, \bar d$ &                   &                  & -0.01 & -0.01 & -0.01 & -0.00 & -0.00 & -0.00 & -0.00 &  0.00 &  0.00 &  0.01 \\
 72 & $c, \bar d$      &                   &                  & -0.01 & -0.01 & -0.01 & -0.01 & -0.00 & -0.00 & -0.00 &  0.00 &  0.00 &  0.01 \\
 73 & $g, \bar s$      &                   &                  &  3.61 &  3.67 &  3.72 &  3.78 &  3.81 &  3.83 &  3.83 &  3.76 &  3.65 &  3.39 \\
 74 & $\bar b, \bar s$ & $\bar d, \bar s$  &                  & -0.02 & -0.02 & -0.01 & -0.02 & -0.01 & -0.01 & -0.01 & -0.00 & -0.00 &  0.00 \\
 75 & $d, \bar s$      & $b, \bar s$       &                  & -0.03 & -0.03 & -0.02 & -0.02 & -0.02 & -0.01 & -0.01 &  0.00 &  0.01 &  0.02 \\
 76 & $\bar s, \bar s$ &                   &                  & -0.02 & -0.02 & -0.02 & -0.01 & -0.01 & -0.01 & -0.01 & -0.00 & -0.00 &  0.00 \\
 77 & $s, \bar s$      &                   &                  & -0.01 & -0.01 & -0.01 & -0.01 & -0.01 & -0.00 & -0.00 &  0.00 &  0.00 &  0.01 \\
 78 & $\bar u, \bar s$ &                   &                  & -0.01 & -0.01 & -0.01 & -0.01 & -0.01 & -0.00 & -0.00 &  0.00 &  0.00 &  0.01 \\
 79 & $u, \bar s$      &                   &                  & -0.05 & -0.04 & -0.04 & -0.03 & -0.03 & -0.02 & -0.01 &  0.00 &  0.01 &  0.03 \\
 80 & $\bar c, \bar s$ &                   &                  & -0.01 & -0.01 & -0.01 & -0.01 & -0.00 & -0.00 &  0.00 &  0.00 &  0.01 &  0.01 \\
 81 & $c, \bar s$      &                   &                  & -0.01 & -0.01 & -0.01 & -0.01 & -0.01 & -0.00 & -0.00 &  0.00 &  0.00 &  0.01 \\
 82 & $g, \bar b$      &                   &                  &  1.60 &  1.64 &  1.68 &  1.72 &  1.76 &  1.78 &  1.80 &  1.79 &  1.73 &  1.62 \\
 83 & $\bar s, \bar b$ & $\bar d, \bar b$  &                  & -0.01 & -0.01 & -0.01 & -0.01 & -0.01 & -0.01 & -0.01 & -0.00 & -0.00 &  0.00 \\
\bottomrule
\end{tabularx}

\end{table}
\begin{table}[!p]
  \centering
  \tiny  
  \begin{tabularx}{\textwidth}{rXXX@{}S@{}S@{}S@{}S@{}S@{}S@{}S@{}S@{}S@{}S}
  \toprule
  \multicolumn{4}{l}{Table~\ref{tab:B10} continues from previous page into next page} &    1  &     2 &     3 &     4 &     5 &     6 &    7  &     8 &     9 &    10 \\
  \midrule
 84 & $d, \bar b$      & $s, \bar b$       &                  & -0.02 & -0.02 & -0.01 & -0.01 & -0.01 & -0.01 & -0.00 & -0.00 &  0.00 &  0.01 \\
 85 & $\bar b, \bar b$ &                   &                  & -0.00 & -0.00 & -0.00 & -0.00 & -0.00 & -0.00 & -0.00 & -0.00 & -0.00 &  0.00 \\
 86 & $b, \bar b$      &                   &                  &  0.89 &  0.80 &  0.71 &  0.63 &  0.54 &  0.46 &  0.38 &  0.30 &  0.23 &  0.17 \\
 87 & $\bar c, \bar b$ & $\bar u, \bar b$  &                  & -0.01 & -0.01 & -0.01 & -0.01 & -0.00 & -0.00 & -0.00 &  0.00 &  0.00 &  0.00 \\
 88 & $u, \bar b$      & $c, \bar b$       &                  & -0.03 & -0.02 & -0.02 & -0.02 & -0.01 & -0.01 & -0.01 & -0.00 &  0.00 &  0.01 \\
 89 & $u, g$           &                   &                  &  5.86 &  6.10 &  6.34 &  6.60 &  6.93 &  7.26 &  7.66 &  8.14 &  8.71 &  9.62 \\
 90 & $\gamma, g$      &                   &                  & -0.01 & -0.01 & -0.01 & -0.01 & -0.00 & -0.00 & -0.00 & -0.00 & -0.00 & -0.00 \\
 91 & $u, \bar b$      & $u, \bar s$       &                  & -0.01 & -0.01 &  0.00 & -0.00 &  0.01 &  0.01 &  0.02 &  0.03 &  0.05 &  0.07 \\
 92 & $u, s$           & $u, b$            &                  & -0.01 & -0.00 & -0.00 &  0.01 &  0.01 &  0.02 &  0.03 &  0.04 &  0.06 &  0.09 \\
 93 & $u, \bar u$      & $c, \bar c$       &                  &  9.66 &  8.98 &  8.25 &  7.53 &  6.78 &  6.07 &  5.34 &  4.65 &  4.02 &  3.50 \\
 94 & $\gamma, \bar c$ & $\gamma, \bar u$  &                  &  0.01 &  0.01 &  0.01 &  0.01 &  0.01 &  0.01 &  0.01 &  0.01 &  0.01 &  0.01 \\
 95 & $g, \bar c$      & $g, \bar u$       &                  & -0.48 & -0.38 & -0.28 & -0.19 & -0.09 & -0.02 &  0.06 &  0.12 &  0.18 &  0.21 \\
 96 & $u, \gamma$      & $c, \gamma$       &                  &  0.02 &  0.01 &  0.02 &  0.02 &  0.02 &  0.02 &  0.02 &  0.02 &  0.02 &  0.02 \\
 97 & $u, g$           & $c, g$            &                  & -0.81 & -0.65 & -0.49 & -0.34 & -0.18 & -0.04 &  0.10 &  0.24 &  0.38 &  0.53 \\
 98 & $c, g$           &                   &                  &  1.98 &  2.02 &  2.02 &  2.05 &  2.05 &  2.06 &  2.05 &  1.98 &  1.92 &  1.76 \\
 99 & $c, \bar b$      & $c, \bar d$       &                  & -0.01 & -0.01 & -0.01 & -0.01 & -0.00 & -0.00 & -0.00 &  0.00 &  0.00 &  0.01 \\
100 & $c, d$           & $c, b$            &                  & -0.02 & -0.01 & -0.01 & -0.01 & -0.01 & -0.00 &  0.00 &  0.00 &  0.01 &  0.02 \\
101 & $d, g$           &                   &                  &  6.24 &  6.40 &  6.65 &  6.85 &  7.11 &  7.37 &  7.67 &  7.98 &  8.32 &  8.78 \\
102 & $d, \bar b$      & $d, \bar s$       &                  & -0.02 & -0.02 & -0.01 & -0.01 & -0.00 & -0.00 &  0.01 &  0.01 &  0.02 &  0.04 \\
103 & $d, s$           & $d, b$            &                  & -0.02 & -0.02 & -0.02 & -0.01 & -0.01 & -0.01 & -0.00 &  0.00 &  0.01 &  0.02 \\
104 & $d, \bar c$      &                   &                  & -0.01 & -0.01 & -0.01 & -0.01 & -0.00 & -0.00 & -0.00 &  0.00 &  0.01 &  0.01 \\
105 & $d, c$           &                   &                  & -0.01 & -0.01 & -0.01 & -0.00 & -0.00 &  0.00 &  0.00 &  0.01 &  0.01 &  0.02 \\
106 & $d, \bar d$      & $s, \bar s$       &                  & 12.68 & 11.72 & 10.70 &  9.69 &  8.63 &  7.61 &  6.58 &  5.60 &  4.68 &  3.85 \\
107 & $\gamma, \bar s$ & $\gamma, \bar d$  &                  & -0.00 & -0.00 & -0.00 & -0.00 & -0.00 & -0.00 &  0.00 &  0.00 &  0.00 &  0.00 \\
108 & $g, \bar s$      & $g, \bar d$       &                  & -0.71 & -0.56 & -0.41 & -0.28 & -0.13 & -0.01 &  0.10 &  0.20 &  0.28 &  0.34 \\
109 & $d, \gamma$      & $s, \gamma$       &                  & -0.00 & -0.00 & -0.00 & -0.00 & -0.00 & -0.00 &  0.00 &  0.00 &  0.00 &  0.00 \\
110 & $d, g$           & $s, g$            &                  & -0.91 & -0.73 & -0.54 & -0.37 & -0.19 & -0.02 &  0.13 &  0.28 &  0.42 &  0.54 \\
111 & $d, \bar d$      & $s, \bar s$       & $b, \bar b$      &  0.09 &  0.07 &  0.06 &  0.05 &  0.05 &  0.04 &  0.03 &  0.03 &  0.03 &  0.02 \\
112 & $g, \bar b$      & $g, \bar s$       & $g, \bar d$      & -0.01 & -0.01 & -0.00 & -0.00 & -0.00 & -0.00 & -0.00 & -0.00 & -0.00 &  0.00 \\
113 & $d, g$           & $s,  g$           & $b,  g$          & -0.01 & -0.01 & -0.01 & -0.00 & -0.00 & -0.00 & -0.00 & -0.00 & -0.00 &  0.00 \\
114 & $d, \gamma$      & $s, \gamma$       & $b, \gamma$      &  0.00 &  0.00 &  0.00 &  0.00 &  0.00 &  0.01 &  0.00 &  0.01 &  0.01 &  0.01 \\
115 & $s, g$           &                   &                  &  3.63 &  3.70 &  3.74 &  3.81 &  3.84 &  3.87 &  3.89 &  3.84 &  3.73 &  3.52 \\
116 & $s, \bar b$      & $s, \bar d$       &                  & -0.02 & -0.02 & -0.01 & -0.01 & -0.01 & -0.01 & -0.00 &  0.00 &  0.01 &  0.01 \\
117 & $s, d$           & $s,  b$           &                  & -0.04 & -0.03 & -0.03 & -0.02 & -0.02 & -0.02 & -0.01 & -0.01 &  0.00 &  0.01 \\
118 & $s, \bar u$      &                   &                  & -0.01 & -0.01 & -0.01 & -0.01 & -0.01 & -0.01 & -0.00 & -0.00 &  0.00 &  0.01 \\
119 & $s, u$           &                   &                  & -0.04 & -0.04 & -0.04 & -0.03 & -0.02 & -0.02 & -0.01 &  0.00 &  0.02 &  0.04 \\
120 & $b, g$           &                   &                  &  1.57 &  1.64 &  1.69 &  1.72 &  1.77 &  1.78 &  1.79 &  1.78 &  1.74 &  1.62 \\
121 & $b, \bar s$      & $b, \bar d$       &                  & -0.01 & -0.01 & -0.01 & -0.01 & -0.01 & -0.00 & -0.00 &  0.00 &  0.00 &  0.00 \\
122 & $b, d$           & $b,  s$           &                  & -0.02 & -0.02 & -0.02 & -0.02 & -0.01 & -0.01 & -0.01 & -0.01 & -0.00 &  0.00 \\
123 & $b, \bar c$      & $b, \bar u$       &                  & -0.01 & -0.01 & -0.01 & -0.01 & -0.00 & -0.00 & -0.00 & -0.00 &  0.00 &  0.00 \\
124 & $b, u$           & $b,  c$           &                  & -0.03 & -0.02 & -0.02 & -0.01 & -0.02 & -0.00 & -0.00 &  0.00 &  0.01 &  0.02 \\
125 & $\gamma, \bar b$ &                   &                  & -0.00 & -0.00 & -0.00 & -0.00 & -0.00 & -0.00 &  0.00 &  0.00 &  0.00 &  0.00 \\
126 & $b, \gamma$      &                   &                  & -0.00 & -0.00 & -0.00 & -0.00 & -0.00 & -0.00 & -0.00 &  0.00 &  0.00 &  0.00 \\
127 & $\bar u, g$      &                   &                  &  3.46 &  3.54 &  3.60 &  3.66 &  3.71 &  3.77 &  3.79 &  3.78 &  3.71 &  3.55 \\
128 & $\bar u, \bar b$ & $\bar u, \bar s$  &                  & -0.01 & -0.01 & -0.01 & -0.01 & -0.00 &  0.00 &  0.00 &  0.01 &  0.01 &  0.02 \\
129 & $\bar u, s$      & $\bar u, b$       &                  & -0.01 & -0.01 & -0.01 & -0.01 & -0.00 & -0.00 &  0.00 &  0.00 &  0.01 &  0.01 \\
130 & $\bar c, c$      & $\bar u, u$       &                  &  9.67 &  8.96 &  8.26 &  7.53 &  6.80 &  6.07 &  5.34 &  4.64 &  4.02 &  3.49 \\
131 & $\gamma, u$      & $\gamma, c$       &                  &  0.02 &  0.02 &  0.01 &  0.02 &  0.02 &  0.02 &  0.02 &  0.02 &  0.02 &  0.02 \\
132 & $g, u$           & $g,  c$           &                  & -0.81 & -0.64 & -0.49 & -0.35 & -0.19 & -0.04 &  0.10 &  0.24 &  0.38 &  0.53 \\
133 & $\bar c, \gamma$ & $\bar u, \gamma$  &                  &  0.01 &  0.01 &  0.01 &  0.01 &  0.01 &  0.01 &  0.01 &  0.01 &  0.01 &  0.01 \\
134 & $\bar c, g$      & $\bar u, g$       &                  & -0.48 & -0.38 & -0.28 & -0.19 & -0.10 & -0.01 &  0.06 &  0.13 &  0.17 &  0.21 \\
135 & $\bar c, g$      &                   &                  &  1.98 &  2.02 &  2.03 &  2.04 &  2.06 &  2.06 &  2.04 &  2.00 &  1.91 &  1.75 \\
136 & $\bar c, \bar b$ & $\bar c, \bar d$  &                  & -0.01 & -0.01 & -0.01 & -0.00 & -0.00 & -0.00 &  0.00 &  0.00 &  0.01 &  0.01 \\
137 & $\bar c, d$      & $\bar c, b$       &                  & -0.02 & -0.02 & -0.01 & -0.01 & -0.01 & -0.00 & -0.00 &  0.00 &  0.01 &  0.01 \\
138 & $\bar d, g$      &                   &                  &  4.66 &  4.72 &  4.85 &  4.95 &  5.03 &  5.12 &  5.18 &  5.20 &  5.18 &  5.03 \\
139 & $\bar d, \bar b$ & $\bar d, \bar s$  &                  & -0.01 & -0.02 & -0.02 & -0.01 & -0.01 & -0.01 & -0.01 & -0.00 & -0.00 &  0.00 \\
140 & $\bar d, s$      & $\bar d, b$       &                  & -0.02 & -0.01 & -0.01 & -0.01 & -0.01 & -0.01 & -0.00 &  0.00 &  0.01 &  0.01 \\
141 & $\bar d, \bar c$ &                   &                  & -0.01 & -0.01 & -0.01 & -0.00 & -0.00 & -0.00 & -0.00 &  0.00 &  0.00 &  0.01 \\
142 & $\bar d, c$      &                   &                  & -0.01 & -0.01 & -0.01 & -0.01 & -0.00 & -0.00 & -0.00 &  0.00 &  0.00 &  0.00 \\
143 & $\bar s, s$      & $\bar d, d$       &                  & 12.67 & 11.72 & 10.67 &  9.68 &  8.65 &  7.60 &  6.57 &  5.60 &  4.68 &  3.86 \\
144 & $\gamma, d$      & $\gamma, s$       &                  & -0.00 & -0.00 & -0.00 & -0.00 & -0.00 & -0.00 &  0.00 &  0.00 &  0.00 &  0.00 \\
145 & $g, d$           & $g, s$            &                  & -0.90 & -0.73 & -0.55 & -0.36 & -0.18 & -0.02 &  0.13 &  0.28 &  0.41 &  0.54 \\
146 & $\bar s, \gamma$ & $\bar d, \gamma$  &                  & -0.00 & -0.00 & -0.00 & -0.00 & -0.00 & -0.00 &  0.00 &  0.00 &  0.00 &  0.00 \\
147 & $\bar s, g$      & $\bar d, g$       &                  & -0.72 & -0.57 & -0.41 & -0.27 & -0.14 & -0.01 &  0.10 &  0.20 &  0.28 &  0.34 \\
148 & $\bar b,  b$     & $\bar s, s$       & $\bar d, d$      &  0.09 &  0.07 &  0.06 &  0.05 &  0.05 &  0.04 &  0.03 &  0.03 &  0.03 &  0.02 \\
149 & $g,  d$          & $g, s$            & $g, b$           & -0.01 & -0.01 & -0.01 & -0.00 & -0.00 & -0.00 & -0.00 & -0.00 & -0.00 &  0.00 \\
150 & $\bar b, g$      & $\bar s, g$       & $\bar d, g$      & -0.01 & -0.01 & -0.00 & -0.00 & -0.00 & -0.00 & -0.00 & -0.00 & -0.00 &  0.00 \\
151 & $\bar b, \gamma$ & $\bar s, \gamma$  & $\bar d, \gamma$ &  0.00 &  0.00 &  0.00 &  0.00 &  0.00 &  0.00 &  0.00 &  0.00 &  0.00 &  0.00 \\
152 & $\bar s, g$      &                   &                  &  3.62 &  3.67 &  3.73 &  3.77 &  3.81 &  3.84 &  3.82 &  3.77 &  3.65 &  3.39 \\
153 & $\bar s, \bar b$ & $\bar s, \bar d$  &                  & -0.02 & -0.02 & -0.02 & -0.01 & -0.01 & -0.01 & -0.01 & -0.00 &  0.00 &  0.00 \\
154 & $\bar s, d$      & $\bar s, b$       &                  & -0.03 & -0.03 & -0.03 & -0.02 & -0.02 & -0.01 & -0.00 &  0.00 &  0.01 &  0.02 \\
155 & $\bar s, \bar u$ &                   &                  & -0.01 & -0.01 & -0.01 & -0.01 & -0.00 & -0.00 & -0.00 &  0.00 &  0.00 &  0.01 \\
156 & $\bar s, u$      &                   &                  & -0.05 & -0.04 & -0.04 & -0.03 & -0.03 & -0.02 & -0.02 & -0.00 &  0.01 &  0.03 \\
157 & $\bar b, g$      &                   &                  &  1.60 &  1.62 &  1.70 &  1.72 &  1.77 &  1.79 &  1.80 &  1.78 &  1.73 &  1.62 \\
158 & $\bar b, \bar s$ & $\bar b, \bar d$  &                  & -0.01 & -0.01 & -0.01 & -0.01 & -0.01 & -0.01 & -0.00 & -0.00 & -0.00 &  0.00 \\
159 & $\bar b, d$      & $\bar b, s$       &                  & -0.02 & -0.01 & -0.01 & -0.01 & -0.01 & -0.01 & -0.00 & -0.00 &  0.01 &  0.01 \\
160 & $\bar b, \bar c$ & $\bar b, \bar u$  &                  & -0.01 & -0.01 & -0.01 & -0.01 & -0.00 & -0.00 & -0.00 &  0.00 &  0.00 &  0.00 \\
161 & $\bar b, u$      & $\bar b, c$       &                  & -0.03 & -0.03 & -0.02 & -0.02 & -0.02 & -0.01 & -0.01 & -0.00 &  0.01 &  0.01 \\
162 & $\gamma, b$      &                   &                  & -0.00 & -0.00 & -0.00 & -0.00 & -0.00 & -0.00 &  0.00 &  0.00 &  0.00 &  0.00 \\
163 & $\bar b, \gamma$ &                   &                  & -0.00 & -0.00 & -0.00 & -0.00 & -0.00 & -0.00 &  0.00 &  0.00 &  0.00 &  0.00 \\
164 & $\gamma,  d$     & $\gamma, s$       & $\gamma, b$      &  0.00 &  0.00 &  0.00 &  0.00 &  0.00 &  0.01 &  0.01 &  0.01 &  0.01 &  0.01 \\
165 & $\gamma, \bar b$ & $\gamma, \bar s$  & $\gamma, \bar d$ &  0.00 &  0.00 &  0.00 &  0.00 &  0.00 &  0.00 &  0.00 &  0.00 &  0.00 &  0.00 \\
166 & $\gamma, \gamma$ &                   &                  &  0.00 &  0.00 &  0.00 &  0.00 &  0.00 &  0.00 &  0.00 &  0.00 &  0.00 &  0.00 \\
\bottomrule
\end{tabularx}

\end{table}
\begin{table}[!p]
  \centering
  \tiny 
  \begin{tabularx}{\textwidth}{rXXX@{}S@{}S@{}S@{}S@{}S@{}S@{}S@{}S@{}S}
  \toprule
  \multicolumn{4}{l}{Table~\ref{tab:B10} continues from previous page into next page} &    11 &    12 &    13 &    14 &    15 &    16 &   17  &    18 &    19 \\
  \midrule
  1 & $g,  u$          &                   &                  & 10.86 & 11.98 & 12.76 & 13.27 & 14.33 & 15.35 & 16.44 & 17.46 & 19.36 \\
  2 & $\bar d, u$      &                   &                  &  0.15 &  0.18 &  0.20 &  0.23 &  0.23 &  0.26 &  0.29 &  0.33 &  0.34 \\
  3 & $d, u$           &                   &                  &  0.43 &  0.55 &  0.65 &  0.71 &  0.83 &  1.00 &  1.20 &  1.57 &  2.30 \\
  4 & $\bar b, u$      & $\bar s, u$       &                  &  0.09 &  0.09 &  0.12 &  0.10 &  0.11 &  0.12 &  0.14 &  0.12 &  0.11 \\
  5 & $s, u$           & $b, u$            &                  &  0.10 &  0.12 &  0.14 &  0.15 &  0.15 &  0.16 &  0.18 &  0.19 &  0.21 \\
  6 & $\bar u, u$      &                   &                  &  0.07 &  0.08 &  0.10 &  0.09 &  0.11 &  0.11 &  0.12 &  0.12 &  0.14 \\
  7 & $u, u$           &                   &                  &  0.32 &  0.41 &  0.49 &  0.59 &  0.72 &  0.88 &  1.21 &  1.61 &  2.52 \\
  8 & $\bar c, u$      &                   &                  &  0.04 &  0.05 &  0.06 &  0.06 &  0.06 &  0.06 &  0.07 &  0.06 &  0.07 \\
  9 & $c, u$           &                   &                  &  0.02 &  0.06 &  0.05 &  0.03 &  0.04 &  0.05 &  0.06 &  0.05 &  0.05 \\
 10 & $g, \gamma$      &                   &                  & -0.00 & -0.00 & -0.00 & -0.00 & -0.00 & -0.00 & -0.00 & -0.00 & -0.00 \\
 11 & $g, g$           &                   &                  & -1.79 & -1.38 & -1.19 & -1.01 & -0.87 & -0.72 & -0.54 & -0.42 & -0.26 \\
 12 & $g, c$           &                   &                  &  1.54 &  1.35 &  1.19 &  1.14 &  0.98 &  0.84 &  0.72 &  0.55 &  0.43 \\
 13 & $\bar b, c$      & $\bar d, c$       &                  &  0.01 &  0.01 &  0.01 &  0.01 &  0.01 &  0.01 &  0.01 &  0.01 &  0.00 \\
 14 & $d, c$           & $b, c$            &                  &  0.02 &  0.03 &  0.03 &  0.03 &  0.03 &  0.03 &  0.03 &  0.03 &  0.03 \\
 15 & $\bar s, c$      &                   &                  &  0.01 &  0.01 &  0.00 &  0.01 &  0.01 &  0.01 &  0.00 &  0.00 &  0.00 \\
 16 & $s, c$           &                   &                  &  0.01 &  0.01 &  0.01 &  0.01 &  0.01 &  0.01 &  0.01 &  0.01 &  0.01 \\
 17 & $\bar u, c$      &                   &                  &  0.01 &  0.01 &  0.01 &  0.01 &  0.01 &  0.00 &  0.01 &  0.00 &  0.00 \\
 18 & $u, c$           &                   &                  &  0.03 &  0.04 &  0.05 &  0.04 &  0.04 &  0.05 &  0.05 &  0.05 &  0.05 \\
 19 & $\bar c, c$      &                   &                  &  0.00 &  0.00 &  0.00 &  0.00 &  0.00 &  0.00 &  0.00 &  0.00 &  0.00 \\
 20 & $c, c$           &                   &                  &  0.00 &  0.00 &  0.00 &  0.00 &  0.00 &  0.00 &  0.00 &  0.00 &  0.00 \\
 21 & $g, d$           &                   &                  &  9.33 &  9.74 &  9.94 & 10.08 & 10.29 & 10.30 & 10.25 & 10.02 &  9.31 \\
 22 & $\bar d, d$      &                   &                  &  0.06 &  0.08 &  0.08 &  0.09 &  0.09 &  0.09 &  0.10 &  0.09 &  0.07 \\
 23 & $d, d$           &                   &                  &  0.14 &  0.15 &  0.20 &  0.21 &  0.22 &  0.29 &  0.33 &  0.37 &  0.45 \\
 24 & $\bar b, d$      & $\bar s, d$       &                  &  0.05 &  0.05 &  0.05 &  0.05 &  0.05 &  0.05 &  0.05 &  0.04 &  0.03 \\
 25 & $s, d$           & $b, d$            &                  &  0.03 &  0.03 &  0.03 &  0.03 &  0.03 &  0.03 &  0.04 &  0.03 &  0.03 \\
 26 & $\bar u, d$      &                   &                  &  0.07 &  0.08 &  0.09 &  0.09 &  0.09 &  0.10 &  0.11 &  0.10 &  0.08 \\
 27 & $u, d$           &                   &                  &  0.43 &  0.54 &  0.63 &  0.72 &  0.79 &  1.00 &  1.26 &  1.54 &  2.33 \\
 28 & $\bar c, d$      &                   &                  &  0.02 &  0.02 &  0.02 &  0.04 &  0.02 &  0.02 &  0.02 &  0.02 &  0.02 \\
 29 & $c, d$           &                   &                  &  0.02 &  0.02 &  0.03 &  0.03 &  0.03 &  0.03 &  0.03 &  0.02 &  0.02 \\
 30 & $g, s$           &                   &                  &  3.21 &  2.92 &  2.73 &  2.54 &  2.35 &  2.09 &  1.79 &  1.59 &  1.18 \\
 31 & $\bar b, s$      & $\bar d, s$       &                  &  0.01 &  0.02 &  0.02 &  0.01 &  0.02 &  0.01 &  0.01 &  0.01 &  0.01 \\
 32 & $d, s$           & $b, s$            &                  &  0.01 &  0.02 &  0.02 &  0.02 &  0.02 &  0.03 &  0.03 &  0.03 &  0.03 \\
 33 & $\bar s, s$      &                   &                  &  0.01 &  0.01 &  0.01 &  0.01 &  0.01 &  0.01 &  0.00 &  0.00 &  0.00 \\
 34 & $s, s$           &                   &                  &  0.01 &  0.01 &  0.01 &  0.01 &  0.01 &  0.01 &  0.00 &  0.01 &  0.01 \\
 35 & $\bar u, s$      &                   &                  &  0.01 &  0.01 &  0.01 &  0.01 &  0.01 &  0.01 &  0.01 &  0.01 &  0.00 \\
 36 & $u, s$           &                   &                  &  0.06 &  0.07 &  0.08 &  0.09 &  0.10 &  0.10 &  0.11 &  0.13 &  0.14 \\
 37 & $\bar c, s$      &                   &                  &  0.01 &  0.01 &  0.01 &  0.01 &  0.01 &  0.01 &  0.01 &  0.00 &  0.01 \\
 38 & $c, s$           &                   &                  &  0.01 &  0.01 &  0.01 &  0.01 &  0.01 &  0.01 &  0.01 &  0.00 &  0.01 \\
 39 & $g,  b$          &                   &                  &  1.45 &  1.28 &  1.16 &  1.08 &  0.97 &  0.79 &  0.61 &  0.47 &  0.28 \\
 40 & $\bar s, b$      & $\bar d, b$       &                  &  0.01 &  0.01 &  0.01 &  0.01 &  0.01 &  0.01 &  0.00 &  0.00 &  0.00 \\
 41 & $d, b$           & $s, b$            &                  &  0.00 &  0.00 &  0.01 &  0.01 &  0.01 &  0.01 &  0.01 &  0.01 &  0.00 \\
 42 & $\bar b, b$      &                   &                  &  0.12 &  0.09 &  0.09 &  0.07 &  0.06 &  0.05 &  0.04 &  0.03 &  0.01 \\
 43 & $b, b$           &                   &                  &  0.00 &  0.00 &  0.00 &  0.00 & -0.00 &  0.00 &  0.00 &  0.00 &  0.00 \\
 44 & $\bar c, b$      & $\bar u, b$       &                  &  0.00 &  0.00 &  0.00 &  0.00 &  0.00 &  0.00 &  0.00 &  0.00 &  0.00 \\
 45 & $u, b$           & $c, b$            &                  &  0.03 &  0.03 &  0.04 &  0.03 &  0.04 &  0.04 &  0.04 &  0.05 &  0.04 \\
 46 & $g, \bar u$      &                   &                  &  3.27 &  3.00 &  2.79 &  2.64 &  2.38 &  2.11 &  1.75 &  1.40 &  0.89 \\
 47 & $\bar d, \bar u$ &                   &                  &  0.04 &  0.05 &  0.05 &  0.04 &  0.04 &  0.04 &  0.03 &  0.03 &  0.01 \\
 48 & $d, \bar u$      &                   &                  &  0.07 &  0.08 &  0.09 &  0.10 &  0.10 &  0.11 &  0.10 &  0.10 &  0.08 \\
 49 & $\bar b, \bar u$ & $\bar s, \bar u$  &                  &  0.02 &  0.02 &  0.02 &  0.02 &  0.02 &  0.01 &  0.01 &  0.01 &  0.00 \\
 50 & $s, \bar u$      & $b, \bar u$       &                  &  0.01 &  0.01 &  0.02 &  0.01 &  0.01 &  0.01 &  0.01 &  0.01 &  0.01 \\
 51 & $\bar u, \bar u$ &                   &                  &  0.01 &  0.01 &  0.01 &  0.01 &  0.01 &  0.01 &  0.01 &  0.01 &  0.00 \\
 52 & $u, \bar u$                          &                  &  0.07 &  0.09 &  0.09 &  0.11 &  0.11 &  0.10 &  0.12 &  0.11 &  0.09 \\
 53 & $\bar c, \bar u$ &                   &                  &  0.00 &  0.00 &  0.00 &  0.00 &  0.00 &  0.00 &  0.00 &  0.00 &  0.00 \\
 54 & $c, \bar u$      &                   &                  &  0.01 &  0.01 &  0.01 &  0.01 &  0.01 &  0.01 &  0.01 &  0.00 &  0.00 \\
 55 & $g, \bar c$      &                   &                  &  1.53 &  1.33 &  1.21 &  1.11 &  0.99 &  0.85 &  0.70 &  0.62 &  0.39 \\
 56 & $\bar b, \bar c$ & $\bar d, \bar c$  &                  &  0.01 &  0.01 &  0.01 &  0.01 &  0.01 &  0.01 &  0.01 &  0.01 &  0.01 \\
 57 & $d, \bar c$      & $b, \bar c$       &                  &  0.02 &  0.02 &  0.02 &  0.02 &  0.02 &  0.03 &  0.02 &  0.02 &  0.04 \\
 58 & $\bar s, \bar c$ &                   &                  &  0.01 &  0.01 &  0.01 &  0.01 &  0.01 &  0.01 &  0.00 &  0.00 &  0.00 \\
 59 & $s, \bar c$      &                   &                  &  0.01 &  0.01 &  0.01 &  0.01 &  0.01 &  0.01 &  0.01 &  0.01 &  0.00 \\
 60 & $\bar u, \bar c$ &                   &                  &  0.00 &  0.00 &  0.00 &  0.00 &  0.00 &  0.00 &  0.00 &  0.00 &  0.00 \\
 61 & $u, \bar c$      &                   &                  &  0.04 &  0.05 &  0.05 &  0.06 &  0.05 &  0.07 &  0.07 &  0.07 &  0.07 \\
 62 & $\bar c, \bar c$ &                   &                  &  0.00 &  0.00 &  0.00 &  0.00 &  0.00 &  0.00 &  0.00 &  0.00 &  0.00 \\
 63 & $c, \bar c$      &                   &                  &  0.00 &  0.00 &  0.00 &  0.00 &  0.00 &  0.00 & -0.00 &  0.00 &  0.00 \\
 64 & $g, \bar d$      &                   &                  &  4.74 &  4.40 &  4.21 &  3.89 &  3.70 &  3.26 &  2.81 &  2.31 &  1.72 \\
 65 & $\bar d, \bar d$ &                   &                  &  0.02 &  0.02 &  0.02 &  0.02 &  0.02 &  0.02 &  0.02 &  0.02 &  0.01 \\
 66 & $d, \bar d$      &                   &                  &  0.07 &  0.07 &  0.09 &  0.08 &  0.10 &  0.09 &  0.09 &  0.09 &  0.08 \\
 67 & $\bar b, \bar d$ & $\bar s, \bar d$  &                  &  0.01 &  0.01 &  0.01 &  0.01 &  0.01 &  0.00 &  0.01 &  0.00 & -0.00 \\
 68 & $s, \bar d$      & $b, \bar d$       &                  &  0.02 &  0.02 &  0.02 &  0.01 &  0.02 &  0.01 &  0.02 &  0.01 &  0.01 \\
 69 & $\bar u, \bar d$ &                   &                  &  0.04 &  0.04 &  0.04 &  0.04 &  0.04 &  0.04 &  0.03 &  0.02 &  0.01 \\
 70 & $u, \bar d$      &                   &                  &  0.15 &  0.17 &  0.18 &  0.23 &  0.24 &  0.27 &  0.28 &  0.33 &  0.34 \\
 71 & $\bar c, \bar d$ &                   &                  &  0.01 &  0.01 &  0.01 &  0.01 &  0.01 &  0.01 &  0.01 &  0.01 &  0.00 \\
 72 & $c, \bar d$      &                   &                  &  0.00 &  0.01 &  0.01 &  0.01 &  0.01 &  0.01 &  0.00 &  0.00 &  0.00 \\
 73 & $g, \bar s$      &                   &                  &  3.03 &  2.71 &  2.49 &  2.27 &  2.06 &  1.71 &  1.38 &  1.04 &  0.54 \\
 74 & $\bar b, \bar s$ & $\bar d, \bar s$  &                  &  0.00 &  0.01 &  0.01 &  0.01 &  0.00 &  0.00 &  0.00 &  0.00 &  0.00 \\
 75 & $d, \bar s$      & $b, \bar s$       &                  &  0.03 &  0.03 &  0.03 &  0.04 &  0.04 &  0.03 &  0.03 &  0.03 &  0.02 \\
 76 & $\bar s, \bar s$ &                   &                  &  0.00 &  0.00 &  0.00 &  0.01 &  0.00 &  0.00 &  0.00 &  0.00 &  0.00 \\
 77 & $s, \bar s$      &                   &                  &  0.01 &  0.01 &  0.01 &  0.01 &  0.01 &  0.01 &  0.01 &  0.00 & -0.00 \\
 78 & $\bar u, \bar s$ &                   &                  &  0.01 &  0.01 &  0.01 &  0.01 &  0.01 &  0.01 &  0.00 &  0.01 &  0.00 \\
 79 & $u, \bar s$      &                   &                  &  0.04 &  0.05 &  0.04 &  0.07 &  0.06 &  0.07 &  0.08 &  0.07 &  0.06 \\
 80 & $\bar c, \bar s$ &                   &                  &  0.01 &  0.01 &  0.01 &  0.01 &  0.01 &  0.01 &  0.01 &  0.00 &  0.00 \\
 81 & $c, \bar s$      &                   &                  &  0.01 &  0.01 &  0.01 &  0.01 &  0.00 &  0.01 &  0.00 &  0.00 &  0.00 \\
 82 & $g, \bar b$      &                   &                  &  1.45 &  1.29 &  1.17 &  1.05 &  0.94 &  0.80 &  0.63 &  0.45 &  0.27 \\
 83 & $\bar s, \bar b$ & $\bar d, \bar b$  &                  &  0.00 &  0.00 &  0.00 &  0.00 &  0.00 &  0.00 &  0.00 &  0.00 &  0.00 \\
\bottomrule
\end{tabularx}

\end{table}
\begin{table}[!p]
  \centering
  \tiny
  \renewcommand{\arraystretch}{0.9}
  \begin{tabularx}{\textwidth}{rXXX@{}S@{}S@{}S@{}S@{}S@{}S@{}S@{}S@{}S}
  \toprule
  \multicolumn{4}{l}{Table~\ref{tab:B10} continues from previous page}                 &    11 &    12 &    13 &    14 &    15 &    16 &   17  &    18 &    19 \\
  \midrule
 84 & $d, \bar b$      & $s, \bar b$       &                  &  0.01 &  0.01 &  0.02 &  0.02 &  0.01 &  0.02 &  0.02 &  0.02 &  0.01 \\
 85 & $\bar b, \bar b$ &                   &                  &  0.00 &  0.00 &  0.00 &  0.00 &  0.00 &  0.00 &  0.00 &  0.00 &  0.00 \\
 86 & $b, \bar b$      &                   &                  &  0.12 &  0.09 &  0.08 &  0.07 &  0.06 &  0.05 &  0.04 &  0.03 &  0.01 \\
 87 & $\bar c, \bar b$ & $\bar u, \bar b$  &                  &  0.01 &  0.01 &  0.00 &  0.01 &  0.01 &  0.01 &  0.00 &  0.00 &  0.00 \\
 88 & $u, \bar b$      & $c, \bar b$       &                  &  0.02 &  0.02 &  0.02 &  0.03 &  0.03 &  0.03 &  0.03 &  0.03 &  0.03 \\
 89 & $u, g$           &                   &                  & 10.90 & 11.96 & 12.73 & 13.58 & 14.15 & 15.28 & 16.41 & 17.65 & 18.83 \\
 90 & $\gamma, g$      &                   &                  & -0.00 & -0.00 & -0.00 & -0.00 & -0.00 & -0.00 & -0.00 & -0.00 & -0.00 \\
 91 & $u, \bar b$      & $u, \bar s$       &                  &  0.08 &  0.10 &  0.12 &  0.11 &  0.11 &  0.12 &  0.13 &  0.11 &  0.11 \\
 92 & $u, s$           & $u, b$            &                  &  0.11 &  0.12 &  0.14 &  0.14 &  0.16 &  0.16 &  0.18 &  0.18 &  0.22 \\
 93 & $u, \bar u$      & $c, \bar c$       &                  &  3.19 &  3.18 &  3.18 &  3.25 &  3.31 &  3.59 &  3.88 &  4.30 &  4.80 \\
 94 & $\gamma, \bar c$ & $\gamma, \bar u$  &                  &  0.01 &  0.01 &  0.01 &  0.01 &  0.01 &  0.01 &  0.01 &  0.01 &  0.01 \\
 95 & $g, \bar c$      & $g, \bar u$       &                  &  0.22 &  0.21 &  0.20 &  0.19 &  0.17 &  0.15 &  0.13 &  0.10 &  0.06 \\
 96 & $u, \gamma$      & $c, \gamma$       &                  &  0.02 &  0.02 &  0.02 &  0.03 &  0.02 &  0.04 &  0.03 &  0.05 &  0.10 \\
 97 & $u, g$           & $c, g$            &                  &  0.68 &  0.79 &  0.88 &  0.95 &  1.00 &  1.10 &  1.17 &  1.25 &  1.25 \\
 98 & $c, g$           &                   &                  &  1.52 &  1.36 &  1.20 &  1.11 &  0.98 &  0.85 &  0.73 &  0.57 &  0.43 \\
 99 & $c, \bar b$      & $c, \bar d$       &                  &  0.01 &  0.01 &  0.01 &  0.01 &  0.01 &  0.01 &  0.01 &  0.01 &  0.01 \\
100 & $c, d$           & $c, b$            &                  &  0.03 &  0.03 &  0.03 &  0.03 &  0.03 &  0.03 &  0.03 &  0.03 &  0.03 \\
101 & $d, g$           &                   &                  &  9.36 &  9.71 &  9.99 & 10.18 & 10.29 & 10.29 & 10.43 & 10.07 &  9.54 \\
102 & $d, \bar b$      & $d, \bar s$       &                  &  0.05 &  0.05 &  0.06 &  0.06 &  0.05 &  0.05 &  0.05 &  0.04 &  0.03 \\
103 & $d, s$           & $d, b$            &                  &  0.03 &  0.03 &  0.03 &  0.03 &  0.04 &  0.03 &  0.03 &  0.03 &  0.03 \\
104 & $d, \bar c$      &                   &                  &  0.02 &  0.02 &  0.02 &  0.02 &  0.02 &  0.02 &  0.02 &  0.02 &  0.02 \\
105 & $d, c$           &                   &                  &  0.02 &  0.03 &  0.02 &  0.02 &  0.03 &  0.03 &  0.03 &  0.02 &  0.02 \\
106 & $d, \bar d$      & $s, \bar s$       &                  &  3.29 &  3.06 &  2.98 &  2.95 &  2.93 &  2.98 &  3.01 &  3.26 &  3.35 \\
107 & $\gamma, \bar s$ & $\gamma, \bar d$  &                  &  0.00 &  0.00 &  0.00 &  0.00 &  0.00 &  0.00 &  0.00 &  0.00 &  0.00 \\
108 & $g, \bar s$      & $g, \bar d$       &                  &  0.36 &  0.35 &  0.33 &  0.32 &  0.30 &  0.27 &  0.22 &  0.17 &  0.11 \\
109 & $d, \gamma$      & $s, \gamma$       &                  &  0.00 &  0.00 &  0.00 &  0.00 &  0.00 &  0.00 &  0.00 &  0.00 &  0.00 \\
110 & $d, g$           & $s, g$            &                  &  0.64 &  0.69 &  0.72 &  0.74 &  0.78 &  0.77 &  0.77 &  0.72 &  0.61 \\
111 & $d, \bar d$      & $s, \bar s$       & $b, \bar b$      &  0.02 &  0.02 &  0.02 &  0.02 &  0.02 &  0.02 &  0.02 &  0.02 &  0.03 \\
112 & $g, \bar b$      & $g, \bar s$       & $g, \bar d$      &  0.00 &  0.00 &  0.00 &  0.00 &  0.00 &  0.00 &  0.00 &  0.00 &  0.00 \\
113 & $d, g$           & $s,  g$           & $b,  g$          &  0.00 &  0.00 &  0.00 &  0.00 &  0.00 &  0.00 &  0.00 &  0.00 &  0.00 \\
114 & $d, \gamma$      & $s, \gamma$       & $b, \gamma$      &  0.01 &  0.01 &  0.01 &  0.01 &  0.01 &  0.01 &  0.01 &  0.01 &  0.02 \\
115 & $s, g$           &                   &                  &  3.21 &  2.93 &  2.72 &  2.55 &  2.37 &  2.08 &  1.79 &  1.57 &  1.27 \\
116 & $s, \bar b$      & $s, \bar d$       &                  &  0.01 &  0.02 &  0.01 &  0.01 &  0.02 &  0.01 &  0.01 &  0.01 &  0.01 \\
117 & $s, d$           & $s,  b$           &                  &  0.01 &  0.02 &  0.02 &  0.03 &  0.02 &  0.03 &  0.03 &  0.03 &  0.03 \\
118 & $s, \bar u$      &                   &                  &  0.01 &  0.01 &  0.01 &  0.01 &  0.01 &  0.01 &  0.01 &  0.01 &  0.00 \\
119 & $s, u$           &                   &                  &  0.06 &  0.08 &  0.07 &  0.08 &  0.09 &  0.10 &  0.11 &  0.13 &  0.16 \\
120 & $b, g$           &                   &                  &  1.45 &  1.30 &  1.16 &  1.05 &  0.94 &  0.80 &  0.67 &  0.47 &  0.25 \\
121 & $b, \bar s$      & $b, \bar d$       &                  &  0.01 &  0.01 &  0.01 &  0.01 &  0.01 &  0.01 &  0.00 &  0.00 &  0.00 \\
122 & $b, d$           & $b,  s$           &                  &  0.01 &  0.01 &  0.01 &  0.01 &  0.01 &  0.01 &  0.01 &  0.01 &  0.00 \\
123 & $b, \bar c$      & $b, \bar u$       &                  &  0.00 &  0.00 &  0.01 &  0.00 &  0.00 &  0.00 &  0.00 &  0.00 &  0.00 \\
124 & $b, u$           & $b,  c$           &                  &  0.03 &  0.03 &  0.04 &  0.03 &  0.04 &  0.04 &  0.04 &  0.04 &  0.03 \\
125 & $\gamma, \bar b$ &                   &                  &  0.00 &  0.00 &  0.00 &  0.00 &  0.00 &  0.00 &  0.00 &  0.00 &  0.00 \\
126 & $b, \gamma$      &                   &                  &  0.00 &  0.00 &  0.00 &  0.00 &  0.00 & -0.00 &  0.00 &  0.00 &  0.00 \\
127 & $\bar u, g$      &                   &                  &  3.26 &  2.97 &  2.79 &  2.58 &  2.38 &  2.10 &  1.79 &  1.42 &  0.85 \\
128 & $\bar u, \bar b$ & $\bar u, \bar s$  &                  &  0.02 &  0.02 &  0.02 &  0.02 &  0.02 &  0.01 &  0.01 &  0.01 &  0.00 \\
129 & $\bar u, s$      & $\bar u, b$       &                  &  0.01 &  0.02 &  0.01 &  0.01 &  0.01 &  0.01 &  0.01 &  0.01 &  0.01 \\
130 & $\bar c, c$      & $\bar u, u$       &                  &  3.20 &  3.20 &  3.19 &  3.22 &  3.32 &  3.57 &  3.82 &  4.33 &  5.01 \\
131 & $\gamma, u$      & $\gamma, c$       &                  &  0.02 &  0.02 &  0.02 &  0.03 &  0.03 &  0.03 &  0.05 &  0.05 &  0.10 \\
132 & $g, u$           & $g,  c$           &                  &  0.68 &  0.79 &  0.86 &  0.94 &  1.00 &  1.11 &  1.19 &  1.24 &  1.23 \\
133 & $\bar c, \gamma$ & $\bar u, \gamma$  &                  &  0.01 &  0.01 &  0.01 &  0.01 &  0.01 &  0.01 &  0.01 &  0.01 &  0.01 \\
134 & $\bar c, g$      & $\bar u, g$       &                  &  0.22 &  0.21 &  0.20 &  0.19 &  0.18 &  0.16 &  0.13 &  0.10 &  0.06 \\
135 & $\bar c, g$      &                   &                  &  1.53 &  1.33 &  1.22 &  1.11 &  1.00 &  0.85 &  0.70 &  0.55 &  0.44 \\
136 & $\bar c, \bar b$ & $\bar c, \bar d$  &                  &  0.01 &  0.01 &  0.01 &  0.01 &  0.01 &  0.01 &  0.01 &  0.01 &  0.01 \\
137 & $\bar c, d$      & $\bar c, b$       &                  &  0.02 &  0.03 &  0.02 &  0.03 &  0.02 &  0.03 &  0.02 &  0.03 &  0.02 \\
138 & $\bar d, g$      &                   &                  &  4.74 &  4.42 &  4.15 &  3.91 &  3.67 &  3.29 &  2.79 &  2.36 &  1.75 \\
139 & $\bar d, \bar b$ & $\bar d, \bar s$  &                  &  0.01 &  0.01 &  0.01 &  0.00 &  0.01 &  0.01 &  0.00 &  0.00 &  0.00 \\
140 & $\bar d, s$      & $\bar d, b$       &                  &  0.02 &  0.02 &  0.02 &  0.02 &  0.02 &  0.02 &  0.01 &  0.01 &  0.01 \\
141 & $\bar d, \bar c$ &                   &                  &  0.01 &  0.01 &  0.01 &  0.01 &  0.01 &  0.01 &  0.01 &  0.00 &  0.01 \\
142 & $\bar d, c$      &                   &                  &  0.00 &  0.01 &  0.01 &  0.00 &  0.01 &  0.00 &  0.00 &  0.00 &  0.00 \\
143 & $\bar s, s$      & $\bar d, d$       &                  &  3.29 &  3.05 &  2.99 &  2.94 &  2.93 &  2.96 &  3.05 &  3.12 &  3.25 \\
144 & $\gamma, d$      & $\gamma, s$       &                  &  0.00 &  0.00 &  0.00 &  0.00 &  0.00 &  0.00 &  0.00 &  0.00 &  0.00 \\
145 & $g, d$           & $g, s$            &                  &  0.64 &  0.69 &  0.74 &  0.74 &  0.77 &  0.77 &  0.76 &  0.72 &  0.62 \\
146 & $\bar s, \gamma$ & $\bar d, \gamma$  &                  &  0.00 &  0.00 &  0.00 &  0.00 &  0.00 &  0.00 &  0.00 &  0.00 &  0.00 \\
147 & $\bar s, g$      & $\bar d, g$       &                  &  0.36 &  0.35 &  0.34 &  0.32 &  0.30 &  0.26 &  0.22 &  0.17 &  0.11 \\
148 & $\bar b,  b$     & $\bar s, s$       & $\bar d, d$      &  0.02 &  0.02 &  0.02 &  0.02 &  0.02 &  0.02 &  0.02 &  0.02 &  0.03 \\
149 & $g,  d$          & $g, s$            & $g, b$           &  0.00 &  0.00 &  0.00 &  0.00 &  0.00 &  0.00 &  0.00 &  0.00 &  0.00 \\
150 & $\bar b, g$      & $\bar s, g$       & $\bar d, g$      &  0.00 &  0.00 &  0.00 &  0.00 &  0.00 &  0.00 &  0.00 &  0.00 &  0.00 \\
151 & $\bar b, \gamma$ & $\bar s, \gamma$  & $\bar d, \gamma$ &  0.00 &  0.00 &  0.00 &  0.00 &  0.00 &  0.00 &  0.00 &  0.00 &  0.00 \\
152 & $\bar s, g$      &                   &                  &  3.02 &  2.72 &  2.45 &  2.29 &  2.04 &  1.73 &  1.38 &  1.04 &  0.51 \\
153 & $\bar s, \bar b$ & $\bar s, \bar d$  &                  &  0.00 &  0.01 &  0.01 &  0.00 &  0.01 &  0.00 &  0.00 &  0.00 &  0.00 \\
154 & $\bar s, d$      & $\bar s, b$       &                  &  0.03 &  0.03 &  0.03 &  0.04 &  0.04 &  0.04 &  0.04 &  0.03 &  0.02 \\
155 & $\bar s, \bar u$ &                   &                  &  0.01 &  0.01 &  0.01 &  0.01 &  0.01 &  0.01 &  0.01 &  0.00 &  0.00 \\
156 & $\bar s, u$      &                   &                  &  0.04 &  0.05 &  0.05 &  0.06 &  0.07 &  0.06 &  0.08 &  0.07 &  0.06 \\
157 & $\bar b, g$      &                   &                  &  1.45 &  1.28 &  1.17 &  1.07 &  0.96 &  0.80 &  0.63 &  0.48 &  0.29 \\
158 & $\bar b, \bar s$ & $\bar b, \bar d$  &                  &  0.00 &  0.00 &  0.00 &  0.01 & -0.00 &  0.00 &  0.00 &  0.00 &  0.00 \\
159 & $\bar b, d$      & $\bar b, s$       &                  &  0.01 &  0.01 &  0.02 &  0.01 &  0.02 &  0.02 &  0.02 &  0.02 &  0.01 \\
160 & $\bar b, \bar c$ & $\bar b, \bar u$  &                  &  0.01 &  0.01 &  0.01 &  0.01 &  0.01 &  0.01 &  0.00 &  0.00 &  0.00 \\
161 & $\bar b, u$      & $\bar b, c$       &                  &  0.02 &  0.03 &  0.02 &  0.03 &  0.02 &  0.03 &  0.04 &  0.02 &  0.03 \\
162 & $\gamma, b$      &                   &                  &  0.00 &  0.00 &  0.00 &  0.00 &  0.00 &  0.00 &  0.00 &  0.00 &  0.00 \\
163 & $\bar b, \gamma$ &                   &                  &  0.00 &  0.00 &  0.00 &  0.00 &  0.00 &  0.00 &  0.00 &  0.00 &  0.00 \\
164 & $\gamma,  d$     & $\gamma, s$       & $\gamma, b$      &  0.01 &  0.01 &  0.01 &  0.01 &  0.01 &  0.01 &  0.01 &  0.01 &  0.02 \\
165 & $\gamma, \bar b$ & $\gamma, \bar s$  & $\gamma, \bar d$ &  0.00 &  0.00 &  0.00 &  0.00 &  0.00 &  0.00 &  0.00 &  0.00 &  0.00 \\
166 & $\gamma, \gamma$ &                   &                  &  0.00 &  0.00 &  0.00 &  0.00 &  0.00 &  0.00 &  0.00 &  0.00 &  0.00 \\
\bottomrule
\end{tabularx}

  \caption{The 166 parton luminosities contributing to the predictions,
    accurate to NLO QCD+EW, to the Z $p_{\mathrm{T}}$ distribution
    measured by the CMS experiment at a centre-of-mass energy of
    \SI{13}{\tera\electronvolt}~\cite{Sirunyan:2019bzr}. The bin edges are:
    20, 22, 26, 28, 32, 37, 43, 52, 65, 85, 120, 160, 190, 220, 250, 300,
    400, 500, 800, 1500~GeV.}
  \label{tab:B10}
\end{table}

In this appendix, we present the complete breakdown of parton luminosities
entering the NLO QCD+EW predictions of the various measurements
considered in section~\ref{sec:results}. For each parton luminosity,
we indicate its percentage contribution to the cross section in a given bin.
In particular:
\begin{itemize}
\item Table~\ref{tab:B1} collects the 35 parton luminosities
  contributing to the Drell--Yan lepton-pair production measured by the
  ATLAS experiment at a centre-of-mass energy of
  \SI{7}{\tera\electronvolt}~\cite{Aad:2013iua}.
  The 13 bins are for the invariant mass of the lepton pair,
  $M_{\ell\bar\ell}$, with edges at 116, 130, 150, 170, 190,
  210, 230, 250, 300, 400, 500, 700, 1000 and 1500 GeV.
\item Tables~\ref{tab:B2}--\ref{tab:B7} collect the 35 parton
  luminosities contributing to the double differential Drell--Yan lepton-pair
  production measured by the CMS experiment at a centre-of-mass energy of
  \SI{7}{\tera\electronvolt}~\cite{Aad:2013iua}. The 132
  bins are split across tables~\ref{tab:B2}--\ref{tab:B7}, where each table
  corresponds to a bin in the invariant mass of the lepton pair:
  table~\ref{tab:B2} to  $\SI{20}{\giga\electronvolt}<M_{\ell\bar\ell}<\SI{30}{\giga\electronvolt}$; table~\ref{tab:B3} to  $\SI{30}{\giga\electronvolt}<M_{\ell\bar\ell}<\SI{45}{\giga\electronvolt}$; table~\ref{tab:B4} to  $\SI{45}{\giga\electronvolt}<M_{\ell\bar\ell}<\SI{60}{\giga\electronvolt}$; table~\ref{tab:B5} to  $\SI{60}{\giga\electronvolt}<M_{\ell\bar\ell}<\SI{120}{\giga\electronvolt}$; table~\ref{tab:B6} to $\SI{120}{\giga\electronvolt}<M_{\ell\bar\ell}<\SI{200}{\giga\electronvolt}$; and table~\ref{tab:B7} to $\SI{120}{\giga\electronvolt}<M_{\ell\bar\ell}<\SI{1500}{\giga\electronvolt}$. In each table, the bins are in the rapidity of the lepton pair, $0.0<y_{\ell\bar\ell}<2.4$, and each of them has a width of 0.1 (0.2 in the largest invariant mass range).
\item Table~\ref{tab:B8} collects the 37 parton luminosities
  contributing to the top-quark pair distributions measured by the ATLAS and
  CMS experiments at a centre-of-mass energy of
  \SI{8}{\tera\electronvolt}~\cite{Aad:2015mbv,Khachatryan:2015oqa}.
  The four distributions, differential in the transverse momentum of the top
  quark, $p_\mathrm{T}^\mathrm{t}$, in the rapidity of the top quark,
  $y_\mathrm{t}$, in the invariant mass of the top-quark pair,
  $m_{\mathrm{t}\bar{\mathrm{t}}}$, and in the rapidity of the top-quark pair,
  $y_{\mathrm{t}\bar{\mathrm{t}}}$ are shown. The bin edges are, respectively:
  0, 60, 100, 150, 200, 260, 320, 400, 500~GeV; 0.0, 0.4, 0.8, 1.2, 1.6, 2.5;
  345, 400, 470, 550, 650, 800, 1100, 1600~GeV;
  and 0.0, 0.3, 0.6, 0.9, 1.3, 2.5.
\item Table~\ref{tab:B10} collects the 166
  parton luminosities contributing to the Z $p_{\mathrm{T}}$ distribution
  measured by the CMS experiment at a centre-of-mass energy of
  \SI{13}{\tera\electronvolt}~\cite{Sirunyan:2019bzr}. The bin edges are:
  20, 22, 26, 28, 32, 37, 43, 52, 65, 85, 120, 160, 190, 220, 250, 300,
  400, 500, 800, 1500~GeV.
\end{itemize}

\section{Low statistics and complementary results}
\label{app:add_plots}

\begin{figure}[!t]
    \centering
    \includegraphics[width=0.46\textwidth]{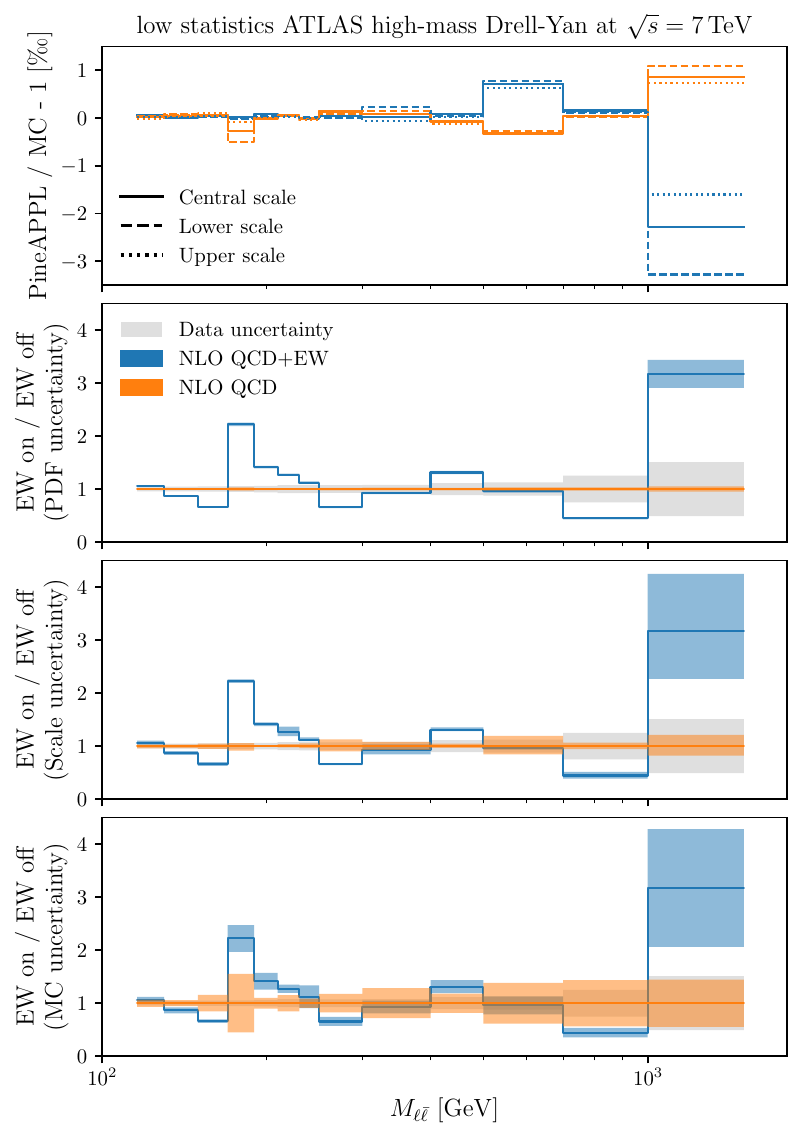}\\
    \caption{Same as figure~\ref{fig:atlaszhighmass49fb}, but for a
    low-statistics MC run.}
    \label{fig:atlaszhighmass49fb-lowstat}
\end{figure}

\begin{figure}[!p]
    \centering
    \includegraphics[width=0.46\textwidth]{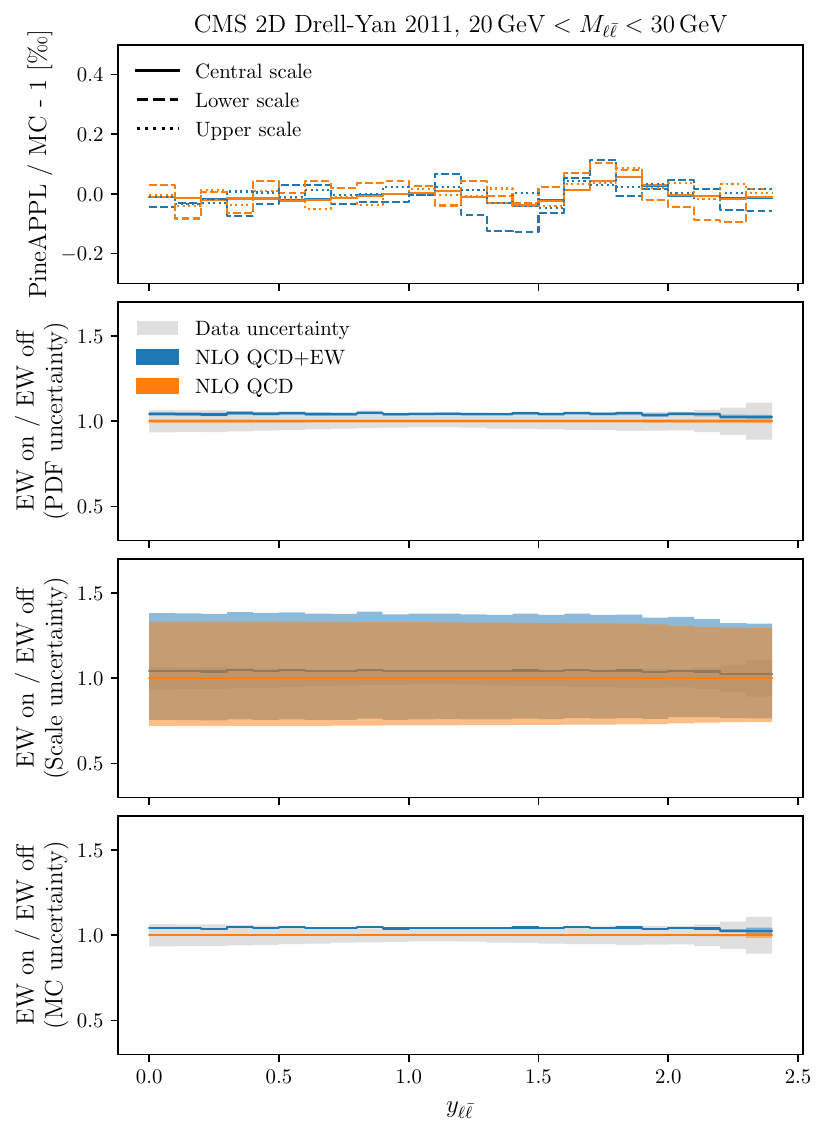}%
    \includegraphics[width=0.46\textwidth]{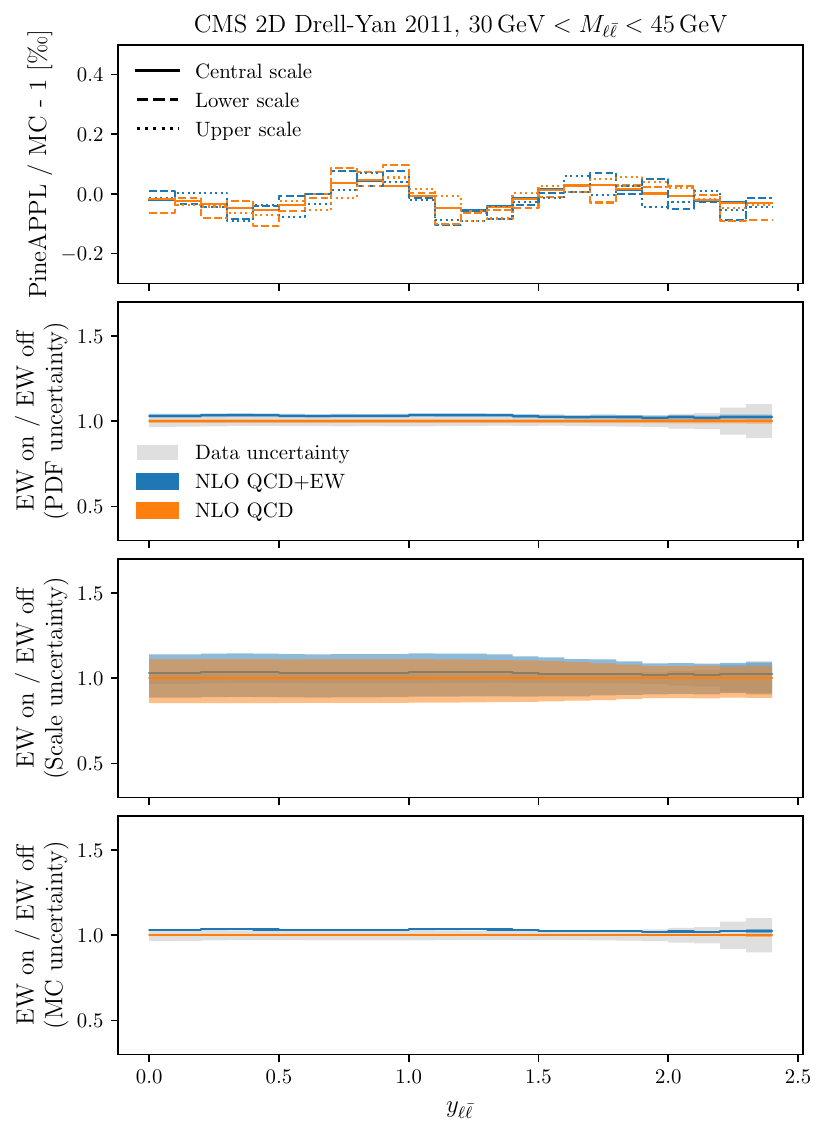}\\
    \caption{Same as figure~\ref{fig:cmsdy2d11_bins3456}, but for the two missing
      lepton-pair invariant mass bins,
      $\SI{20}{\giga\electronvolt}<M_{\ell\bar\ell}<\SI{30}{\giga\electronvolt}$ and
      $\SI{30}{\giga\electronvolt}<M_{\ell\bar\ell}<\SI{45}{\giga\electronvolt}$.}
    \label{fig:cmsdy2d11_bins12}
\end{figure}

\begin{figure}[!p]
    \centering
    \includegraphics[width=0.46\textwidth]{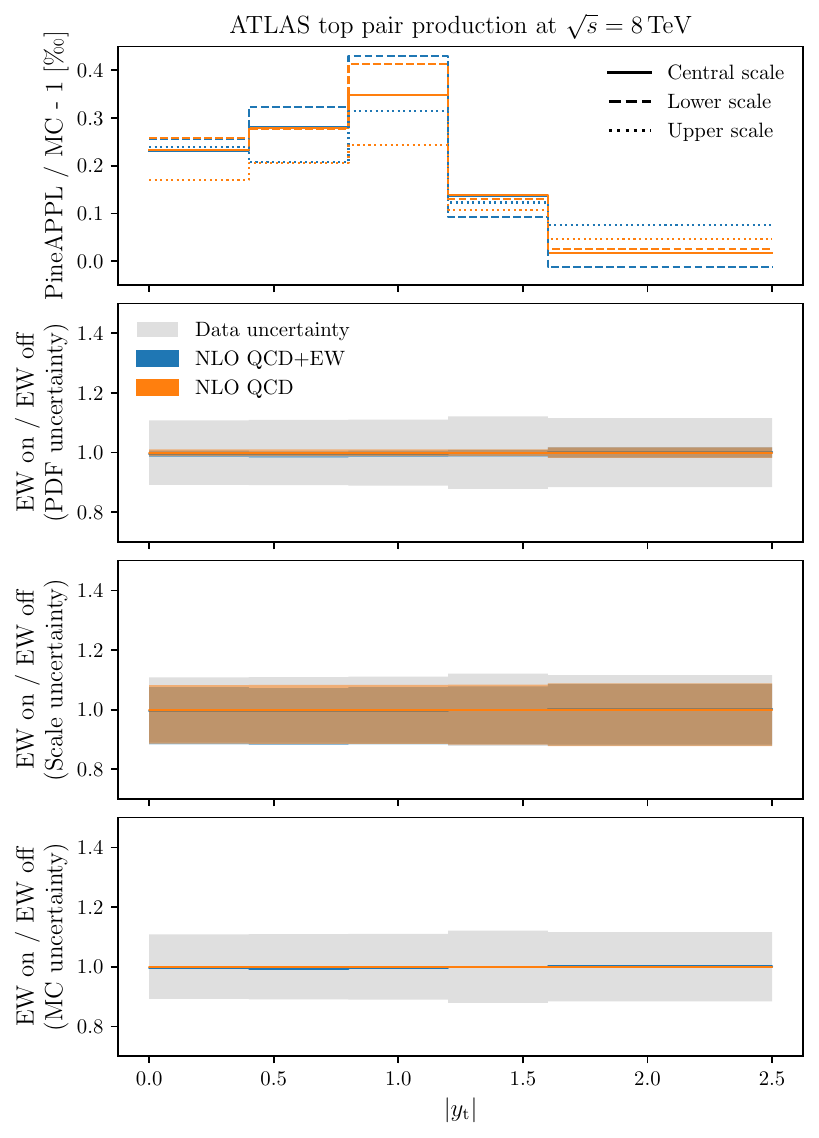}
    \includegraphics[width=0.46\textwidth]{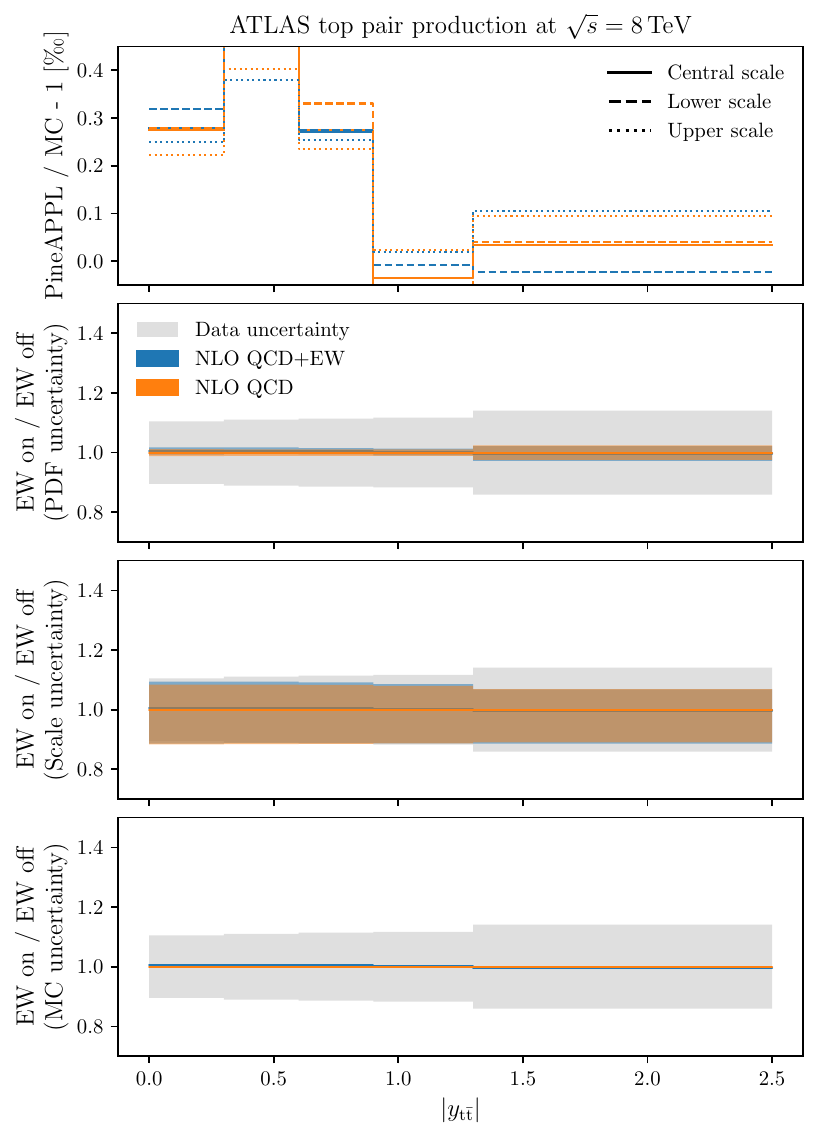}\\
    \caption{Same as figure~\ref{fig:atlastop}, but for the distributions
      in the rapidity of the top quark, $y_{\mathrm{t}}$, and in the rapidity of
      the top-quark pair, $y_{\mathrm{t}\bar{\mathrm{t}}}$.}
    \label{fig:atlastop-rap}
\end{figure}

In this appendix we collect some additional plots, in the same format of those
presented in figures~\ref{fig:atlaszhighmass49fb}--\ref{fig:cmsZ13TeV},
specifically: figure~\ref{fig:atlaszhighmass49fb-lowstat} is the same as
figure~\ref{fig:atlaszhighmass49fb}, but for a low-statistic MC run;
figure~\ref{fig:cmsdy2d11_bins12} is the same as
figure~\ref{fig:cmsdy2d11_bins3456}, but for the two missing lepton-pair
invariant mass bins,
$\SI{20}{\giga\electronvolt}<M_{\ell\bar\ell}<\SI{30}{\giga\electronvolt}$ and
$\SI{30}{\giga\electronvolt}<M_{\ell\bar\ell}<\SI{45}{\giga\electronvolt}$;
and figure~\ref{fig:atlastop-rap} is the same as figure~\ref{fig:atlastop},
but for the distributions in the rapidity of the top quark, $y_{\mathrm{t}}$,
and in the rapidity of the top-quark pair, $y_{\mathrm{t}\bar{\mathrm{t}}}$. In this
case the factorisation and renormalisation scales are kept equal to
$H_\mathrm{T}/4$, see section~\ref{sec:toppair}.

From figure~\ref{fig:atlaszhighmass49fb-lowstat}, we observe that the validation
of the \textsc{PineAPPL} result remains successful: its difference with respect
to the MC result is at most \SI{3}{\permille}, as usual irrespective of the
accuracy of the theory and of the choice of scale. As expected, however, the
result is largely unreliable to make any conclusion about the size of the EW
corrections: large fluctuations are seen in the predictions, and the MC
uncertainty dominates over the PDF and scale uncertainties.

Figure~\ref{fig:cmsdy2d11_bins12} demonstrates that the two bins at the
lowest invariant mass of the Drell--Yan lepton-pair measured by the CMS
experiment display very similar features as the bin at immediately larger
values of invariant mass, see figure~\ref{fig:cmsdy2d11_bins3456}: the EW
correction enhances the cross section by \SIrange{2}{3}{\percent} across all the
rapidity range; the amount of this shift is slightly larger than the PDF
uncertainty, but is largely overshot by the scale uncertainty; the MC
uncertainty remains comparatively negligible.

Finally, figure~\ref{fig:atlastop-rap} allows us to further validate the
\textsc{PineAPPL} result against the MC result for the top-quark
rapidity distributions, and to explicitly check that the size of the EW
corrections in this case is negligible with respect to the companion
top-quark transverse momentum and top-quark pair invariant mass distributions,
see figure~\ref{fig:atlastop}, consistently with what was already observed in
ref.~\cite{Czakon:2017wor}.

\bibliographystyle{JHEP}
\bibliography{paper}

\end{document}